\pdfoutput=1
\documentclass[aps,prd,amsmath,floats,floatfix, twocolumn,
superscriptaddress,showpacs,nofootinbib]{revtex4-2}

\usepackage[T1]{fontenc}
\usepackage[utf8]{inputenc}
\usepackage{lmodern}
\usepackage{verbatim}
\usepackage[caption=false]{subfig}
\usepackage{multirow}
\usepackage[dvipsnames, usenames]{xcolor}
\definecolor{linkcolor}{rgb}{0.0,0.3,0.5}
\usepackage[hypertexnames=false, unicode, colorlinks=true, linkcolor=linkcolor,
citecolor=linkcolor, filecolor=linkcolor,urlcolor=linkcolor,
pdfusetitle]{hyperref}

\usepackage[all]{hypcap}
\usepackage{graphicx}
\usepackage{comment}

\usepackage{xspace}
\usepackage{amssymb}
\usepackage[normalem]{ulem} %for \sout
\usepackage{bm} % boldmath
\usepackage{tensor}
\usepackage{mathrsfs} 
\usepackage{microtype}
\usepackage{tensor} 
\usepackage[english]{babel}
\usepackage{blindtext}

\graphicspath{%
  {figs/}%
}
\newcommand{\etal}{\textit{et al.\ }}

\newcommand{\ind}{{SXS:BBH:0305}}
\newcommand{\indnew}{{SXS:BBH:1936}}

\begin{document}
\title{Quasinormal-mode filters: a new approach to analyze the gravitational-wave ringdown of binary black-hole mergers}

\newcommand\Caltech{\affiliation{TAPIR 350-17, California Institute of
    Technology, 1200 E California Boulevard, Pasadena, CA 91125, USA}}
\newcommand{\Olemiss}{\affiliation{Department of Physics and Astronomy,
    The University of Mississippi, University, MS 38677, USA}}
\newcommand{\Columbia}{\affiliation{Department of Physics and Astronomy, Columbia University, New York, NY 10027, USA}}
\newcommand{\ANU}{\affiliation{OzGrav-ANU, Centre for Gravitational Astrophysics, College of Science,
The Australian National University, ACT 2601, Australia}}
\newcommand{\Cornell}{\affiliation{Cornell Center for Astrophysics and Planetary
     Science, Cornell University, Ithaca, New York 14853, USA}}
\newcommand{\MaxPlanck}{\affiliation{Max Planck Institute for Gravitational
     Physics (Albert Einstein Institute), Am M{\"u}hlenberg 1, D-14476 Potsdam,
     Germany}}

\author{Sizheng Ma}
\email{sma@caltech.edu}
\Caltech

\author{Keefe Mitman}
\Caltech

\author{Ling Sun}
\ANU

% \author{Leo C. Stein}
% \Olemiss

% \author{Macarena Lagos}
% \Columbia

% \author{Lam Hui}
% \Columbia

\author{Nils Deppe} \Caltech
\author{Fran\c{c}ois H\'{e}bert} \Caltech
\author{Lawrence E.~Kidder} \Cornell
\author{Jordan Moxon} \Caltech
\author{William Throwe} \Cornell
\author{Nils L.~Vu} \MaxPlanck

\author{Yanbei Chen}
\Caltech

% Because hyperref only gets the *last* author, we need to be explicit.
\hypersetup{pdfauthor={Ma et al.}}

\date{\today}

%==========================================================================
\begin{abstract}
We propose two frequency-domain filters to analyze ringdown signals of binary black hole mergers. The first rational filter is constructed based on a set of (arbitrary) quasi-normal modes (QNMs) of the remnant black holes, whereas the second full filter comes from the transmissivity of the remnant black holes. The two filters can remove corresponding QNMs from original time-domain ringdowns, while changing early inspiral signals in a trivial way --- merely a time and phase shift. After filtering out dominant QNMs, we can visualize the existence of various subdominant effects. For example, by applying our filters to a GW150914-like numerical relativity (NR) waveform, we find second-order effects in the $(l=4,m=4),(l=5,m=4)$ and $(l=5,m=5)$ harmonics; the spherical-spheroidal mixing mode in the $(l=2,m=2)$ harmonic; and a mixing mode in the $(l=2,m=1)$ harmonic due to a gravitational recoil. In another NR simulation where two component spins are anti-aligned with the orbital angular momentum, we also find retrograde modes. 
% Additionally, we propose to use the rational filter to estimate the start time of a QNM. 
The filters are sensitive to the remnant properties (i.e., mass and spin) and thus have a potential application to future data analyses and parameter estimations. We also investigate the stability of the full filter. Its connection to the instability of QNM spectra is discussed.
\end{abstract}
\maketitle

\section{Introduction}
Ringdown is the final stage of a gravitational wave (GW) signal emitted by a binary black hole (BBH) coalescence. It is associated with the oscillations of the remnant black hole (BH), and contains rich information of the system. With an increasing number of GW events \cite{LIGOScientific:2016dsl,LIGOScientific:2018mvr,LIGOScientific:2020ibl,LIGOScientific:2021djp} observed by ground-based detectors \cite{LIGO2014,Virgo2014,KAGRA}, comprehensive studies of the ringdown signal and its rich features become crucial to understanding the geometry of extreme spacetimes and testing General Relativity (GR). 
%it seems timely and necessary to have a better understanding of the ringdown signal.

A standard description of the ringdown comes from the BH perturbation (BHP) theory. The perturbation of a single BH has been an important topic for decades \cite{Kokkotas:1999bd,Nollert_1999,Cardoso:2016ryw,Berti:2009kk}. GWs emitted by the BH during ringdown are characterized by a set of quasinormal modes (QNMs) \footnote{Except for the late gravitational tail \cite{PhysRevD.5.2419,PhysRevD.5.2439}.}, which are complex and dissipative by their nature. As a consequence, the time-domain evolution of each QNM is a damped sinusoid. Due to the no-hair theorem \cite{penrose2002golden,Chrusciel:2012jk,PhysRevLett.26.331,PhysRev.164.1776}, QNMs predicted by GR are completely determined by the mass and spin of the BH. Therefore, measuring the frequency and decay rate of a QNM from a ringdown signal would allow people to infer the mass and spin of the BH, as pointed out by Echeverria \cite{PhysRevD.40.3194}. This method is dubbed \emph{BH spectroscopy}. The idea was then generalized by Dreyer \etal \cite{Dreyer:2003bv} and Berti \etal \cite{Berti:2005ys,Berti:2007zu}, and they showed that one could test the no-hair theorem if multiple modes are observed at the same time. Later on, a lot of effort has been made to investigate BH spectroscopy under different scenarios \cite{Gossan:2011ha,Caudill:2011kv,Meidam:2014jpa,Bhagwat:2016ntk,Berti:2016lat,Baibhav:2017jhs,Maselli:2017kvl,Yang:2017zxs,DaSilvaCosta:2017njq,Baibhav:2018rfk,Carullo:2018sfu,Brito:2018rfr,Nakano:2018vay,Cabero:2019zyt,Bhagwat:2019bwv,Ota:2019bzl,Bustillo:2020buq,JimenezForteza:2020cve,Isi:2021iql,Finch:2021qph}. In particular, the studies by Cardoso \etal \cite{Cardoso:2016ryw,Cardoso:2017njb,Cardoso:2016rao}, Foit \etal \cite{Foit:2016uxn} and Laghi \etal \cite{Laghi:2020rgl} implied that QNMs could reflect the quantum nature of BHs or other exotic compact objects (ECOs); hence one can use this fact to test GR and constrain modified gravity \cite{Blazquez-Salcedo:2016enn,Glampedakis:2017dvb,Silva:2022srr,Maselli:2019mjd}. Since the detection of GW150914 \cite{LIGOScientific:2016aoc}, BH spectroscopy with real observational data has become available. Carullo \etal \cite{Carullo:2019flw} studied the late-time portion of the ringdown of GW150914 and found no evidence for the presence of more than one QNM. Then Isi \etal \cite{Isi:2019aib} extended the analysis to the peak of the strain and showed evidence of at least one overtone, with $3.6\sigma$ confidence. This led to a test of the no-hair theorem at the $\sim20\%$ level. Recently, Cotesta \etal \cite{Cotesta:2022pci} raised an opposing viewpoint that the search for the first overtone in the ringdown of GW150914 might be impacted by noises, therefore the conclusion still remains controversial \cite{Isi:2022mhy,Finch:2022ynt}. On the other hand, Capano \etal \cite{Capano:2021etf} studied the QNM spectrum of GW190521 \cite{LIGOScientific:2020iuh} and found the $l=m=3$ harmonic. More GW events were used to perform BH spectroscopy in Refs.~\cite{LIGOScientific:2020tif,LIGOScientific:2021sio,Ghosh:2021mrv}.

The inspiral-merger-ringdown (IMR) consistency test is another important extension of BH spectroscopy. One can infer the properties of binaries separately from the inspiral waves and the ringdown waves, and check whether they are consistent with the predictions of GR. The idea was proposed originally by Hughes \etal \cite{Hughes:2004vw}, and more careful analyses were carried out later \cite{Luna:2006gw,Nakano:2015uja,Ghosh:2017gfp,Ghosh:2016qgn}. So far, no deviation from GR has been found in observational data \cite{LIGOScientific:2016lio,LIGOScientific:2017bnn,LIGOScientific:2019fpa,Breschi:2019wki}. In addition, Refs.~\cite{Cabero:2017avf,Isi:2020tac} used this method to test Hawking's area law \cite{PhysRevLett.26.1344}.

An essential ingredient for BH spectroscopy is to understand how QNMs are excited at merger \cite{PhysRevD.38.1040,PhysRevD.41.2492,Dorband:2006gg,Berti:2006wq,Hadar:2011vj,Hadar:2009ip,Kamaretsos:2011um,Zhang:2013ksa,London:2018gaq,Hughes:2019zmt,Apte:2019txp,Lim:2022veo,Lim:2019xrb,Cano:2021myl} and when the ringdown starts \cite{Bhagwat:2017tkm,Bhagwat:2019dtm,Okounkova:2020vwu}. An accurate investigation for a BBH system during a highly nonlinear regime was not available until the numerical relativity (NR) breakthrough was made in 2005 by Pretorius \cite{Pretorius:2005gq}. Since then, a usual method to study the ringdown of a numerical waveform has been fitting it to the prediction of BHP theory. For example, Buonanno \etal \cite{Buonanno:2006ui} decomposed the ringdown signal into a sum of the fundamental mode and several overtones. Berti \etal \cite{Berti:2007fi,Berti:2007dg} and Kamaretsos \etal \cite{Kamaretsos:2011um} fit the ringdown of unequal-mass, nonspinning systems with only the fundamental mode. London \etal \cite{London:2014cma} carried out a more systematic study for various nonspinning BBHs and identified overtones within the NR waveforms. On the other hand, the fitting was also an important step to calibrate the effective one-body model \cite{Taracchini:2012ig,Pan:2011gk,Taracchini:2013rva,Damour:2014yha,Taracchini:2014zpa}. Later, given the motivation of BH spectroscopy with real observational data, Thrane \etal \cite{Thrane:2017lqn} fit the ringdown of a GW150914-like NR simulation without any overtone, and they found some inconsistency between the QNM model (with fundamental modes only) and NR waveform. This puzzle was resolved by Giesler \etal \cite{Giesler:2019uxc}, where the authors found that the inclusion of overtones could extend the linear regime to the peak strain amplitude. This work sparked another wave for ringdown modeling, including the study for multimode ringdown fitting \cite{Cook:2020otn}, and the impacts of other effects on ringdown signals, such as retrograde modes \footnote{The author of Ref.~\cite{Dhani:2020nik} used the name ``mirror mode'' instead. In this work we will always use ``retrograde mode''.} \cite{Dhani:2020nik}, more overtones \cite{Forteza:2021wfq}, precessing systems \cite{Finch:2021iip}, angular emission patterns \cite{Li:2021wgz}, and the Bondi-van der Burg-Metzner-Sachs freedom \cite{MaganaZertuche:2021syq}.

It is surprising to see that the linear BHP theory is good enough to explain the waveform beyond the peak of the strain, given that the dynamics at the merger are believed to still be violent. Okounkova \cite{Okounkova:2020vwu} provided a possible explanation based on previous Kerrness tests \cite{Bhagwat:2017tkm}: most of the near-zone nonlinearities \footnote{Here we do not consider the wave-zone nonlinearities, say the memory effect, which has been obtained from NR \cite{Mitman:2020pbt,Mitman:2020bjf}.} are absorbed by the event horizon and barely escape to infinity. Nonetheless, it still seems elusive to draw an incontrovertible conclusion, since recent studies \cite{Pook-Kolb:2020jlr,Mourier:2020mwa} showed that multipole moments of dynamical horizon are also compatible with the superposition of linear QNMs soon after the formation of the common horizon. Furthermore, it was shown that applying second-order BHP theory to the close-limit approximation could improve the agreement between the ringdown model and the full numerical waveform --- the improvement was not only limited to the regime near the peak, but also extended to the late portion of the ringdown signal \cite{Gleiser:1996yc}. Then it is natural to ask: \emph{where are the second-order effects?} In the past, the second-order perturbation of a Schwarzschild BH was used by  Tomita \etal \cite{tomita1976nonlinear,tomita1974non} in the process of a gravitational collapse to investigate the stability of the horizon. Cunningham \etal \cite{1980ApJ...236..674C} treated the spin as a small perturbation during the Oppenheimer-Snyder collapse and studied its second-order effect. Later on, second-order perturbation theory was motivated by the close-limit approximation \cite{Price:1994pm}, including the metric perturbation of a Schwarzschild BH \cite{Gleiser:1995gx,Gleiser:1996yc,Gleiser:1998rw,Nicasio:2000ge,Brizuela:2006ne,Brizuela:2007zza,Brizuela:2009qd} and the perturbation of a Kerr BH within the Newman-Penrose formalism \cite{Campanelli:1998jv}. Recently, more comprehensive treatments were used to deal with the perturbation of a Kerr BH and its metric reconstruction \cite{Green:2019nam,Loutrel:2020wbw,Ripley:2020xby}. An important feature of second-order BHP theory is that the master equation has the same potential as the first-order theory, while the source term is quadratic in terms of the first-order perturbations. Accordingly, the time evolution of the second-order perturbations can be influenced by the second-order QNMs, known as ``sum tones'' and ``difference tones'' \cite{Zlochower:2003yh,Ioka:2007ak,Nakano:2007cj,Okuzumi:2008ej,Pazos:2010xf}. For instance, Nakano \etal \cite{Nakano:2007cj} found the existence of a component twice the $(l=2,m=2)$ QNM in the $(l=4,m=4)$ harmonic by looking at a perturbed Schwarzschild BH. So far, very few studies have been done on the second-order effects within the ringdown of a BBH waveform. London \etal \cite{London:2014cma} investigated 68 NR waveforms and presented the evidence of the second-order mode $(l_1, m_1, n_1)\times(l_2, m_2, n_2)=(2,2,0)\times(2,2,0)$ in the $(l=4,m=4)$ harmonic via time-domain fitting. Beyond the second-order effect, Sberna \etal \cite{Sberna:2021eui} showed that the growth of BH mass due to the absorption of the linear QNMs can induce a third-order secular effect.

The time-domain fitting proves to be powerful to extract the physics from ringdown signals. However, one always has to be careful of overfitting --- more QNMs included (e.g., overtones or retrograde modes) may act as additional basis functions to misinterpret other effects. Taking this caveat into consideration, in this paper we propose a complementary tool to analyze a ringdown waveform --- we define two frequency-domain filters that are able to remove any particular QNM from the ringdown. After the dominant mode is filtered out, we can visualize the existence of subdominant effects, including mode mixing, second-order modes, and retrograde modes.

This paper is organized as follows. In Sec.~\ref{sec:filter_definition}, we introduce two types of filters and show their properties. Then in Sec.~\ref{sec:applications}, we apply the two filters to NR waveforms and discuss the results. Section \ref{sec:stability} focuses on the stability of the filter under perturbations. Next, in Sec.~\ref{sec:remnant_depedence}, we discuss how the filter depends on the remnant BH's mass and spin. We also investigate the possibility to use the filter for parameter estimation. Finally, we summarize the results in Sec.~\ref{sec:conclusion}.

Throughout this paper, we use the geometric units with $G = c = 1$. We always use the notation $\omega_{lmn}$ to refer to the $(l,m,n)$ QNM.

% \B{Spheroidal Leakage }

% \B{manifest}

% \B{PN}

% \B{associated with the dominate angular mode $(l=2,m=2)$}

% \B{nonlinear effects: see the introduction in \cite{Brizuela:2009qd}}

% \B{rational filter and full filter}

\section{QNM Filters}
\label{sec:filter_definition}
In this section, we introduce two types of filters for QNMs. In Sec.~\ref{subsec:QNM} we first review briefly the QNM decomposition model of a ringdown signal. Then in Sec.~\ref{sec:first_filter_toy_model} we describe a rational filter, which can remove any particular QNM from a ringdown signal. Two toy models are used to explore the effect of this filter. After understanding the effects of the rational filter, in Sec.~\ref{sec:second_filter_Wronskian} we argue that the inverse of the remnant BH's transmissivity can also serve as a filter. Remarkably, we find that the waveform filtered by this filter has a physical meaning.

\subsection{Decomposing late waveforms into QNMs}
\label{subsec:QNM}

It has been widely accepted that the late part of the GW emitted by binary black-hole mergers can be described as a linear combination of QNMs and a power-law tail, which arise from different features of the retarded gravitational Green's function $G_{lm}$: the QNMs correspond to poles of $G_{lm}$, while the power-law tail arises from integrating along a branch-cut $G_{lm}$ \cite{PhysRevD.34.384}.  
% For particles plunging into a black hole,

In the special case of a high-mass-ratio merger, which can be modeled as an orbiting and then plunging particle, QNM excitations at late times have been computed~\cite{Hadar:2011vj,Hadar:2009ip,Zhang:2013ksa} and further analyzed in terms of multipole and overtone excitations~\cite{Hughes:2019zmt,Oshita:2021iyn,Lim:2019xrb,Lim:2022veo}. As an example, in linear perturbation theory, the gravitational waveform at infinity sourced by the particle can be described by 
\begin{equation}
h(t,r_*) =  \sum_{lm}\int \frac{d\omega }{2\pi} e^{-i\omega t} \int dy  G_{lm}(r_*,y,\omega) S_{lm}(y,\omega),
\end{equation}
where $S_{lm}(y,\omega)$ is the source term, and it has the general form of \begin{equation}
    S_{lm}(y,\omega) = e^{i\omega T(y)}P(\omega,y)
\end{equation}
where $(T(y),y)$ parametrizes the radial trajectory of the particle, and $P(\omega,y)$ is a rational function of $\omega$.  For each $y$, as long at $t>T(y)$ one can close the $\omega$-contour from the lower-half complex plane, hence only collect the poles of the Green's function $G_{lm}$ and a branch-cut contribution which corresponds to power-law tails. Even though the particle's $T(y)$ becomes infinity for $y\rightarrow -\infty$, the source term $P(\omega,y)$ exponentially decays to zero, soon after $y$ becomes negative, i.e., when the particle plunges across the light ring and approaches the horizon.   
%This relies on the property that $S_{lm}$ will no longer source GWs after the particle crosses the light ring and gets close to the horizon.

Gravitational waveforms from collapsing stars and merging {\it comparable-mass} BHs were argued to have similar late-stage properties~\cite{1978ApJ...224..643C,1979ApJ...230..870C,1980ApJ...236..674C,PhysRevD.40.3194,Nichols:2010qi,Nichols:2011ih,flanagan1998measuring}. The regime of QNM decomposition is often referred to as the ``linear regime'', although the decomposition requires both linearity and homogeneity (i.e., the QNMs are homogeneous solutions to the linearized Einstein's equations). 

Now assuming that a ringdown signal $h(\theta,\phi,t)$ is a linear combination of QNMs, starting from $t_0$, namely
% our subsequent discussion will be based primarily on Refs.~\cite{Li:2021wgz,Lim:2019xrb}, and we refer the interested reader to these two papers for more comprehensive reviews. Starting from a time $t_0$,  then it can be decomposed into a sum over a set of QNMs \cite{Li:2021wgz,Lim:2019xrb}
\begin{align}
    &h(\theta,\phi,t)=(h_+-ih_\times)(\theta,\phi,t) \notag \\
    &=\sum_{kmn}\left[A_{kmn}e^{-i\omega_{kmn}(t-t_0)}\tensor[_{-2}]{S}{_{kmn}}(a\omega_{kmn},\theta,\phi)\right. \notag \\
    &\left.+A^\prime_{kmn}e^{i\omega^*_{kmn}(t-t_0)}\tensor[_{-2}]{{S^{*}_{kmn}}}{}(a\omega_{kmn},\pi-\theta,\phi)\right], \label{eq:spheroidal_decom}
\end{align}
where $\times,+$ refer two polarization states of the GW, $a$ is the dimensional spin of the BH, $\omega_{kmn}$ are the frequencies of QNMs, $\tensor[_{-2}]{S}{_{kmn}}(a\omega_{kmn},\theta,\phi)$ are the spin-weighted spheroidal harmonics \cite{Berti:2005gp}, and $(A_{kmn},A^\prime_{kmn})$ are the mode amplitudes. It is usually more convenient to decompose the waveform in terms of spin-weighted spherical harmonics $\tensor[_{-2}]{Y}{_{lm}}(\theta,\phi)$
\begin{align}
    h(\theta,\phi,t)=\sum_{lm}h_{lm}(t)\tensor[_{-2}]{Y}{_{lm}}(\theta,\phi). \label{eq:spherical_decom}
\end{align}
with $h_{lm}$ being the $(l,m)$ spherical multipole harmonic. The mode mixing between the two bases: $\tensor[_{-2}]{S}{_{kmn}}(a\omega_{lmn},\theta,\phi)$ and $\tensor[_{-2}]{Y}{_{lm}}(\theta,\phi)$, is given by \cite{Berti:2014fga,London:2018nxs}
\begin{align}
    \tensor[_{-2}]{S}{_{kmn}}(a\omega_{lmn},\theta,\phi)=\sum_{l}\mu_{mlkn}^*(a\omega_{lmn})\tensor[_{-2}]{Y}{_{lm}}(\theta,\phi). \label{eq:ss_mode_mixing}
\end{align}
By combining Eqs.~\eqref{eq:spheroidal_decom}, \eqref{eq:spherical_decom} and \eqref{eq:ss_mode_mixing}, we obtain the QNM decomposition model for $h_{lm}$:
\begin{align}
    h_{lm}=\sum_{k,n} &\left[ C_{mlkn}e^{-i\omega_{kmn}(t-t_0)} +C^\prime_{mlkn}e^{i\omega^*_{k-mn}(t-t_0)}\right]. \label{h_lm_decom_ss}
\end{align}
Explicit relations between $C_{mlkn}$ and $A_{kmn}$ [Eq.~\eqref{eq:spheroidal_decom}] can be found in Ref.~\cite{Lim:2019xrb}. Note that the second term in Eq.~\eqref{h_lm_decom_ss} corresponds to the retrograde modes, which are also dubbed ``mirror modes'' in Refs.~\cite{Dhani:2020nik,Dhani:2021vac}

\begin{figure}[htb]
        \includegraphics[width=\columnwidth,clip=true]{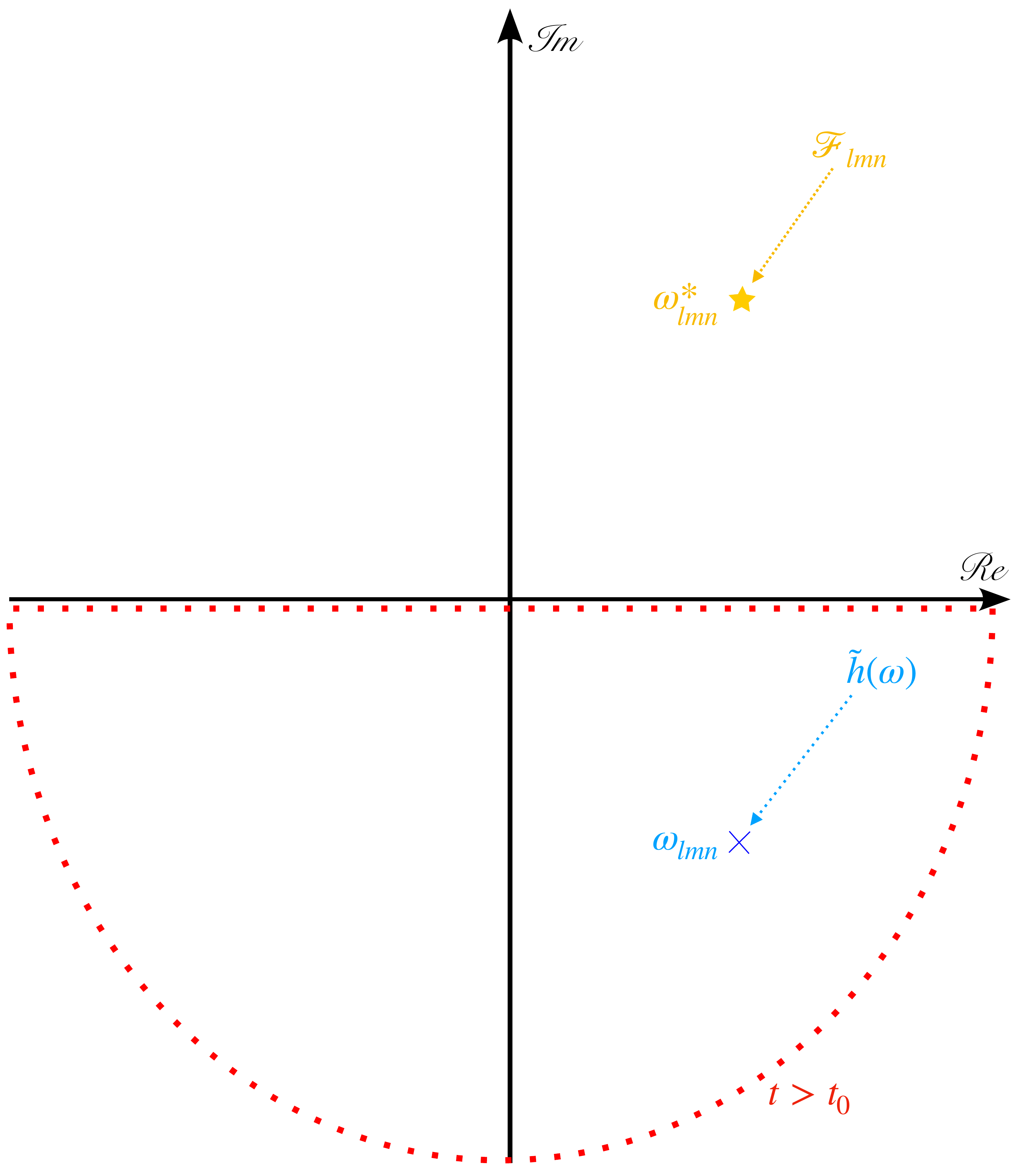}
  \caption{The pole of the original waveform $\tilde{h}(\omega)$ (in blue) and the filtered one $\tilde{h}^{\rm filter}(\omega)$ (in orange). The contour is closed from the upper (lower) plane when $t<t_0$ ($t>t_0$). }
 \label{fig:filter_poles}
\end{figure}

\subsection{The rational filter and two toy models}
\label{sec:first_filter_toy_model}
For simplicity's sake we consider a single QNM signal in the time domain:
\begin{align}
    h(t)=e^{-i\omega_{lmn} (t-t_0)}\Theta(t-t_0), \label{h_t_QNM}
\end{align}
where $\omega_{lmn}$ is the complex frequency of a specific QNM, $\Theta(t-t_0)$ is the Heaviside step function, and $t_0$ refers to the start time of the mode. If we are interested in the regime of $t>t_0$ and want to annihilate the mode content $\omega_{lmn}$ therein, a natural choice is to use a time-domain operator
\begin{align}
    \left(\frac{d}{dt}+i\omega_{lmn}\right)h(t)=\delta(t-t_0), \label{annihilation_time}
\end{align}
with $\delta(t-t_0)$ being the Dirac function. However, this operation can lead to additional numerical noises. Instead, we first transform the signal $h(t)$ in Eq.~\eqref{h_t_QNM} to the frequency domain
\begin{align}
    \tilde{h}(\omega)=\frac{1}{\sqrt{2\pi}}\int h(t)e^{i\omega t}dt,
\end{align}
and obtain
\begin{align}
    \tilde{h}(\omega)=\frac{i}{\sqrt{2\pi}} \frac{e^{i\omega t_0}}{\omega-\omega_{lmn}}.
\end{align}
Then we define a frequency-domain filter $\mathcal{F}_{lmn}$:
\begin{align}
    \mathcal{F}_{lmn}=\frac{\omega-\omega_{lmn}}{\omega-\omega^*_{lmn}}, \label{filter_single}
\end{align}
where $*$ represents the complex conjugate. We remark that the numerator of $\mathcal{F}_{lmn}$ corresponds to the annihilation operator [Eq.~(\ref{annihilation_time})] in the frequency domain, while the denominator is introduced to make $|\mathcal{F}_{lmn}|=1$ (when $\omega$ is real-valued) and therefore ensure that the filter does not diverge at high frequency. Finally, we impose the filter via
\begin{align}
    \tilde{h}^{\rm filter}(\omega)=\mathcal{F}_{lmn}\tilde{h}(\omega),
\end{align}
and transform the filtered signal to the time domain again
\begin{align}
    h^{\rm filter}(t)=\frac{1}{\sqrt{2\pi}}\int\tilde{h}^{\rm filter}(\omega)e^{-i\omega t}d\omega,
\end{align}
which yields
\begin{align}
    h^{\rm filter}(t)=-e^{-i\omega_{lmn}^* (t-t_0)}\Theta(t_0-t).
\end{align}
Notice that the sign of the argument in the Heaviside step function $\Theta$ has changed. This can be understood in terms of the impact of the filter on the pole of the waveform, as shown in Fig.~\ref{fig:filter_poles}. The pole of the original waveform $\tilde{h}(\omega)$ (in blue) lies in the lower half plane, implying its excitation after $t_0$. After imposing the rational filter $\mathcal{F}_{lmn}$, the pole is lifted to the upper panel (in orange). Therefore, the filtered waveform becomes a ring-up signal prior to $t_0$, whereas the original ringdown is removed after that moment.

\begin{figure}[htb]
        \includegraphics[width=\columnwidth,clip=true]{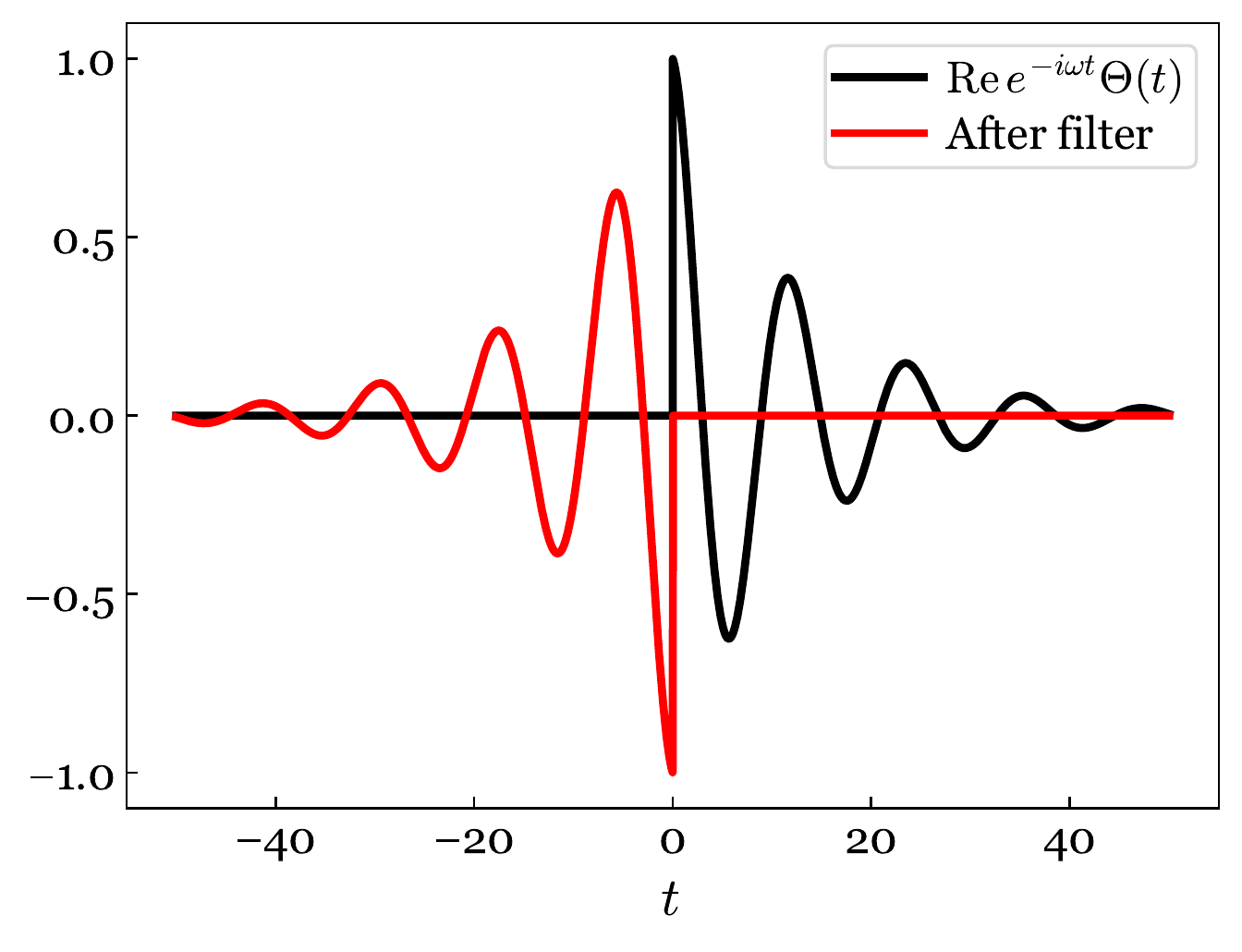}
  \caption{The effect of the frequency-domain filter in Eq.~\eqref{filter_single} on a single QNM signal. The mode is chosen to be the fundamental $(l=2,m=2)$ QNM of a Kerr BH with dimensionless spin 0.69. The signal starts at $t=0$, and it is padded with 0 for $t<0$. After applying the filter, the original signal (its real part is shown as the black curve) is removed from the regime of interest $(t>0)$, whereas an undesired ``flipped ringdown'' is introduced for $t<0$ (red curve). This ``flipped ringdown'' resembles the original signal, but decays backward in time. }
 \label{fig:effect_of_filter}
\end{figure}

To be specific, we consider a toy model in Fig.~\ref{fig:effect_of_filter} to illustrate the effect of the filter. We pick the fundamental $(l=2,m=2)$ QNM of a Kerr BH with dimensionless spin $\chi=0.69$. The QNM frequencies are obtained from the PYTHON package $\textsf{qnm}$ \cite{Stein:2019mop}. The start time $t_0$ is set to 0. Indeed, we can see that within our interested regime $t>t_0$, the filter is able to remove the mode content $\omega_{lmn}$ completely. Meanwhile, the ring-up signal (``flipped ringdown'') is introduced before $t_0$. As we will see, this feature can contaminate GWs at merger, but it will not affect our analysis as long as we focus on the regime $t>t_0$. As for an early, low-frequency inspiral signal, since its frequency $\omega$ is small compared to $\omega_{lmn}$,  we can perform a Taylor expansion around $\omega=0$
\begin{align}
    \mathcal{F}_{lmn}=\exp [-i\phi_{lmn}-i\omega t_{lmn}+\,\mathcal{O}(\omega^2)],
\end{align}
where the two real constants $t_{lmn}$ and $\phi_{lmn}$ are given by
% \begin{subequations}
% 
% \begin{align}
% &\phi_{lmn}(\omega')=\frac{2\text{Im}\left[\omega_{lmn}\right]}{|\omega'-\omega_{lmn}|^{2}}\omega'+i\ln\left(\frac{\omega'-\omega_{lmn}}{\omega^\prime-\omega_{lmn}^*}\right), \\
% &t_{lmn}(\omega')=-\frac{2\text{Im}\left[\omega_{lmn}\right]}{|\omega'-\omega_{lmn}|^{2}}, \label{time_shift_ana}
% \end{align}
% \end{subequations}
\begin{align}
    &\phi_{lmn}=-2\tan^{-1}\frac{\omega_{lmn}^{\rm i}}{\omega_{lmn}^{\rm r}},
    &t_{lmn}=-\frac{2\omega_{lmn}^{\rm i}}{|\omega_{lmn}|^2}, \label{phase_time_shift_ana}
\end{align}
with $\omega_{lmn}^{\rm r}$ and $\omega_{lmn}^{\rm i}$ being the real and imaginary part of $\omega_{lmn}=\omega_{lmn}^{\rm r}+i\omega_{lmn}^{\rm i}$, respectively. Consequently, imposing the filter $\mathcal{F}_{lmn}$ to the low-frequency inspiral signal is equivalent to shifting the original signal in phase and backward in time\footnote{Strictly speaking, Eq.~\eqref{phase_time_shift_ana} is for zero frequency components. An accurate estimation for other low frequencies is not needed in this paper.}. For a Kerr BH with $\chi=0.69$, the $(l=2,m=2)$ fundamental mode leads to $t_{lmn}\sim0.57M_f$, which can be neglected for most of ringdown analyses. 
% \ls{It's a bit unclear in what sense this effect can be neglected. This seems irrelevant as long as we only look at ringdown. Is there a reason/scenario where we need to look at the inspiral with the filter applied? Also, how about the phase shift?} \ls{I mean, the preceding sentence says that the time shift can be neglected for \emph{ringdown} analysis. Given that it only impacts $t<0$, it should not matter for ringdown analysis anyway no matter how large the time shift would be? Or maybe just remove ``ringdown'' from the preceding sentence?}\sma{Yes, that's true. But we always need to ensure we use the same start time for analysis. The waveform is shifted by the filter so it's difficult to say where the peak is.}\kmcomment{slightly disconnected from this discussion, but do we want to explain that the time/phase shift in Eq. (7) is only for the zero-frequency part of ringdown? really the time/phase shift should be
% \begin{align}
% \phi_{Q}(\omega')=\frac{2\text{Im}\left[\omega_{Q}\right]}{|\omega'-\omega_{Q}|^{2}}\omega'+i\log\left(\frac{\omega'-\omega_{Q}}{\omega-\overline{\omega_{Q}}}\right)
% \end{align}
% \begin{align}
% t_{Q}(\omega')=\frac{2\text{Im}\left[\omega_{Q}\right]}{|\omega'-\omega_{Q}|^{2}}.
% \end{align}
% where $\omega'$ is the frequency of the inspiral phase}\sma{I've added a footnote to clarify this.}\kmcomment{We should also make sure to say when we're applying a time/phase shift and when we're not.} \sma{done}
However, if we want to remove a series of QNMs, we need to apply:
\begin{align}
    \mathcal{F}_{\rm tot}=\prod_{lmn}\mathcal{F}_{lmn}, \label{filter_final}
\end{align}
where $n$ stands for the overtone index. Then the time shift $t_{lmn}$ may not be negligible anymore.

\begin{figure}[htb]
  \includegraphics[width=\columnwidth,clip=true]{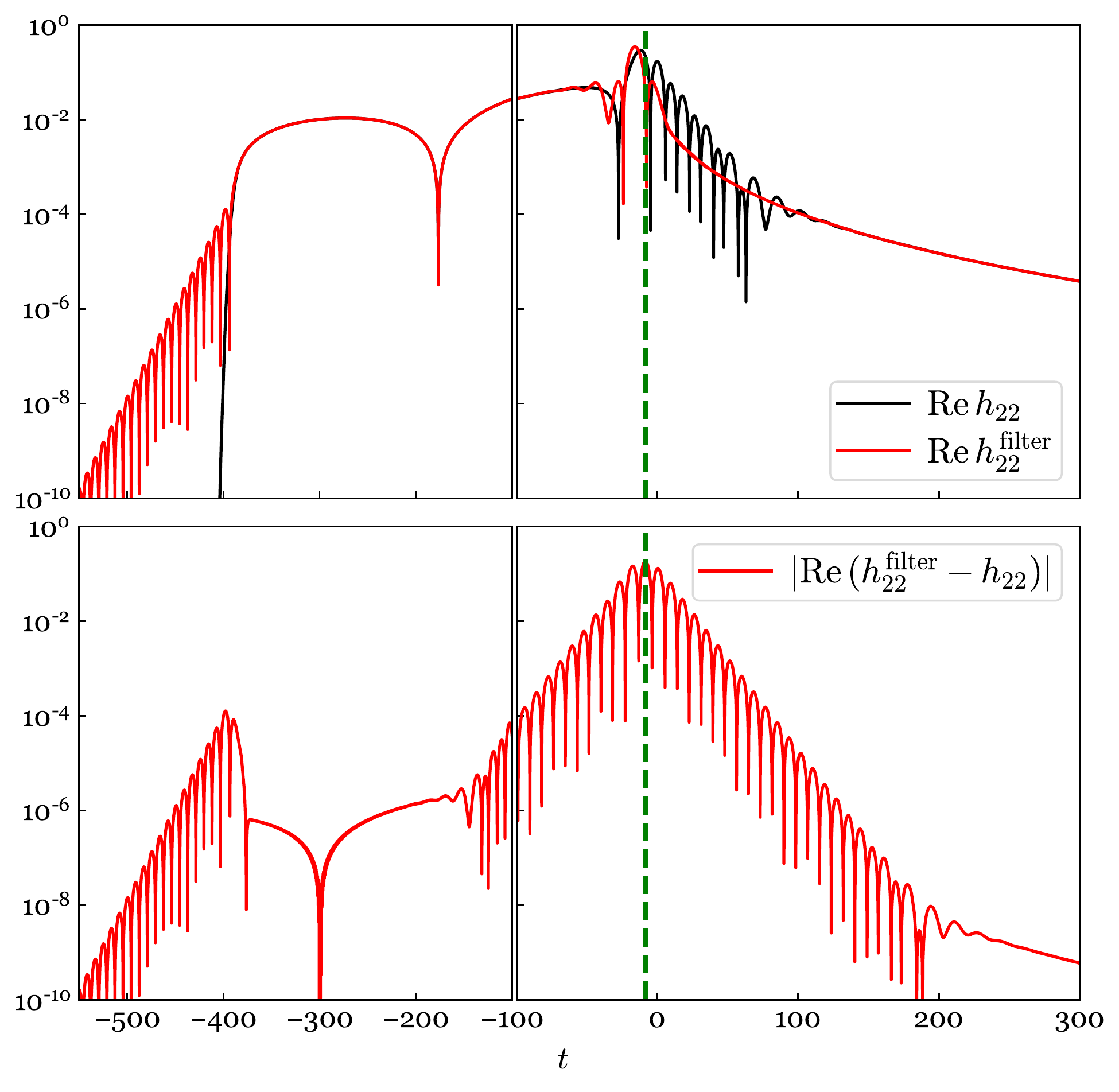}
  \caption{The impact of the filter $\mathcal{F}_{220}$ on the GW emitted by a single, perturbed Schwarzschild BH. In the upper panel, the real part of the filtered waveform (red curve) is compared with the original $h_{22}$ (black curve). Note that here we have undone the time shift induced by the filter by aligning two waveforms in the early regime. In the lower panel, the difference between the two waveforms corresponds to the combination of the ``flipped ringdown'' and the real ringdown (see the black and red curves in Fig.~\ref{fig:effect_of_filter}). Its peak (the vertical dashed line) represents the start time of the ringdown.}
 \label{fig:RWZ}
\end{figure}

We then switch our attention to a more realistic case: a Schwarzschild BH perturbed by an even-parity Gaussian pulse. The Zerilli equation \cite{1969PhDT........13Z} is solved numerically. Figure \ref{fig:RWZ} shows the waveform $h_{22}$ at future null infinity. We see $h_{22}$ (the black curve) consists of the excitation, ringdown, and tail regime. After applying the filter $\mathcal{F}_{220}$ (the red curve), the ringdown oscillations are completely removed from the tail beyond a certain time around the merger, yet a few wiggles appear prior to that time. This is due to the nonphysical ``flipped ringdown'' (see the red curve in Fig.~\ref{fig:effect_of_filter}). The difference between the original $h_{22}$ and the filtered waveform, as shown in the lower panel of Fig.~\ref{fig:RWZ}, corresponds to the combination of the ``flipped ringdown'' and the real ringdown (namely the combination of the black and red curves in Fig.~\ref{fig:effect_of_filter}). Note that here we have undone the time shift induced by the filter by aligning two waveforms in the early regime. The peak of the difference (the vertical green dashed line) represents the start time of the ringdown $t_0$ [see Eq.~\eqref{h_t_QNM}]. In addition, we see a generic feature: a new damped sinusoid that decays backward in time shows up before the onset of the original signal. For a BBH waveform, it appears before the entire inspiral regime, thus it does not impact our analysis.

\begin{figure}[htb]
        \includegraphics[width=\columnwidth,clip=true]{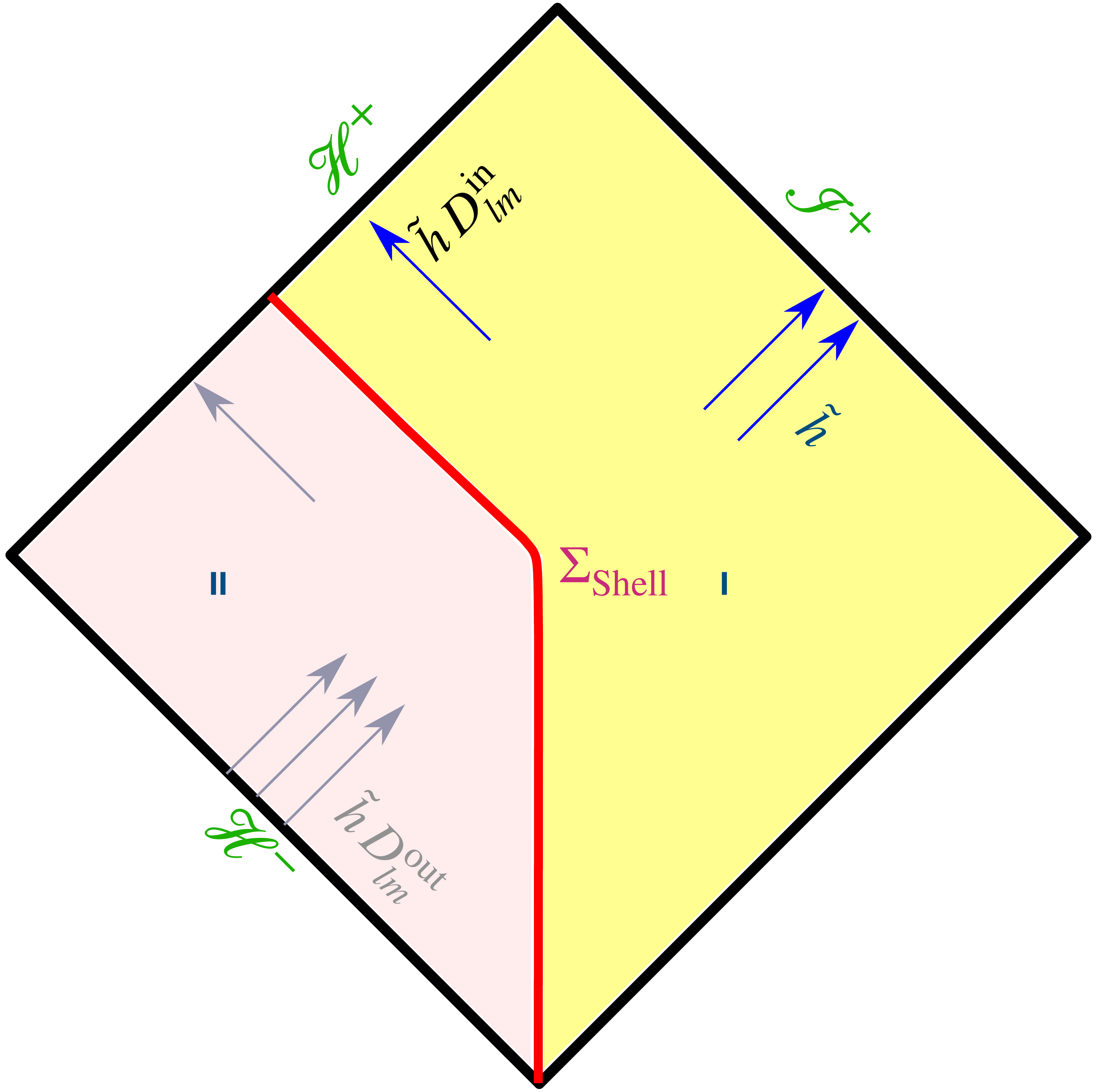}
  \caption{The physical meaning of $\mathcal{F}^D_{lm}$ based on the hybrid approach. The spacetime is split by a time-like world tube $\Sigma_{\rm Shell}$ (red line) into an inner PN regime II and an outer BHP regime I. During the spacetime reconstruction, we take a waveform from NR at null infinity $\mathscr{I}^+$, and evolve it backward into the bulk using BHP theory as if $\Sigma_{\rm Shell}$ were not there. The result is proportional to the up-mode solution to the homogeneous Teukolsky equation. In particular, an image wave $\tilde{h}D^{\rm out}_{lm}$ needs to appear at the past horizon $\mathscr{H}^-$, and it is proportional to the filtered waveform. The image wave is spurious since the entire $\mathscr{H}^-$ lies inside the PN regime II, where the BHP theory does not apply. It exists there as a source to drive the wave in regime I. During the ringdown phase of $\tilde{h}$, the linear QNMs are free ringing of the remnant BH and hence can be annihilated by $D^{\rm out}_{lm}$, whereas nonlinear pieces are driven by some sources and thus cannot be removed. }
 \label{fig:image_wave}
\end{figure}

\begin{figure*}[htb]
        \includegraphics[width=1.9\columnwidth,clip=true]{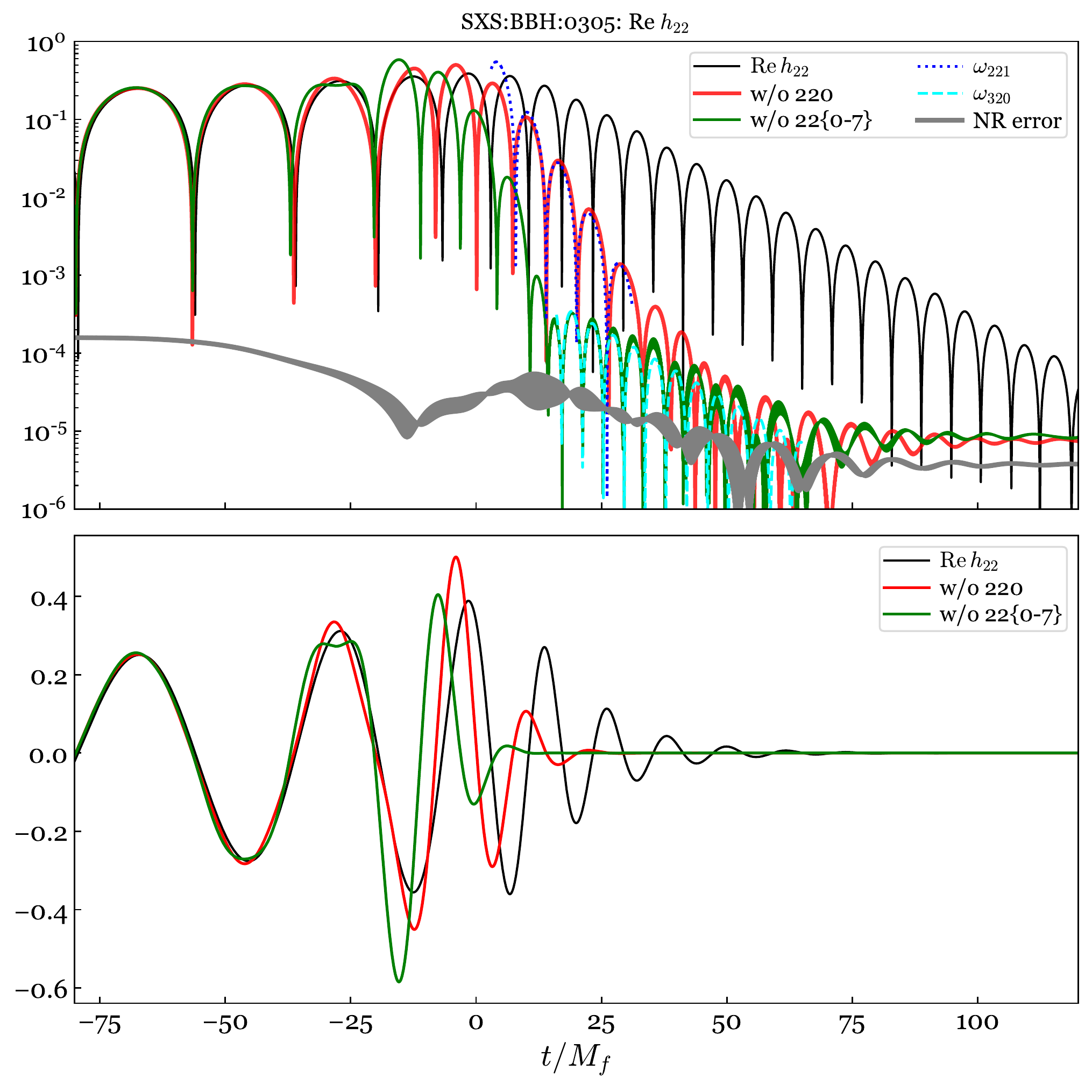}
  \caption{The effect of the filter $\mathcal{F}_{lmn}$ on $h_{22}$ of \ind. Here we have aligned the early inspiral portion between the original signal $h_{22}$ (black) and the filtered waveforms. After removing $\omega_{220}$ from the original waveform, the oscillation in the ringdown of the filtered waveform (red) is consistent with the QNM $\omega_{221}$ (blue). If we further remove $\omega_{22,n=1...7}$, the residual shows the existence of the QNM $\omega_{320}$ (cyan), which is caused by the spherical-spheroidal mixing. For comparison, we evaluate the numerical error of this waveform (gray) by taking the difference between two adjacent numerical resolutions.}
 \label{fig:GW150914_h22_final}
\end{figure*}

\subsection{The full filter: the inverse of BH transmissivity}
\label{sec:second_filter_Wronskian}
Following Teukolsky's approach for the linear perturbation of a Kerr BH with dimensional spin $a$ \cite{PhysRevLett.29.1114,1973ApJ...185..635T}, we first write 
\begin{align}
    \Psi=\rho^{-4}\psi_4=R_{lm}(r,\omega)\tensor[_{-2}]{S}{_{lm}}(a\omega,\theta,\phi)e^{i\omega t},
\end{align}
where $\rho=-(r-ia\cos\theta)^{-1}$, $(t,r,\theta,\phi)$ is the Boyer-Lindquist coordinate system, and $\psi_4$ is the Weyl scalar. The radial function $R_{lm}(r,\omega)$ satisfies the radial Teukolsky equation \cite{PhysRevLett.29.1114,1973ApJ...185..635T}. The \textit{up}-mode solution $R_{lm}^{\rm up}$ to the homogeneous Teukolsky equation
is of particular interest to us. Its asymptotic behavior near future null infinity and the horizon is given by \cite{Teukolsky:1974yv}
\begin{align}
&R^{\rm up}_{lm} \sim 
\begin{cases}
r^3 e^{i\omega r_*} ,\quad  & r_*\rightarrow +\infty, \\
\\
 D^{\rm out}_{lm} e^{i \omega  r_*} + \Delta^2 D^{\rm in}_{lm} e^{-i\omega r_*},   & r_*\rightarrow -\infty,
\end{cases} 
\label{up-mode}%
\end{align}
with $\Delta=r^2-2r+a^2$ and $r_*$ being the tortoise radius. Fig. \ref{fig:image_wave} exhibits the physical meaning of the up-mode --- a wave is emitted from the past horizon $\mathscr{H}^-$ and it gets reflected and transmitted by the BH potential. The transmissivity and reflectivity are given by $1/D^{\rm out}_{lm}$ and $D^{\rm in}_{lm}/D^{\rm out}_{lm}$, respectively. As for a QNM of the BH, its mode frequency $\omega_{lmn}$ satisfies
\begin{align}
     D^{\rm out}_{lm}(\omega_{lmn})=0.
\end{align}
Therefore, we can write
\begin{align}
    D^{\rm out}_{lm}\sim \prod_{n}(\omega-\omega_{lmn}).\label{Dout_expansion}
\end{align}
Comparing Eq.~\eqref{Dout_expansion} with the filter in Eq.~\eqref{filter_final} [also Eq.~\eqref{filter_single}], we see $D^{\rm out}_{lm}$ serves the same role as $\mathcal{F}_{\rm tot}$: it can remove all $\omega_{lmn}$'s that are associated with the indices $(l,m)$ at once. In practice, since $D^{\rm out}_{lm}$ diverges as $\omega\to0$ \cite{Mano:1996vt}, we instead define a filter
\begin{align}
    \mathcal{F}^D_{lm}=\frac{D^{\rm out}_{lm}}{D^{\rm out *}_{lm}}, \label{D_filter_phase}
\end{align}
which is a direct analogue of Eq.~\eqref{filter_single} to ensure $|\mathcal{F}^D_{lm}|=1$ when $\omega$ is real-valued. Below we will call $\mathcal{F}^D_{lm}$ the full filter.

Interestingly, unlike the filter in Eq.~\eqref{filter_single} that was introduced purely phenomenologically, the current filtered waveform $\tilde{h}\, D^{\rm out}_{lm}$ bears a physical meaning. To be concrete, in Ref.~\cite{Ma:2022xmp}, some use the \textit{hybrid approach} \cite{Nichols:2010qi,Nichols:2011ih} to reconstruct the spacetime near merging compact objects based on NR waveforms at future null infinity $\mathscr{I}^+$. Below we give a brief introduction and refer the interested readers to Refs.~\cite{Nichols:2010qi,Nichols:2011ih,Ma:2022xmp} for more details. The hybrid method is an approximated, \textit{ab initio} waveform mode. For a BBH merger spacetime in Fig.~\ref{fig:image_wave}, the spacetime is split by a time-like world tube $\Sigma_{\rm Shell}$ into an inner strong-gravity region II and an outer weak-gravity region I, where the strong-gravity metric in II is given by the post-Newtonian (PN) theory while the one in I is provided by BHP theory. The hybrid method evolves two metrics jointly and they communicate via boundary conditions on the world tube $\Sigma_{\rm Shell}$. Note that close to the merger, the PN theory may break down, but the errors stay within the BH potential as long as the shell $\Sigma_{\rm Shell}$ falls rapidly enough into the future horizon $\mathscr{H}^+$. As a result, the hybrid method was able to predict a reasonable inspiral-merger-ringdown waveform for a BBH system \cite{Nichols:2010qi,Nichols:2011ih}. 

In Ref.~\cite{Ma:2022xmp}, on the other hand, we reversed the process --- we started with a NR waveform at $\mathscr{I}^+$ and evolved it backward into the bulk (the region I) using BHP theory. This process allows us to construct the entire spacetime as if the worldtube were not there. It turns out that the solution is proportional to the up-mode solution in Eq.~\eqref{up-mode}, and the coefficient is determined by the NR waveform $\tilde{h}$ at $\mathscr{I}^+$. As shown in Fig.~\ref{fig:image_wave}, the process leads to an outgoing wave $\tilde{h}\, D^{\rm out}_{lm}$ at the past horizon $\mathscr{H}^-$, although it is not real because the entire $\mathscr{H}^-$ lies inside the strong-gravity region, where BHP theory does not apply. Nevertheless, we can think of the filtered waveform $\tilde{h}\, D^{\rm out}_{lm}$ as an \textit{image wave}, which is akin to the image charge in electrodynamics. The image wave exists there to drive the signal in region I --- acting as a source --- by providing a desired boundary condition on $\Sigma_{\rm Shell}$. In particular, during the ringdown phase of $\tilde{h}$, a linear QNM corresponds to the free ringing of the BH, and thus there is no corresponding source term. Consequently, it can be annihilated by $D^{\rm out}_{lm}$, which is consistent with our phenomenological construction in Sec.~\ref{sec:first_filter_toy_model}. In contrast, second-order effects (during the ringdown phase) \cite{Gleiser:1995gx,Loutrel:2020wbw,Ripley:2020xby} are driven by sources, and hence cannot be removed by $D^{\rm out}_{lm}$. The filtered waveform $\tilde{h}\, D^{\rm out}_{lm}$ represents the image wave (an effective source) for the second-order effects.

\begin{table}[tbh]
    \centering
    \caption{A list of NR simulations (nonprecessing) used in this paper. The
first column is the SXS identifier \cite{Boyle:2019kee}. The second column is the mass ratio $q>1$. The third column gives the number of quasicirular orbits that the systems undergo before the merger. The fourth and fifth columns correspond to the initial spin components along the direction of the orbital angular momentum (the $z-$axis). The remnant mass $(M_f)$, as a fraction of the total system mass $M_{\rm tot}$, and spin $(\chi_f)$ are in the final two columns. The waveform \ind~is a GW150914-like system.}
    \begin{tabular}{c c c c c c c} \hline\hline
ID& \multirow{2}{*}{$q$} & \multirow{2}{*}{$N_{\rm cycle}$} & \multirow{2}{*}{$\chi_{1}^z$} & \multirow{2}{*}{$\chi_{2}^z$} & \multirow{2}{*}{$\frac{M_f}{M_{\rm tot}}$} & \multirow{2}{*}{$\chi_f$} \\ 
SXS:BBH: & & & & & & \\ \hline
0305 & 1.2 & 15.2 & $0.33$ & $-0.44$ & $0.952$ & $0.692$ \\ \hline
1107 & 10.0 & 30.4 & $\sim10^{-6}$ & $\sim10^{-8}$ & 0.992 & $0.261$ \\ \hline
1936 & 4.0 & 16.5 & $-0.8$ & $-0.8$ & 0.985 & $0.022$ \\
 \hline\hline
     \end{tabular}
     \label{table:SXS_runs}
\end{table}

\section{Applications of the filters}
\label{sec:applications}
In this section, we use three NR simulations, SXS:BBH:0305, 1107 and 1936, in the Simulating eXtreme Spacetimes (SXS) catalog \cite{Boyle:2019kee} as examples to demonstrate the applications of the filters. As summarized in Table \ref{table:SXS_runs}, these three waveforms are for nonprecessing systems: the initial individual spins $\chi^z_{1,2}$ are (anti-)aligned with the orbital angular momentum (along the $z$-axis), and the mass ratio between the primary BH and the secondary BH is denoted by $q$, i.e., $q>1$. The systems undergo $N_{\rm cycle}$ quasicircular orbits before the merger. The remnants are Kerr BHs with mass $M_f$ and spin $\chi_f$. In particular, \ind~is a GW150914-like system \cite{LIGOScientific:2016aoc}. We want to emphasize again that our rational filter leads to a time shift backwards in time. For the sake of comparison, in this section we always undo the time shift by aligning the early portions of waveforms (i.e., minimizing their mismatch).

\subsection{The GW150914-like system: \ind}
In this subsection, we investigate several $(l,m)$ harmonics of SXS:BBH:0305. Sec.~\ref{sec:305_h22} focuses on $h_{22}$, where we show that the $\omega_{320}$ QNM mixes into $h_{22}$ due to the spherical-spheroidal mixing \cite{Berti:2014fga,Kelly:2012nd,Dhani:2021vac}. Sec.~\ref{sec:305_nonlinearity} focuses on second-order effects in $h_{44},h_{54}$ and $h_{55}$ contributed by the quadratic couplings $h_{22}^2$ and $h_{22}h_{33}$, respectively. Finally in Sec.~\ref{sec:305_h21}, we study the leakage of the $\omega_{220}$ mode into the harmonic $h_{21}$ due to the gravitational recoil \cite{Kelly:2012nd,Boyle:2015nqa}. 
% Finally, in Sec.~\ref{sec:305_start_time}, we use the method outlined in Sec.~\ref{sec:first_filter_toy_model} to determine the start time of QNMs in $h_{22}$.

\begin{figure}[htb]
        \includegraphics[width=\columnwidth,clip=true]{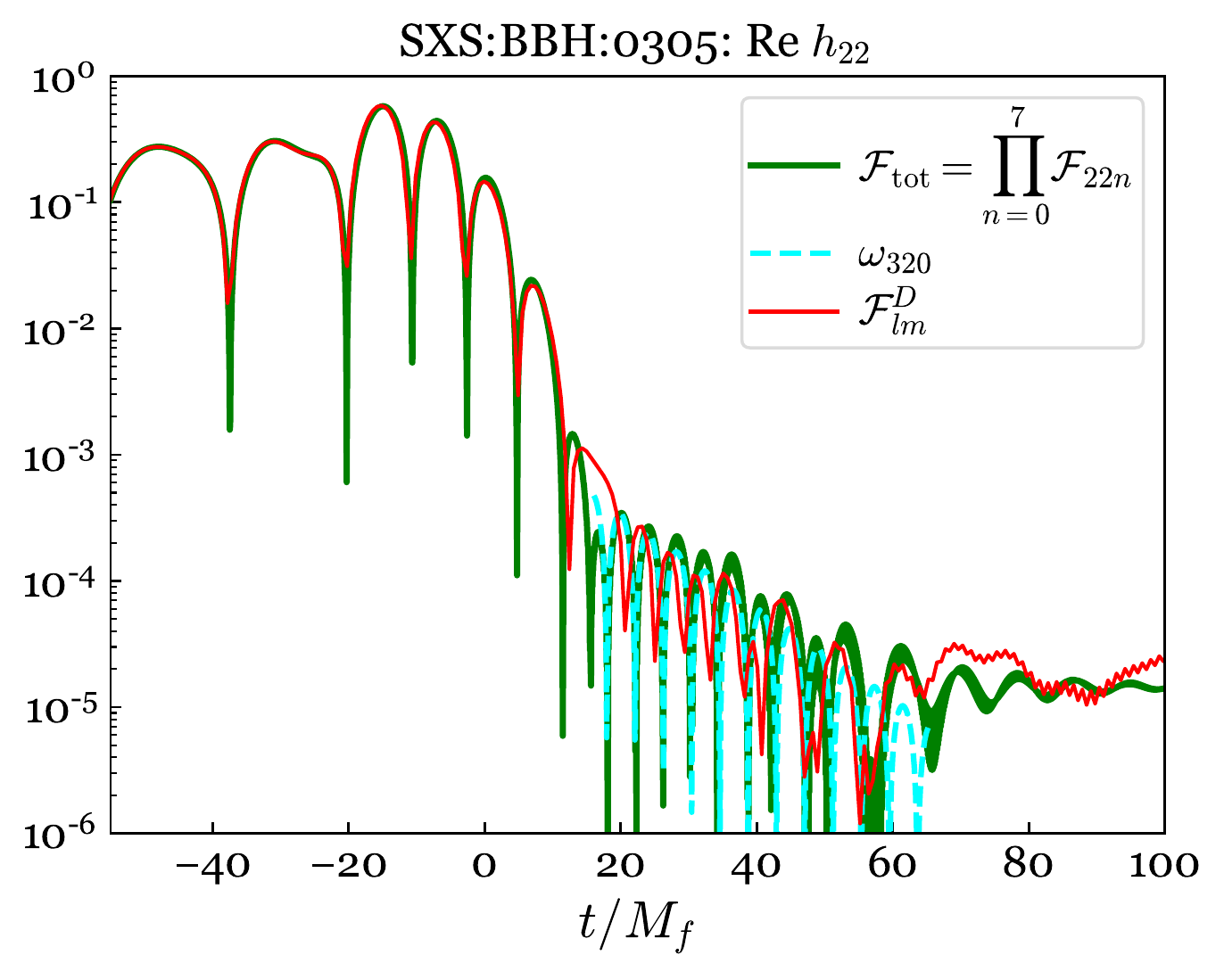}
  \caption{A comparison between the full filter $\mathcal{F}^D_{lm}$ [Eq.~\eqref{D_filter_phase}] and the rational filter $\mathcal{F}_{\rm tot}$ [Eq.~\eqref{filter_final}] associated with $\omega_{22,n=0...7}$. The latter one is more accurate to reveal the existence of the QNM $\omega_{320}$ in $h_{22}$ of \ind. We attribute the inaccuracy of the full filter to the numerical noise when we interpolate the value of $D^{\rm out}_{lm}$ from the Black Hole Perturbation Toolkit.}
 \label{fig:two_filters_comparison}
\end{figure}

\begin{figure*}[htb]
    \subfloat[SXS:BBH:0305: Re $h_{44}$]{\includegraphics[width=\columnwidth,clip=true]{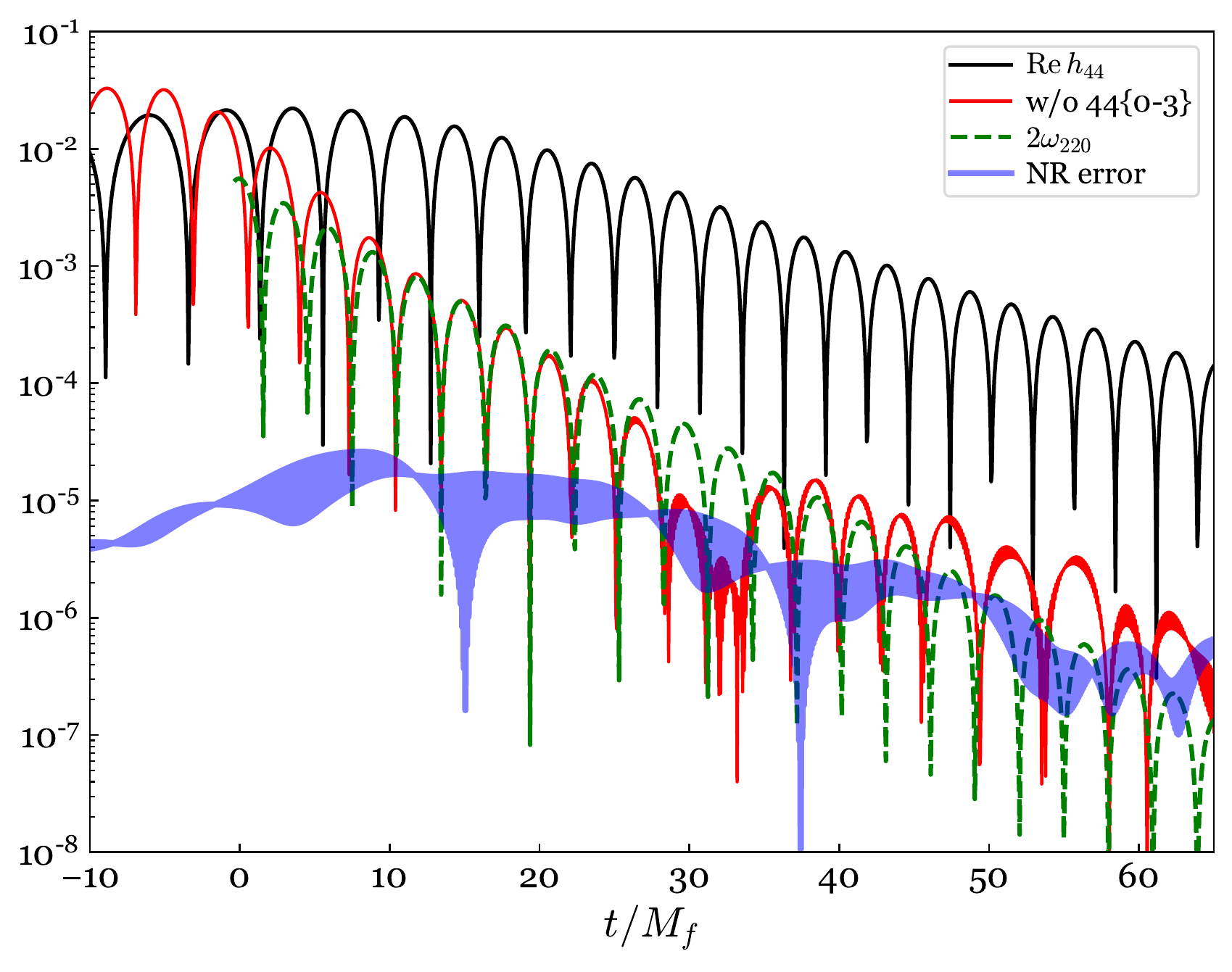}} \\
    \subfloat[SXS:BBH:0305: Re $h_{54}$]{\includegraphics[width=\columnwidth,clip=true]{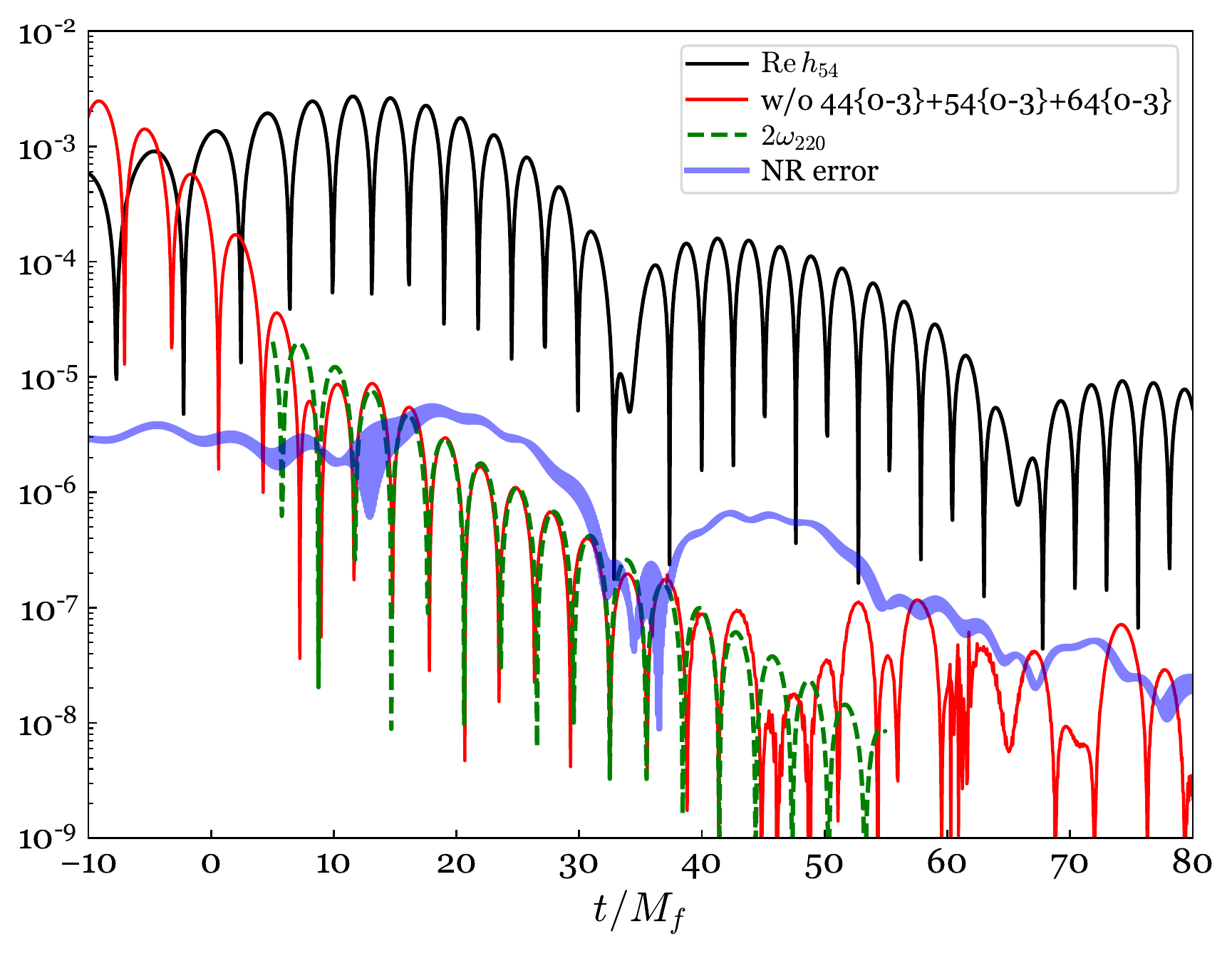}}
    \subfloat[SXS:BBH:0305: Re $h_{55}$]{\includegraphics[width=\columnwidth,clip=true]{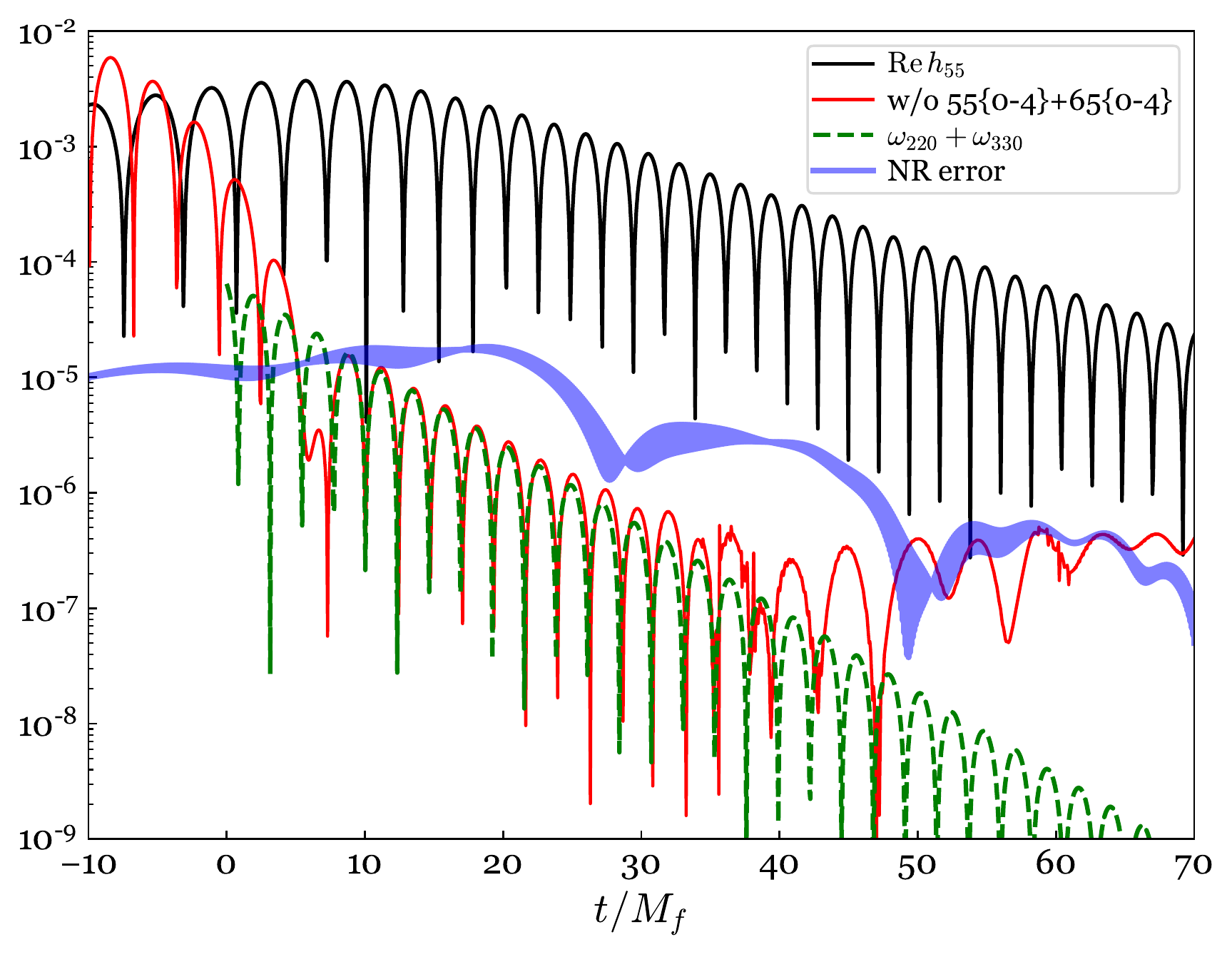}}
  \caption{Second-order modes in $h_{44}$ (top), $h_{54}$ (bottom left), $h_{55}$ (bottom right)  of \ind. After removing linear QNMs and relevant spherical-spheroidal mixing modes from original waveforms (black curves), filtered waveforms (red curves) contain oscillations that are consistent with the sum tone of $2\omega_{220}$ or $\omega_{220}+\omega_{330}$ (green dashed curves). As for the harmonics $h_{55}$ and $h_{54}$, the comparison is done in the superrest frame to avoid other mixing modes.}
 \label{fig:nonlinearity}
\end{figure*}

\subsubsection{$h_{22}$: the spherical-spheroidal mixing}
\label{sec:305_h22}
It was found that the harmonic $h_{22}$ of \ind~can be modeled as a superposition of $\omega_{22,n=0...7}$ up to the peak strain amplitude \cite{Giesler:2019uxc}. To compare our analysis results using the new method with theirs, we first apply a filter $\mathcal{F}_{l=2,m=2,n=0}$ [Eq.~\eqref{filter_single}] to $h_{22}$. As shown in Fig.~\ref{fig:GW150914_h22_final}, the filtered waveform (the red curve) has a smaller amplitude than $h_{22}$ in the late ringdown regime, and we see that the main residual oscillation is consistent with the frequency and the decay rate of the first overtone $\omega_{221}$ (blue). Here the blue dashed curve is obtained by fitting the filtered waveform within the window of $[12,28]\,M_f$; and the mode amplitude and phase of the first overtone are 0.08 and $-0.57$ rad at $t=12\,M_f$. We note that the amplitude of the first overtone is reduced by $\mathcal{F}_{220}$ \cite{Ma_filter_ligo_data}. One needs to take the reduction factor into account while comparing with the original amplitude, and we leave this comparison for future work. On the other hand, the result serves as strong evidence to support that $\mathcal{F}_{lmn}$ is indeed able to annihilate the corresponding $(l,m,n)$ QNM. 

Next we continue to remove $\omega_{22,n=1...7}$ based on the conclusion in Ref.~\cite{Giesler:2019uxc}, and obtain the green curve in Fig.~\ref{fig:GW150914_h22_final}. We can see that the oscillation is consistent with $\omega_{320}$ (cyan) in the window of $[16,65]\,M_f$, whose amplitude and phase are $\sim 4.4\times10^{-4}$ and $-0.79$ rad at $t=16\,M_f$ after the filters. To ensure the oscillation is physical rather than numerical artifacts, we compute the numerical (truncation) error of this NR simulation by taking the difference between two adjacent numerical resolutions. We see that the residual in the filtered waveform is still above the numerical noise floor. Therefore, this piece of the dominant residual signal corresponds to the spherical-spheroidal mixing in the remnant Kerr spacetime\footnote{The supertranslation can also make $h_{32}$ leak into $h_{22}$, e.g., Eq.~(8) of Ref.~\cite{Kelly:2012nd}. We have checked that the presence of the mode $\omega_{320}$ is due to the spherical-spheroidal mixing by transforming the waveform to the superrest frame using the technique presented in Ref.~\cite{Mitman_BMS_fixing}. For more on this, see Ref.~\cite{MaganaZertuche:2021syq,Mitman:2021xkq,Mitman_BMS_fixing}.} \cite{Berti:2014fga,Kelly:2012nd,Dhani:2021vac}. Meanwhile, we find the filter shifts the waveform backward in time by $\sim 14.1M_f$, close to the prediction given by Eq.~\eqref{phase_time_shift_ana} 
% \kmcomment{If you use the frequency at or near the peak to compute the time/phase shift you should obtain better agreement}\sma{Gotcha!}:
\begin{align}
\sum_{n=0}^{n=7} t_{l=2,m=2,n}\sim 12.9\,M_f.
\end{align}
In Fig.~\ref{fig:GW150914_h22_final} we have aligned the early inspiral portion between the original signal $h_{22}$ (the black curve) and the filtered waveforms for comparisons.

% In addition, the early portion of the filtered waveform (the gray curve) \ls{again, green curve?} becomes different from the original waveform (the black curve). This is mainly caused by the backward time shift in Eq.~\eqref{phase_time_shift_ana}. In fact, we find that two waveforms can be well aligned if we shift the filtered waveform

% We barely see the $l=3,m=2,n=1,2$ modes.

Then in Fig.~\ref{fig:two_filters_comparison} we investigate the effect of the full filter $\mathcal{F}^D_{lm}$ [Eq.~\eqref{D_filter_phase}], where the value of $D^{\rm out}_{lm}$ is obtained from the Black Hole Perturbation Toolkit \cite{BHPToolkit}. The result is almost identical to that of the rational filter $\mathcal{F}_{\rm tot}$ up to $t\sim 10M_f$, but it is less accurate to reveal the spherical-spheroidal mixing. We attribute the inaccuracy to the numerical noise when we interpolate the value of $D^{\rm out}_{lm}$ from the Black Hole Perturbation Toolkit, and we leave a more precise calculation of $D^{\rm out}_{lm}$ for future studies. In addition, we find a nice property of the full filter $\mathcal{F}^D_{lm}$: it does not give rise to any time shift, as opposed to the rational filter. One could benefit from this feature in real data analyses.

\subsubsection{$h_{44},h_{55},h_{54}$: the second-order QNMs}
\label{sec:305_nonlinearity}
London \etal \cite{London:2014cma} found evidence for the second-order mode in the $h_{44}$ harmonic, contributed by a quadratic coupling $\sim h_{22}^2$. Therefore, it is expected to see the sum tone $2\omega_{220}$ in the ringdown of $h_{44}$. In the upper panel of Fig.~\ref{fig:nonlinearity}, we first remove the linear QNMs $\omega_{44,n=0...3}$ from $h_{44}$, and then fit the filtered waveform with $2\omega_{220}$ in the window of $[12,30]\,M_f$. We can see a decent agreement. The corresponding mode amplitude and phase are $7.9\times10^{-4}$ and $3.1$ rad at $t=12\,M_f$ after the filters. In addition, the signal is larger than the numerical (truncation) error, which is evaluated by computing the difference between two adjacent numerical resolutions. This result shows that the second-order mode does exist in the ringdown regime. Furthermore, we find evidence for the presence of $\omega_{220}+\omega_{221}$ and $2\omega_{221}$ in the ringdown of $h_{44}$ as well, and we leave more discussions in our follow-up work \cite{Mitman_nonlinear}. On the other hand, while we are preparing our manuscript, we notice that Ref.~\cite{Cheung_nonlinear} also carries out comprehensive studies on the second-order modes with a different approach, so we refer the interested reader to Ref.~\cite{Cheung_nonlinear} for more details.

In addition, it is also expected that $h_{55},h_{54}$ can be sourced by $h_{22}h_{33}$ and $h_{22}^2$, respectively. In this case, we find that one has to map the waveforms to the superrest frame \cite{MaganaZertuche:2021syq,Mitman:2021xkq} to reveal these second-order effects. We do this using the technique presented in Ref.~\cite{Mitman_BMS_fixing}, based on the SpECTRE code \cite{Kidder:2016hev,spectrecode}. In the bottom left panel of Fig.~\ref{fig:nonlinearity}, after removing the linear QNMs $\omega_{54,n=0...3}$, as well as $\omega_{44,n=0...3}$ and $\omega_{64,n=0...3}$ caused by the spherical-spheroidal mixing, we find the residual signal of $h_{54}$ is consistent with the sum tone $2\omega_{220}$ in the window of $[10,40]\,M_f$, with an amplitude of $1.2\times10^{-5}$ and a phase of $-2.9$ rad at $t=10\,M_f$. As for $h_{55}$, the bottom right panel of Fig.~\ref{fig:nonlinearity} shows the existence of $\omega_{220}+\omega_{330}$ in $[8,28]\,M_f$, whose amplitude and phase are $1.9\times10^{-5}$ and $-1.96$ rad at $t=8\,M_f$. Nevertheless, we see the amplitudes of these two second-order effects are on the same order of the numerical noise, therefore their existence is not conclusive.

Finally, we want to remark again that the amplitudes of the second-order effects are reduced by the filters. In consequence, the amplitudes obtained from our approach are smaller than their original values. 
% We leave more careful studies for future work.

\begin{figure}[htb]
        \includegraphics[width=\columnwidth,clip=true]{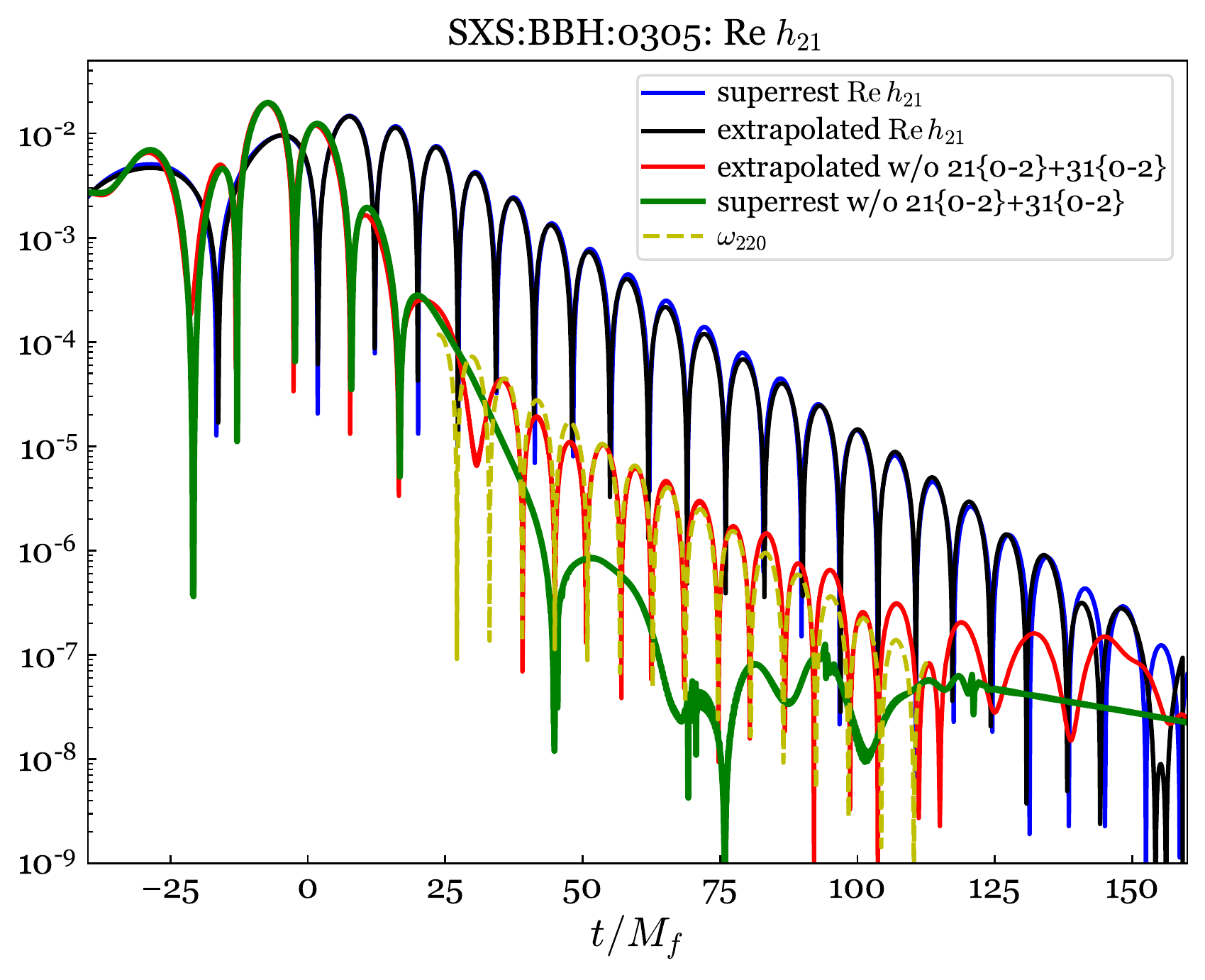}
  \caption{Leakage of the $\omega_{220}$ mode into the $h_{21}$ harmonic due to the gravitational recoil. After removing $\omega_{21,n=0...2}$ and $\omega_{31,n=0,1}$ from the original $h_{21}$ waveform (black curve), the red curve exhibits the presence of the $\omega_{220}$ mode (yellow dashed curve). If we transform the waveform to the superrest frame (blue curve) and repeat our filtering process, the mixing mode $\omega_{220}$ will be completely removed (green curve).}
 \label{fig:GW150914_h21}
\end{figure}

\subsubsection{$h_{21}$: the mode mixing due to a gravitational recoil}
\label{sec:305_h21}
We repeat our process for the harmonic $h_{21}$ of \ind. As shown in Fig.~\ref{fig:GW150914_h21}, after removing the linear QNMs $\omega_{21,n=0...2}$ and the spherical-spheroidal mixing modes $\omega_{31,n=0,1}$, we find the remaining oscillation is consistent with the mode $\omega_{220}$ (the red and yellow dashed curves). We then use $\omega_{220}$ to fit the filtered waveform in the window of $[55,92]\,M_f$. The result is shown as the yellow dashed curves. The corresponding mode amplitude and phase are $9.5\times10^{-6}$ and $-0.62$ rad at $t=55\,M_f$. This leakage is caused by a boost in the orbital plane, and this phenomenon has been discussed by Kelly \etal \cite{Kelly:2012nd} and Boyle \cite{Boyle:2015nqa}. To verify this, we transform the waveform to the superrest frame (the blue curve) \cite{MaganaZertuche:2021syq,Mitman:2021xkq}, where the remnant BH is in the center-of-mass frame. After applying the same filter, we can see the mixing is completely removed (the green curve), while the other portion of the waveform remains unchanged.

We note that the leakage of $\omega_{220}$ into $h_{21}$ is a common phenomenon, especially for high mass-ratio events whose kick velocities are relatively large. Failing to take this effect into account may misinterpret the mixing mode $\omega_{220}$ as retrograde modes \cite{Dhani:2020nik,Dhani:2021vac}. We will explain more details in Sec.~\ref{sec:retrograde}.

\begin{figure}[htb]
        \includegraphics[width=\columnwidth,clip=true]{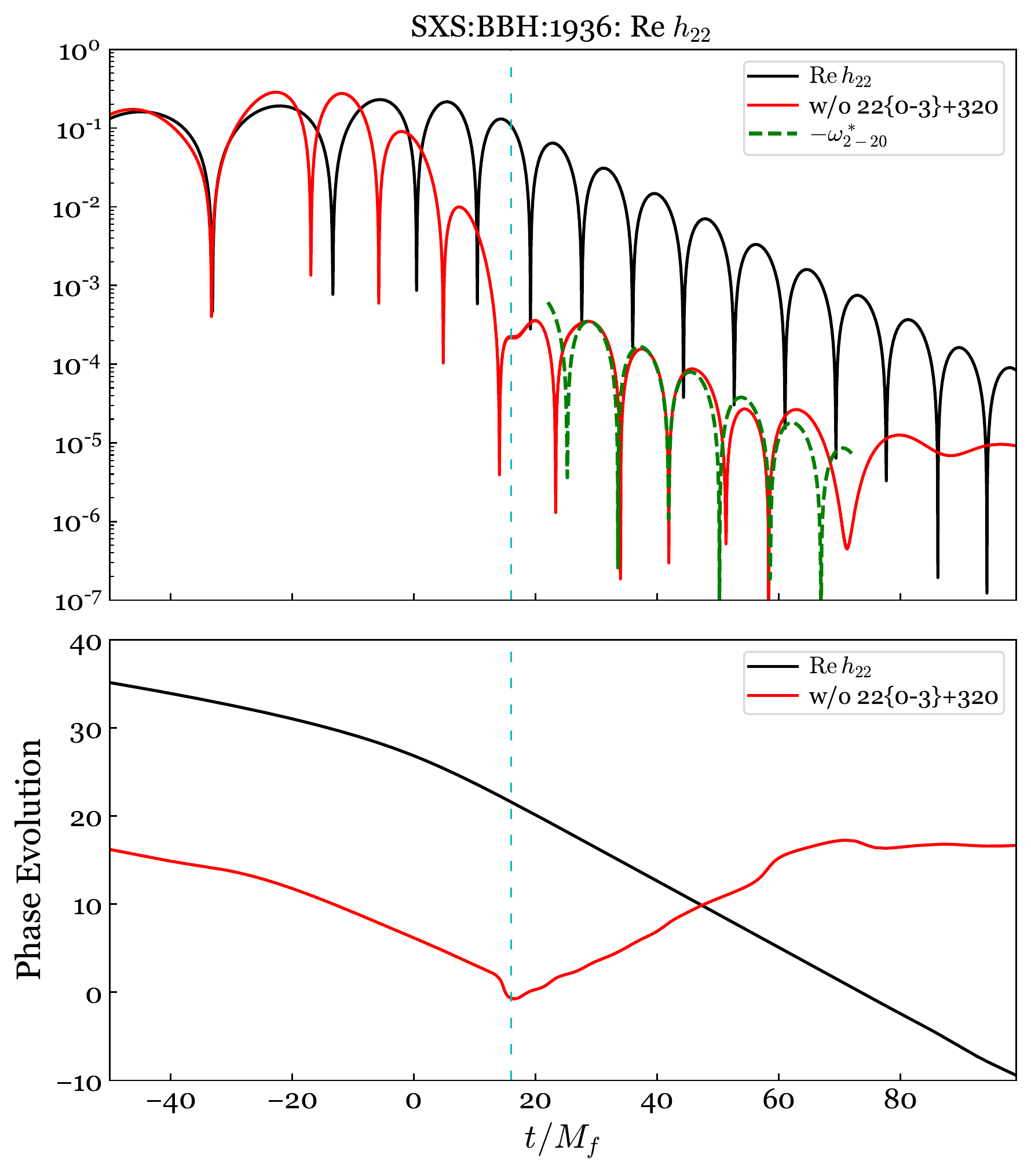}
  \caption{Retrograde mode $-\omega_{2-20}^*$ in the ringdown of \indnew. Top panel: after removing the $\omega_{22,n=0...3}$ modes and the spherical-spheroidal mixing mode $\omega_{320}$ from the original harmonic $h_{22}$ (black curve), we reveal the presence of $-\omega_{2-20}^*$ (green dashed curve) in the residual waveform (red curve). Bottom panel: the phase evolution of the original waveform (black curve) and the filtered waveform (the red curve). The phase of the original waveform decreases monotonically, indicating that the prograde modes are dominant. However, the phase of the filtered waveform starts to grow at the same time as the residual oscillations in the top panel appear, which demonstrates that the residual oscillations are retrograde modes.}
 \label{fig:sxs_1936_retrograde}
\end{figure}

\begin{figure}[htb]
        \includegraphics[width=\columnwidth,clip=true]{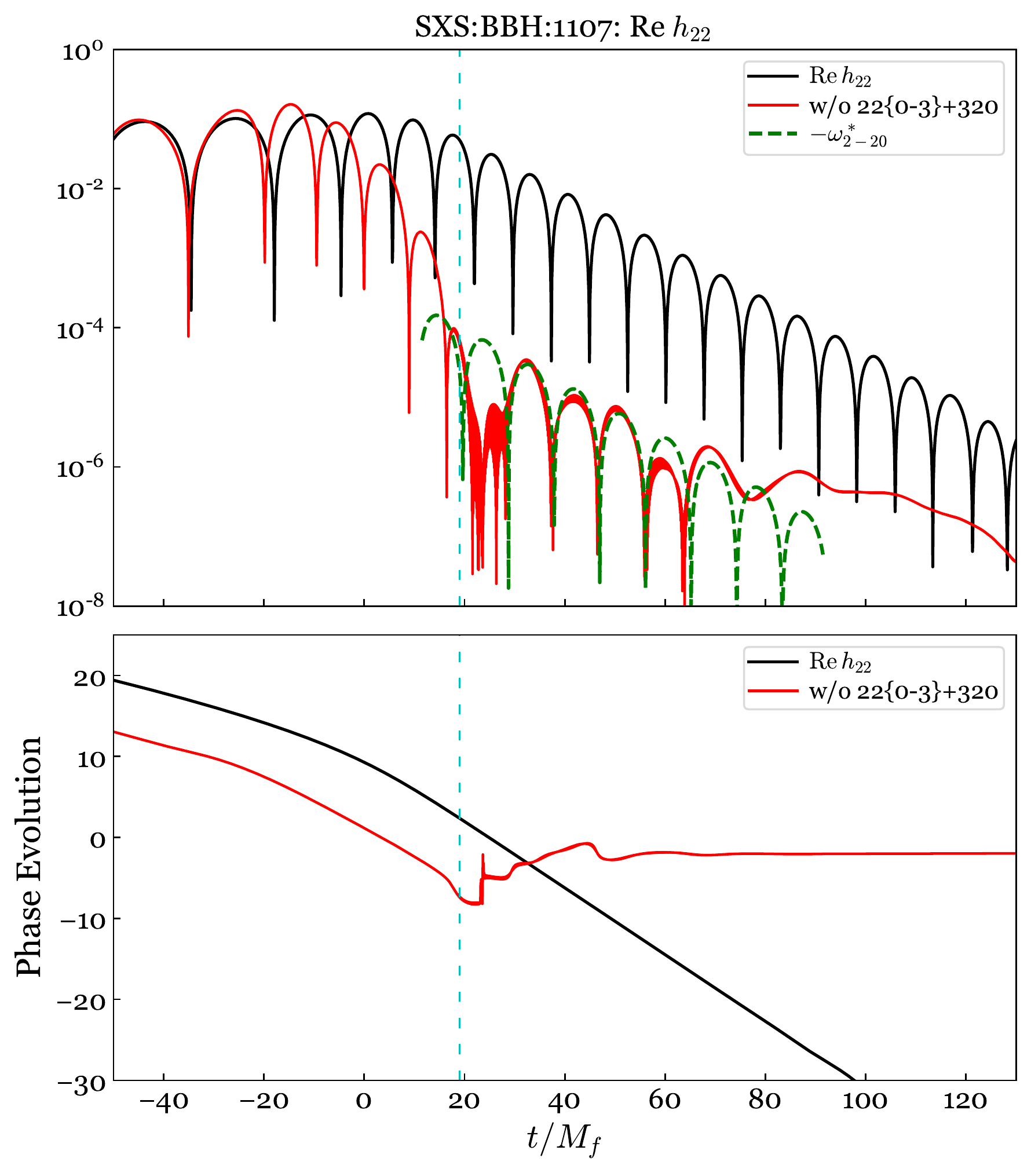}
  \caption{Same as Fig.~\ref{fig:sxs_1936_retrograde}, the retrograde mode $-\omega_{2-20}^*$ in the $h_{22}$ of SXS:BBH:1107.}
 \label{fig:sxs_1107_retrograde}
\end{figure}

\subsection{The retrograde modes}
\label{sec:retrograde}
% As summarized in Ref.~\cite{Li:2021wgz}, a gravitational harmonic $h_{lm}$ can be written as
% \begin{align}
%     h_{lm}=\sum_{n}[C_{lmn}e^{-i\omega_{lmn}t}+C^\prime_{lmn}e^{i\omega^*_{l-mn}t}],
% \end{align}
% where the second term corresponds to the retrograde modes, which are also dubbed ``mirror modes'' in Refs.~\cite{Dhani:2020nik,Dhani:2021vac}. 
It was found that taking into account the retrograde modes [e.g., the second term of Eq.~\eqref{h_lm_decom_ss}] would expand the linear perturbation regime. To partially address the debate on overfitting, we use our rational filter as a complementary tool to visualize the presence of the retrograde modes.

We first take SXS:BBH:1936 with non-negligible retrograde modes (see Appendix A of Ref.~\cite{Ma:2022xmp}). In the top panel of Fig.~\ref{fig:sxs_1936_retrograde}, we remove the prograde modes $\omega_{22,n=0...3}$ and the spherical-spheroidal mixing mode $\omega_{320}$ from the original harmonic $h_{22}$ (the black curve), then the red curve shows the existence of $-\omega_{2-20}^*$ in the residual. In the plot, the green dashed curve is obtained by fitting the filtered waveform with $-\omega_{2-20}^*$ in the window of $[28,60]\,M_f$. Its mode amplitude and phase are $3.9\times 10^{-4}$ and $2.6$ rad at $t=28\,M_f$. To further support our result, we investigate the phase evolution of the waveforms. For a prograde mode, its phase should decrease monotonically over time due to the term $e^{-i\omega_{lmn}t}$ [see the first term of Eq.~\eqref{h_lm_decom_ss}], whereas a retrograde mode's phase should increase due to the term $e^{i\omega^*_{l-mn}t}$ [see the second term in Eq.~\eqref{h_lm_decom_ss}]. In the bottom panel of Fig.~\ref{fig:sxs_1936_retrograde}, we see the phase of the original waveform (the black curve) decreases with time, indicating that the progrades are more dominant. After applying the filter, the decreasing trend terminates at $\sim 16M_f$ after the peak and the phase starts to grow at the same time that the residual oscillations in the top panel appear. This observation confirms the physical origin of the residual oscillations.

Then we look into the case of SXS:BBH:1107 investigated by Dhani \cite{Dhani:2020nik}. As shown in Fig.~\ref{fig:sxs_1107_retrograde}, there are a few cycles in the filtered residual waveform $h_{22}$ (the red curve) that are consistent with the retrograde mode $-\omega^{*}_{2-20}$. Meanwhile, the phase of the filtered waveform also grows within that regime, which serves as more evidence. Nevertheless, the retrograde mode in this case is weaker and noisier than that of \indnew. Furthermore, we find applying retrograde filters (not only the fundamental mode but also overtones) has little impact on the early portion $(t\lesssim 0)$ of the red curve in Fig.~\ref{fig:sxs_1107_retrograde}, meaning there is no strong evidence for the existence of retrograde modes within that regime. 
% \ls{Do you mean the early portion before the merger or the early portion in the ringdown? Is there any impact from retrograde overtone at earlier times?} \sma{I mean $(t\lesssim 0M_f)$. There is no impact from retrograde modes at earlier times.} \ls{From the phase plot, it seems that the retrograde mode starts at the peak; just curious when retrograde mode exactly starts}
As for the harmonic $h_{21}$, we find it has a mixing component from the mode $\omega_{220}$ due to the gravitational recoil, similar to the case discussed in Sec.~\ref{sec:305_h21}. This effect was not taken into consideration by Dhani \cite{Dhani:2020nik}, so we speculate that this could be the cause for the crests and troughs in the mismatch of $h_{21}$, e.g., Fig.~3 of Ref.~\cite{Dhani:2020nik}.

Finally, we want to note that the vertical dashed lines in Figs.~\ref{fig:sxs_1936_retrograde} and \ref{fig:sxs_1107_retrograde} do not necessarily correspond to the start time of the retrograde mode $-\omega_{2-20}^*$ in the original waveforms (the black curves), because of the time shift induced by our rational filter. To undo the time shift, here we simply align the early inspiral portion of the filtered waveforms with the original ones, making the location of the dashed lines less informative.

\begin{figure}[htb]
        \includegraphics[width=\columnwidth,clip=true]{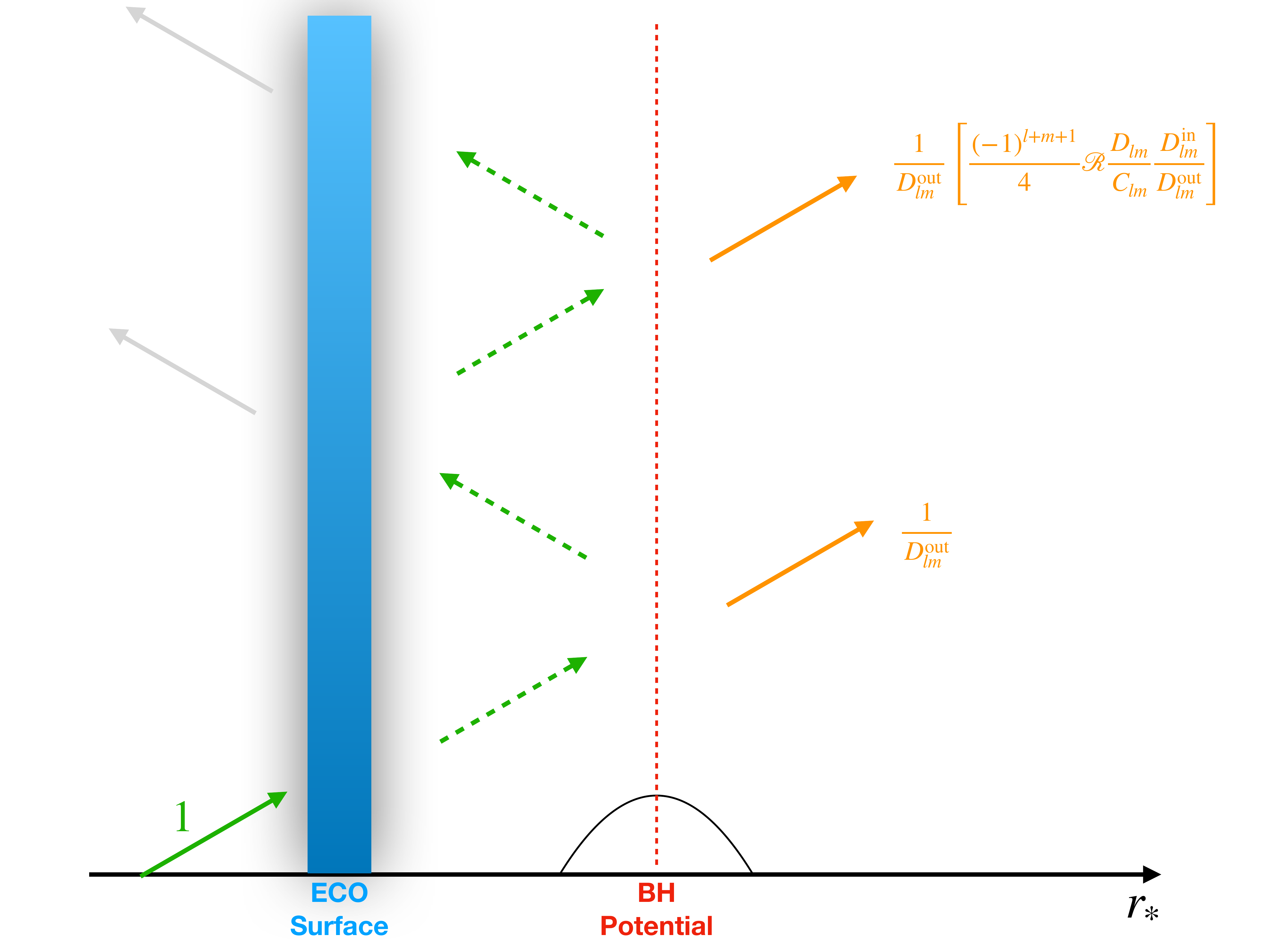}
  \caption{The up-mode solution of an ECO. We assume that a GW emerges from the horizon $(r_*=-\infty)$ and its amplitude is unity. It bounces back and forth within the cavity formed by the ECO surface and the BH potential. The GW seen by an observer at infinity consists of the main transmissive wave $1/D^{\rm out}_{lm}$ and a series of echoes.}
 \label{fig:echo}
\end{figure}

\begin{figure*}[htb]
    \centering
    \subfloat[$\epsilon=10^{-1}$\label{fig:wronskian-a}]{\includegraphics[width=1.0\textwidth]{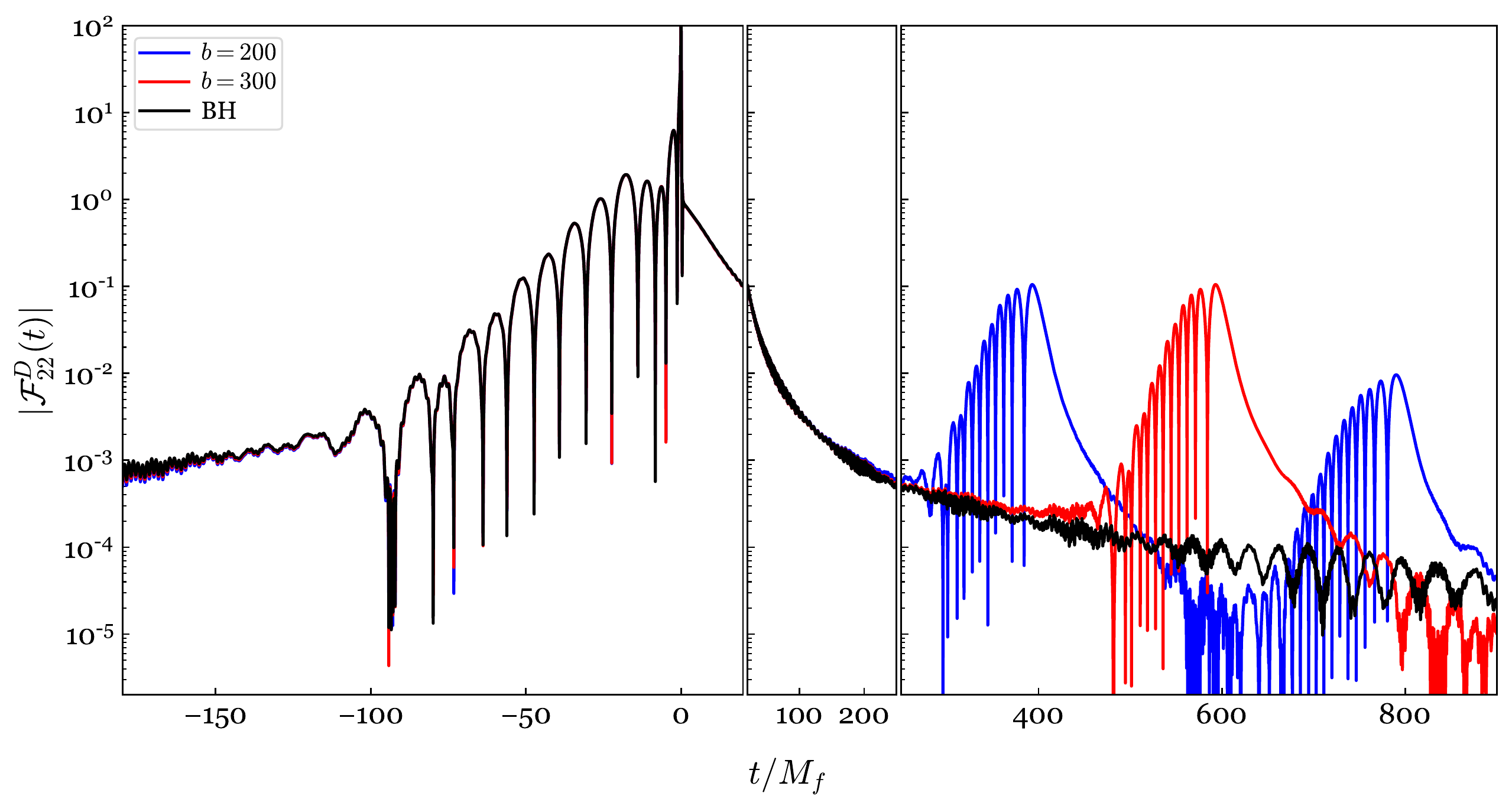}}\\
    \subfloat[$b=200M_f$\label{fig:wronskian-b}]{\includegraphics[width=1.0\textwidth]{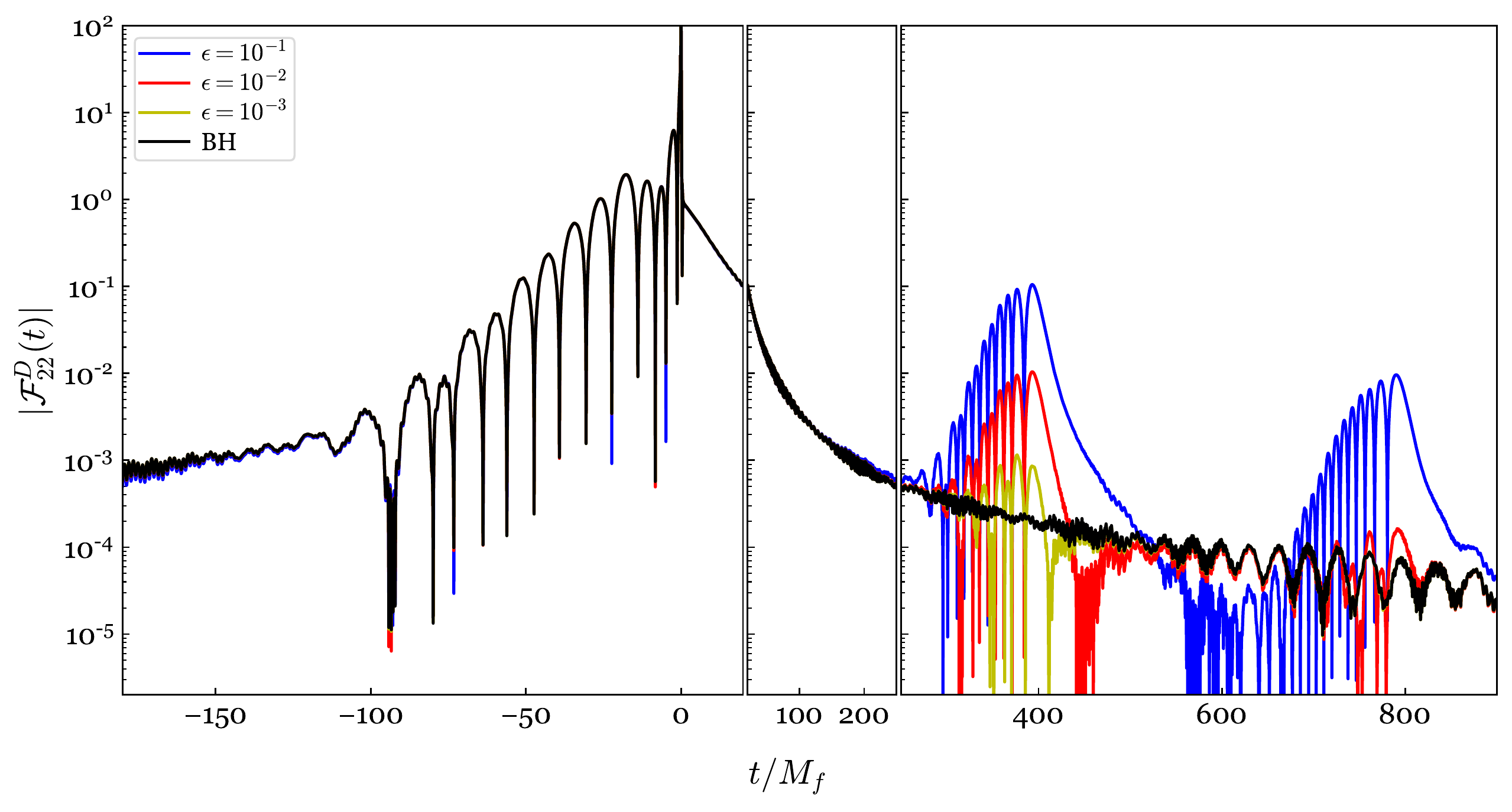}}
  \caption{The filter $\mathcal{F}^{D~{\rm ECO}}_{lm}$ of a nonspinning ECO in the time domain. In the top panel, we set $b$ to $200M_f$ (blue) and $300M_f$ (red), while fixing the value of $\epsilon$ to $10^{-1}$. They are compared with that of a Schwarzschild BH (black). In the bottom panel, we choose $\epsilon=10^{-1},10^{-2},10^{-3}$ (blue, red and yellow) and set $b$ to $200M_f$. In both cases, the original signal (around $t\sim 0$) remains unchanged. The perturbation appears as periodic echoes with the time interval $2b$. The amplitude of the $n$th echo is proportional to $\epsilon^n$.}
 \label{fig:wronskian}
\end{figure*}

\section{The stability of the full filter $\mathcal{F}^D_{lm}$ }
\label{sec:stability}

% \textcolor{RoyalPurple}{
% Notes for instability
% \begin{itemize}
% % \item \cite{Andersson:1996cm}
% \item environmental effects \cite{Cardoso:2016ryw,Barausse:2014pra,Barausse:2014tra}
% \item QNMs are complete when the BH potential is approximated by step functions \cite{Nollert:1996rf}
% \item the Completeness of the Quasinormal Modes of the Poeschl-Teller Potential at very late times\cite{Beyer:1998nu}
% \item (2004) causality; when the signal can be explained by QNMs \cite{Szpak:2004sf}
% \item (1993) QNMs are complete when spatial discontinuiteis exist \cite{Ching:1993gt}
% \item (1998) condition for completeness; exicitation factor; QNMs are not orthogonal and complete under
% this inner product. This has been a longstanding theoretical
% question and it is doubtful whether such an inner product can
% be defined for QNMs that is also of practical use \cite{Nollert:1998ys}
% \end{itemize}
% }

The QNM spectra of BHs have been found to be unstable \cite{Nollert:1996rf,Barausse:2014pra,Barausse:2014tra,Jaramillo:2020tuu,Cheung:2021bol}. In particular, Cheung \etal \cite{Cheung:2021bol} classified the instability into two categories: ``migration instability'' and ``overtaking instability''. For migration instability, the fundamental QNM drifts drastically from its unperturbed value when the perturbation is distant from the BH. This kind of instability is related to the asymptotic behavior of the eigenfunction near the horizon $(e^{-i\omega_{lmn} r_*})$ and infinity $(e^{i\omega_{lmn} r_*})$. Recalling that ${\rm Im}~\omega_{lmn}<0$, the eigenfunction of the QNM increases exponentially as $|r_*|\to\infty$. Any small perturbation of the BH potential at a large $|r_*|$ will lead to a significant change of $\omega_{lmn}$. For overtaking instability, a family of new modes appears near a bumpy BH, trapped between two potential barriers \footnote{They are called  ``matter-driven'' modes by Barausse \etal \cite{Barausse:2014pra,Barausse:2014tra}.}. One of the new modes might have a smaller decay rate than the unperturbed fundamental mode when the perturbation is at a large distance. Consequently, this new mode overtakes the original fundamental mode.

\begin{figure*}[htb]
    \centering
    \subfloat[$\epsilon=10^{-1}$]{\includegraphics[width=1.0\textwidth]{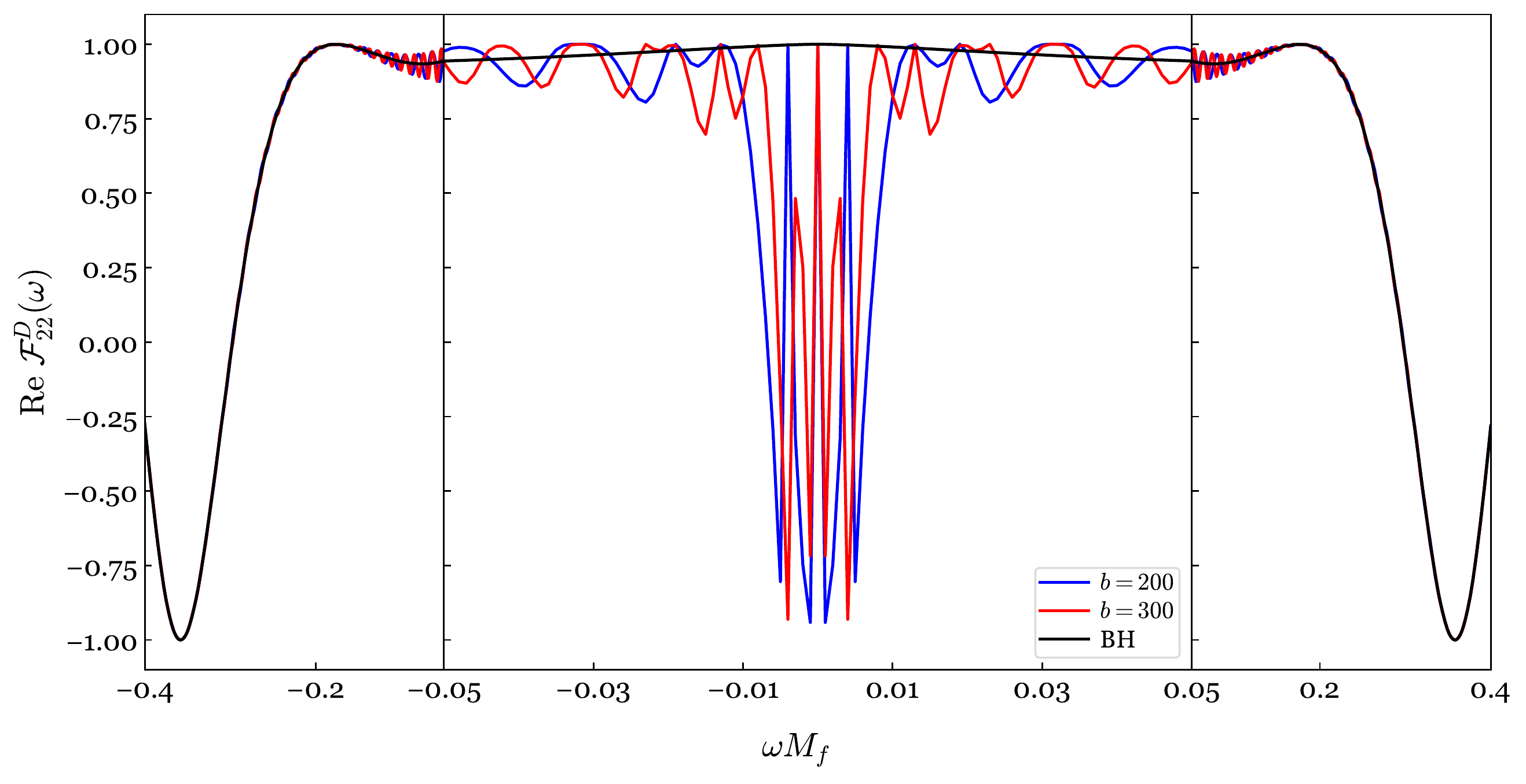}}\\
    \subfloat[$b=200M_f$]{\includegraphics[width=1.0\textwidth]{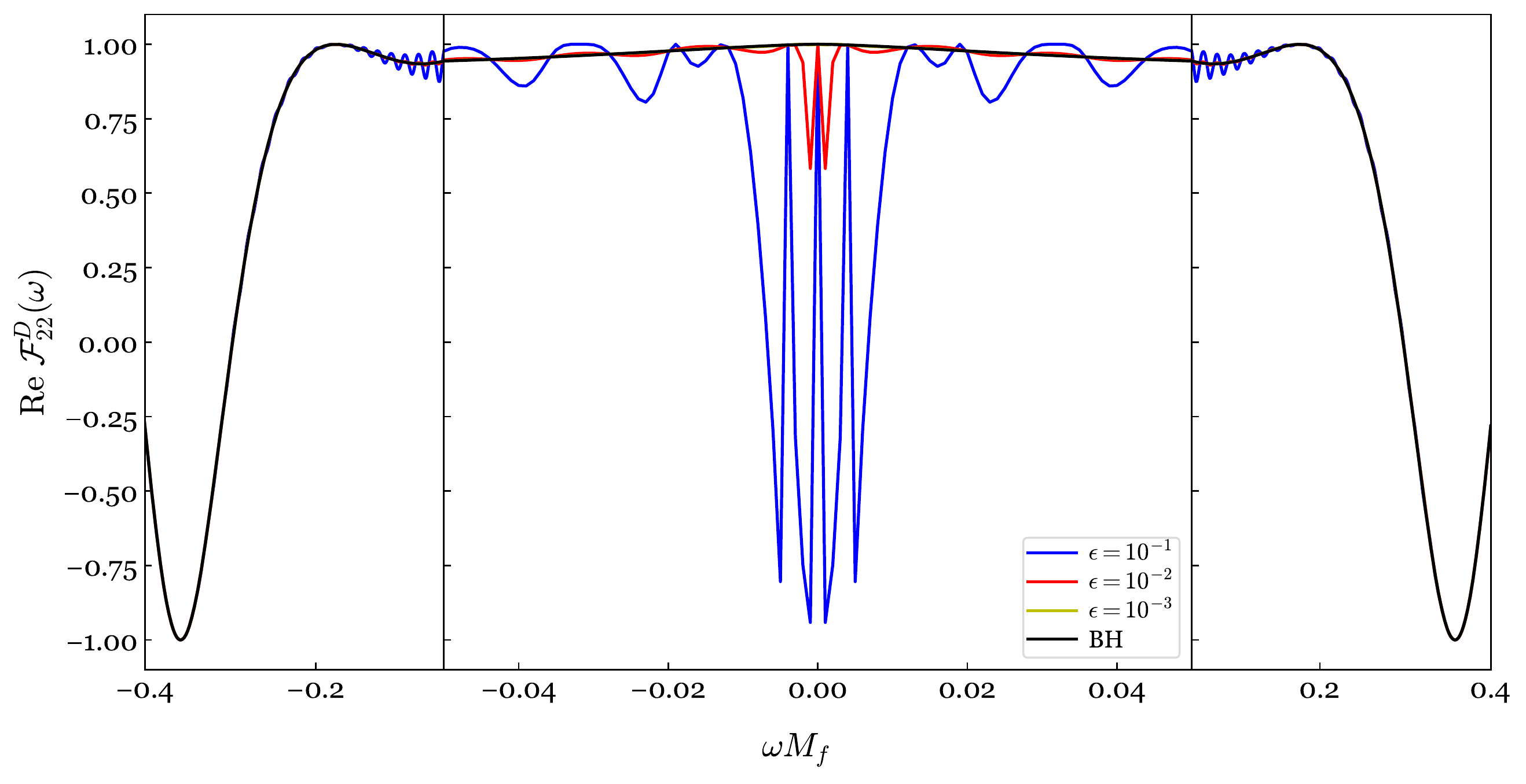}}
  \caption{Same as Fig.~\ref{fig:wronskian}. The real part of $\mathcal{F}_{22}^D$ in the frequency domain.}
 \label{fig:wronskian_freq}
\end{figure*}

The spectral instability, however, may have a limited effect on observational results (e.g., time-domain responses of a BH), as discussed in Refs. \cite{Nollert:1996rf,Barausse:2014pra,Barausse:2014tra,Cardoso:2016rao,Berti:2022hwx,Kyutoku:2022gbr}. Nollert \cite{Nollert:1996rf} and Barausse \etal \cite{Barausse:2014pra,Barausse:2014tra} showed that the prompt time-domain response is independent of perturbations when the perturbations are far from the system, even though the QNMs of the perturbed system are completely different from the ones of an isolated BH. Cardoso \etal \cite{Cardoso:2016rao} drew a similar conclusion while considering near-horizon perturbations. In fact, as pointed out by Hui \etal \cite{Hui:2019aox} and Berti \etal \cite{Berti:2022hwx}, despite the nonlocality of QNMs, one needs to appreciate the causal structure of the system while considering the time-domain signals --- a time response reflects the nature of each single potential bump that is causally connected to the observer, e.g., the prompt ringdown of a regular BH. The QNMs of the perturbed system do not show up until very late times as ``echoes'' \cite{Cardoso:2016oxy} when the initial Cauchy data travels and experiences the entire potential\footnote{We note that QNMs can become complete under some conditions \cite{Beyer:1998nu,Szpak:2004sf,Nollert:1998ys} (see also Refs.~\cite{Nollert:1996rf,Ching:1993gt} for relevant discussions). In particular, Beyer \cite{Beyer:1998nu} showed the completeness of QNMs of the Poeschl-Teller potential at a late time --- a regime where solutions can be expanded with respect to its QNMs.}. Therefore, the time-domain signal is stable in the sense that the original waveform remains unchanged, whereas the additional perturbation appears only as echoes that are well separated from the original signal in time.

% \begin{figure*}[htb]

%   \caption{}
%  \label{fig:Kerr_wronskian}
% \end{figure*}

The instability of the QNM spectra implies that QNMs may not be the most natural basis for ringdowns. One might need to rearrange QNMs into new subsets and sum each subset to form a new basis, in either time or frequency domain. In fact, the Backwards One-Body (BOB) model \cite{McWilliams:2018ztb} is an inspiring example, where the contributions of overtones associated with the same $(l,m)$ harmonic are rearranged and summed into a single time-domain function $\sim{\rm sech} \gamma t$, where $\gamma$ is a constant. One may further postulate that the time-domain function could be treated as the leading term of a new set of basis and the term $\sim{\rm sech} \gamma t$ provided by the BOB model contains most power of the ringdown. Another relevant time-domain basis was discussed by some of us for superkick systems \cite{Ma:2021znq}: it was found that the time-domain basis can even be extended to the inspiral regime for the superkick systems. A direct consequence is the \textit{collective excitation} of QNMs --- the amplitudes of different QNMs are correlated as a result of the time-domain basis being projected to the QNM basis. In fact, such a correlation (universality) has been found in not only the superkick systems \cite{Ma:2021znq}, but also extreme mass-ratio inspirals \cite{Hughes:2019zmt,Lim:2019xrb,Apte:2019txp,Lim:2022veo,Oshita:2021iyn}. 

Based on the above discussions, we want to ask: Do the filters reflect the nature of the system? Can we distinguish a BH from other objects (e.g. a bumpy BH or an exotic compact object) using our filters? In particular, since the full filter $\mathcal{F}^D_{lm}$ contains a collection of the corresponding QNMs $\omega_{lmn}$'s as a result of Eq.~\eqref{Dout_expansion}, is the filter stable or not under perturbations in the BH potential, given the spectral instability? In fact, a similar topic has been investigated recently by Kyutoku \etal \cite{Kyutoku:2022gbr}. The ``phase shift'' introduced by the authors is essentially the phase of our full filter in Eq.~\eqref{D_filter_phase}, and they showed that the phase shift of a Schwarzschild BH is stable when it is perturbed by a small Pöschl-Teller bump. In this work, we continue their studies and adopt another simple model to provide a qualitative answer. More sophisticated discussions are left for future work.

In Fig.~\ref{fig:echo}, we consider an exotic compact object (ECO) whose surface is close to the would-be horizon. The surface can partially reflect GWs and the reflectivity $\mathcal{R}$ is given by
\begin{align}
    \mathcal{R}=\epsilon e^{-2ib}, \label{reflectivity}
\end{align}
where $\epsilon$ is a constant, and $r_*=-b$ is the location of the ECO surface with the factor of two representing the round trip between the ECO surface and the BH potential. By imposing a physical boundary condition based on the membrane paradigm at the ECO surface \cite{Chen:2020htz}, we obtain the up-mode solution [in parallel with Eq.~\eqref{up-mode}]:
\begin{align}
&R^{\rm up~ECO}_{lm} \sim 
\begin{cases}
r^3 e^{i\omega r_*} ,\quad  & r_*\rightarrow +\infty, \\
\\
 \tilde{D}^{\rm out}_{lm} e^{i \omega  r_*} + \Delta^2 \tilde{D}^{\rm in}_{lm} e^{-i\omega r_*},   & r_*\rightarrow -\infty,
\end{cases} 
\label{up-mode-eco}%
\end{align}
with
\begin{align}
    \tilde{D}^{\rm out}_{lm}=D_{lm}^{\rm out}\left[1-(-1)^{l+m+1}\mathcal{R}\frac{D_{lm}}{4C_{lm}}\frac{D_{lm}^{\rm in}}{D_{lm}^{\rm out}}\right], \label{Dout_eco}
\end{align}
where the factor $D_{lm}/C_{lm}$ comes from the Teukolsky-Starobinsky (TS) relation \cite{Starobinsky:1973aij,Teukolsky:1974yv}. We refer interested readers to Appendix \ref{sec:eco_up_mode} for derivation. Note that Eq.~\eqref{Dout_eco} takes a similar form to the Wronskian in Eq.~(5.2) of Ref.~\cite{Hui:2019aox}.

We then define the filter $\mathcal{F}^{D~{\rm ECO}}_{lm}$ for the ECO system:
\begin{align}
    \mathcal{F}^{D~{\rm ECO}}_{lm}=\frac{\tilde{D}^{\rm out}_{lm}}{\tilde{D}^{{\rm out *}}_{lm}}. \label{ECO_filter}
\end{align}
To transform the filter to the time domain, we first need to apply the Planck-taper filter $\mathcal{F}(\omega)$ \cite{McKechan:2010kp} to remove the high-frequency contribution:
\begin{align}
    \mathcal{F}(\omega;\omega_1,\omega_1)=
    \begin{cases}
    0, &  \omega<\omega_1, \\
    \displaystyle \frac{1}{ e^z+1},&\omega_1<\omega<\omega_2, \\
    1, & \omega>\omega_2,
    \end{cases}
    \label{F_planck_filter}
\end{align}
with 
\begin{align}
    z=\frac{\omega_2-\omega_1}{\omega-\omega_2}+\frac{\omega_2-\omega_1}{\omega-\omega_1}.
\end{align}
Figure \ref{fig:wronskian} shows a nonspinning ECO case. The filters for a spinning ECO have the same qualitative feature so we refer readers to Appendix \ref{app:kerr_D} for results. In the absence of perturbations, we see that the black curve assembles the Dirac function $\delta(t)$ near $t=0$ because of the fact that $|\mathcal{F}^{D~{\rm ECO}}_{22}(\omega)|=1$. Most of the signals (i.e., the damped sinusoids) lie on the left side of the Dirac function ($t<0$), and the reason is exactly the same as the flipped ringdown in Fig.~\ref{fig:effect_of_filter}. We also see the tail-like feature at an earlier time.

\begin{figure*}[htb]
        \includegraphics[width=\columnwidth,height=6.8cm,clip=true]{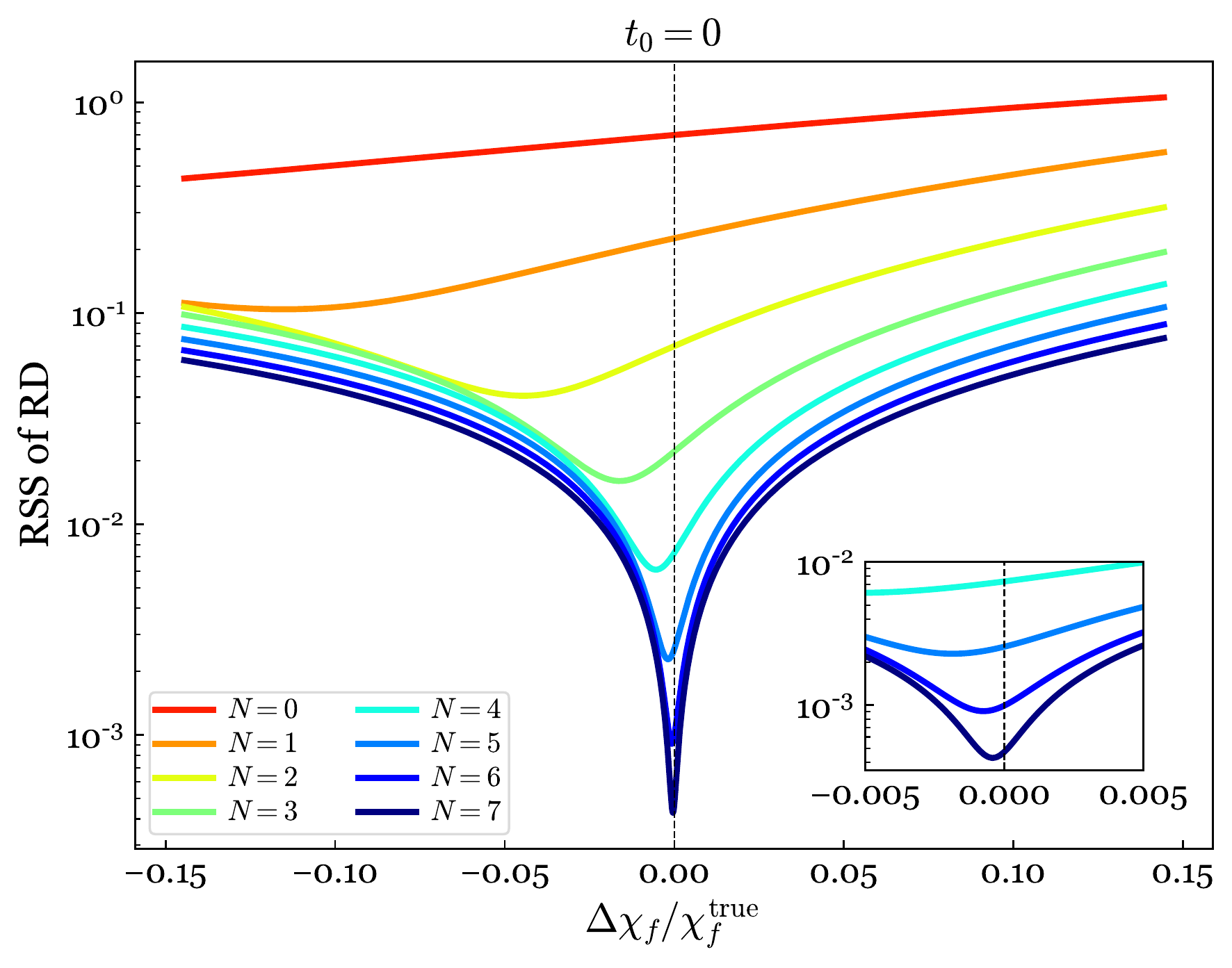}
        \includegraphics[width=\columnwidth,height=6.8cm,clip=true]{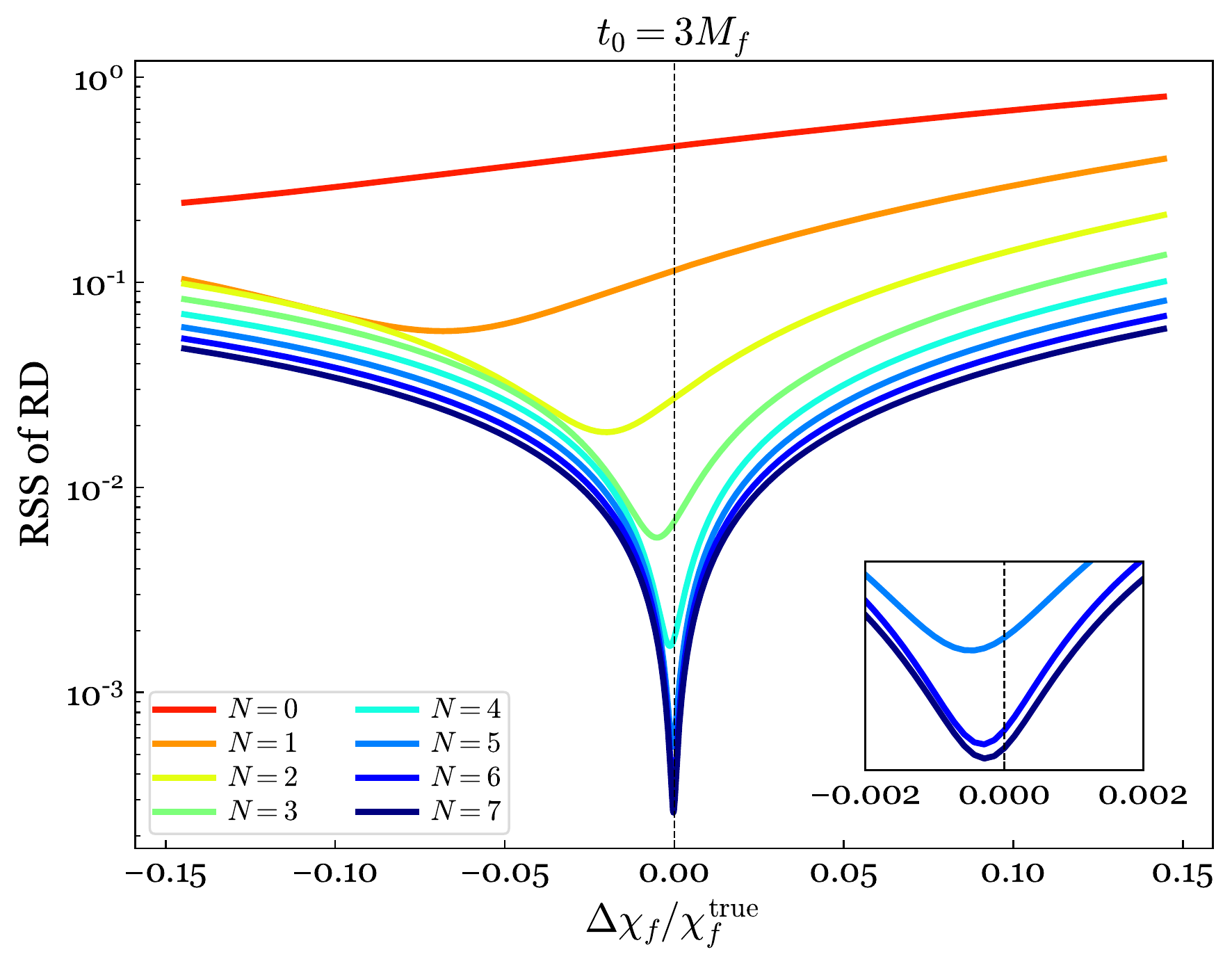}
        \includegraphics[width=\columnwidth,height=6.8cm,clip=true]{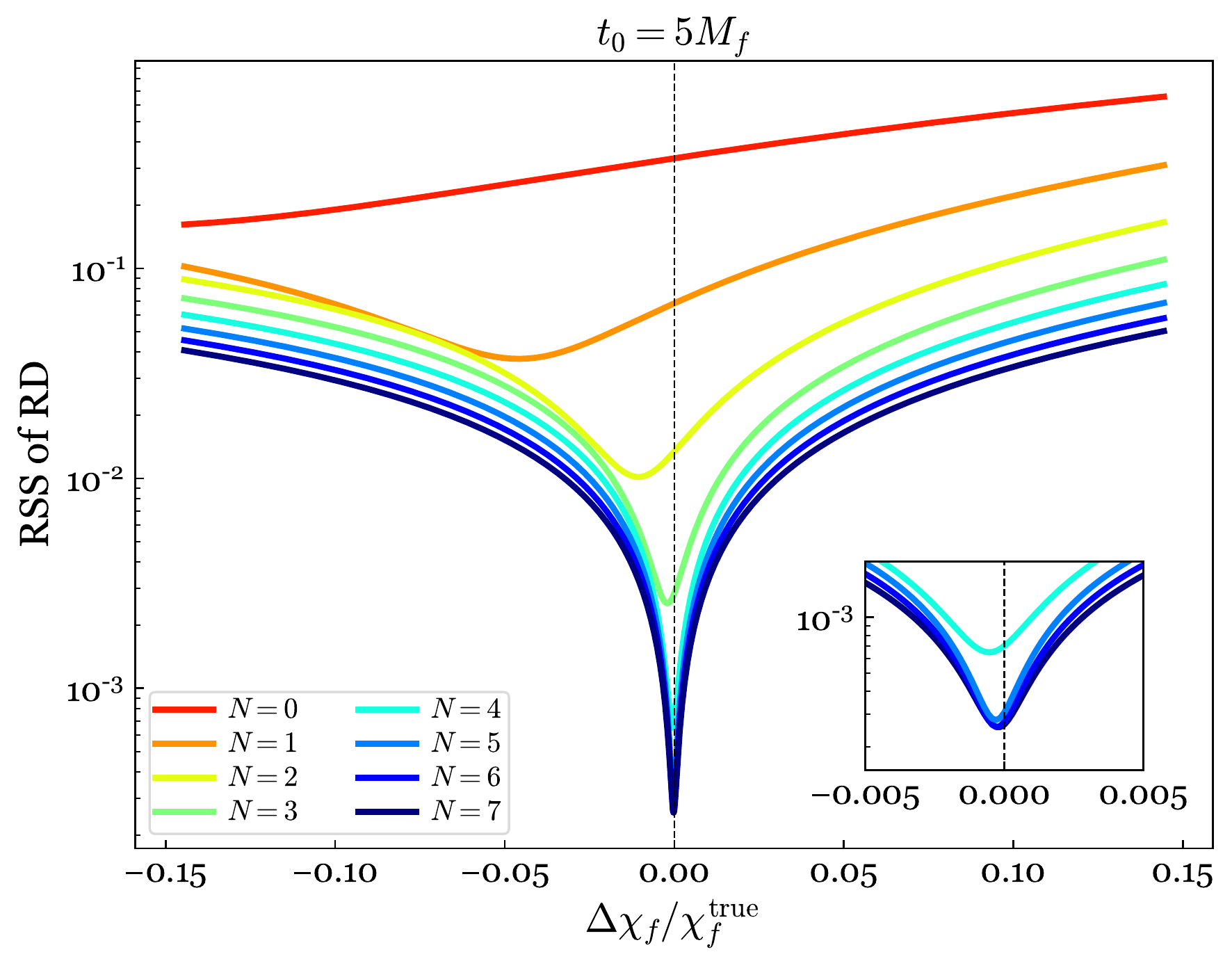}
        \includegraphics[width=\columnwidth,height=6.8cm,clip=true]{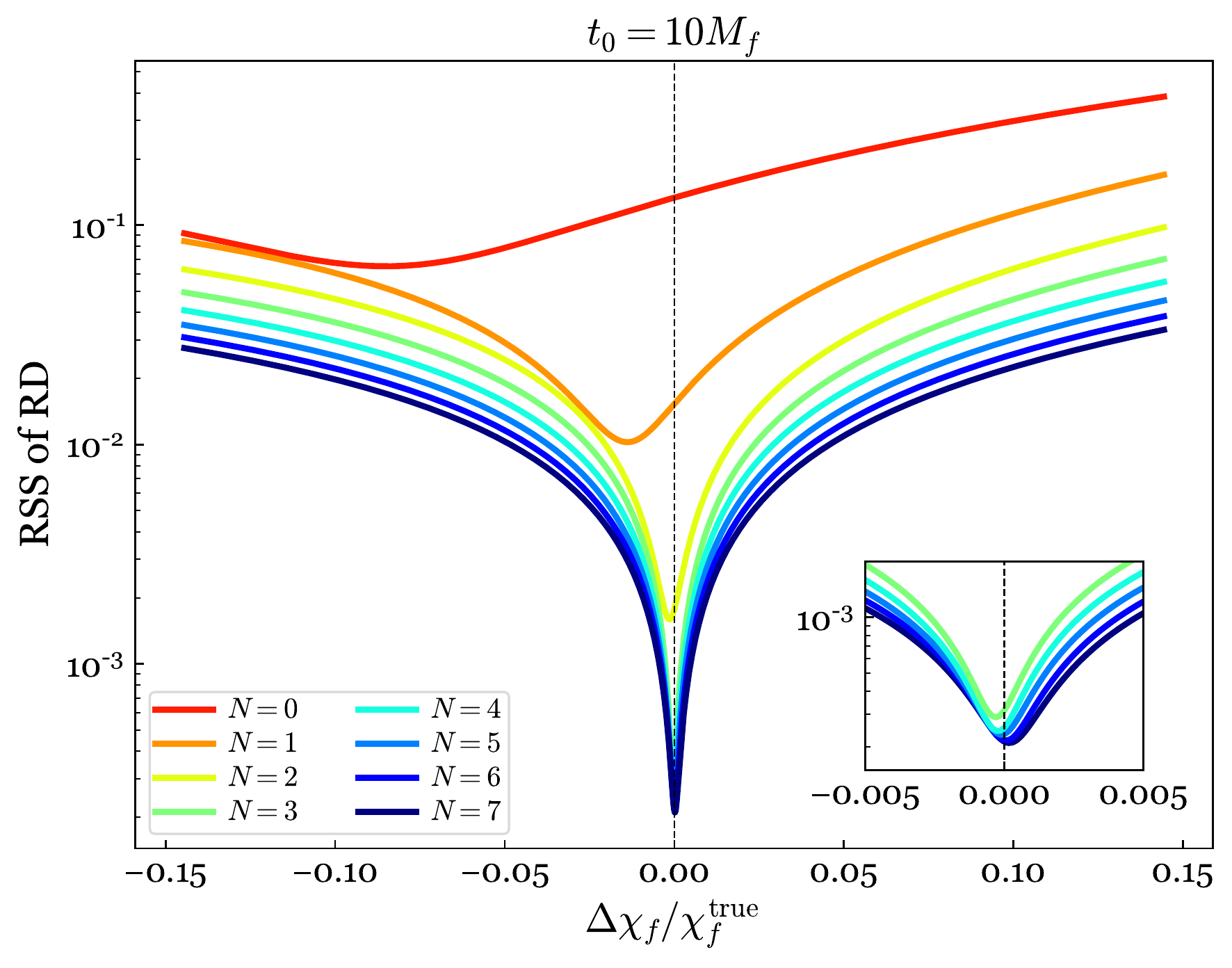}
        \includegraphics[width=\columnwidth,height=6.8cm,clip=true]{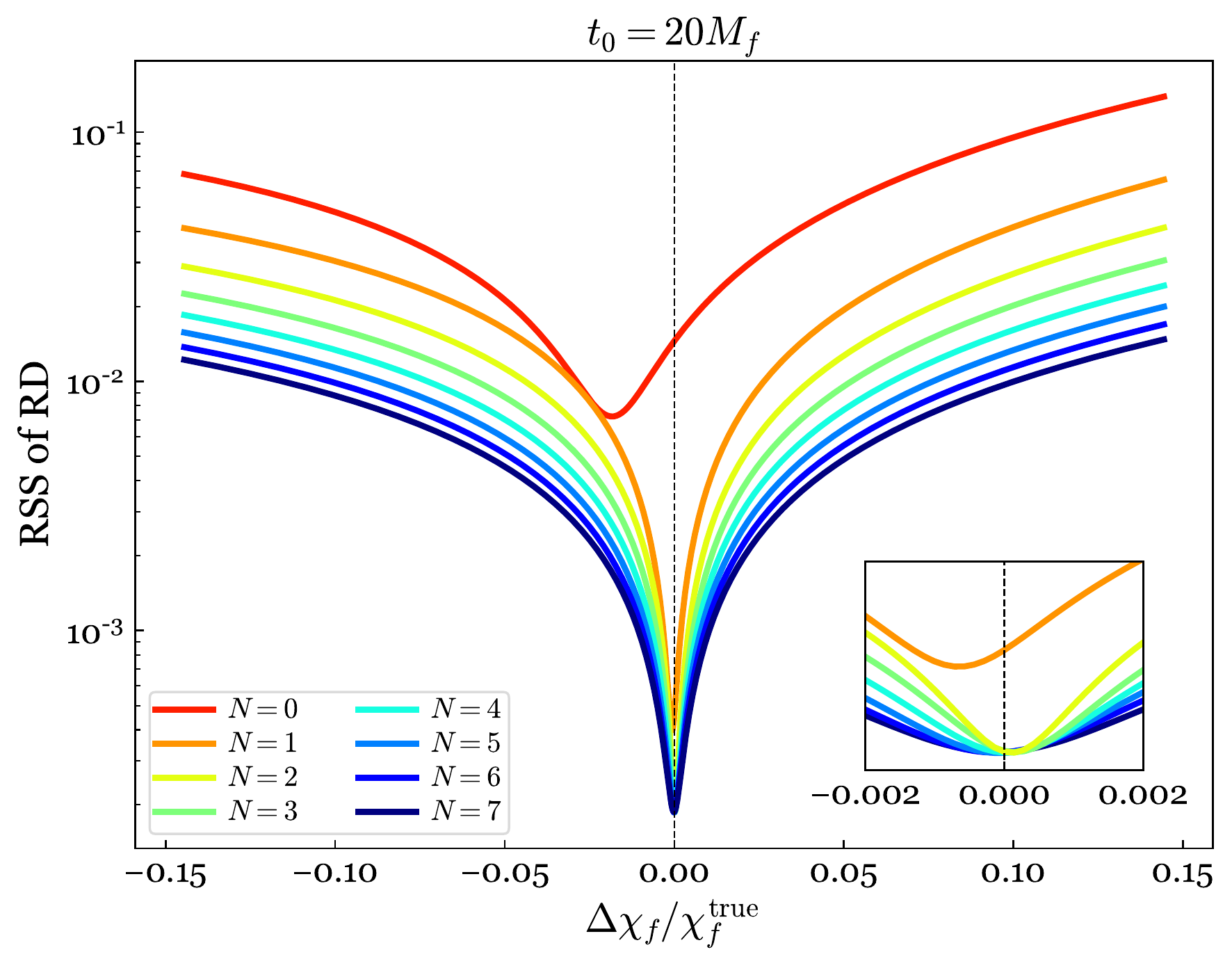}
        \includegraphics[width=\columnwidth,height=6.8cm,clip=true]{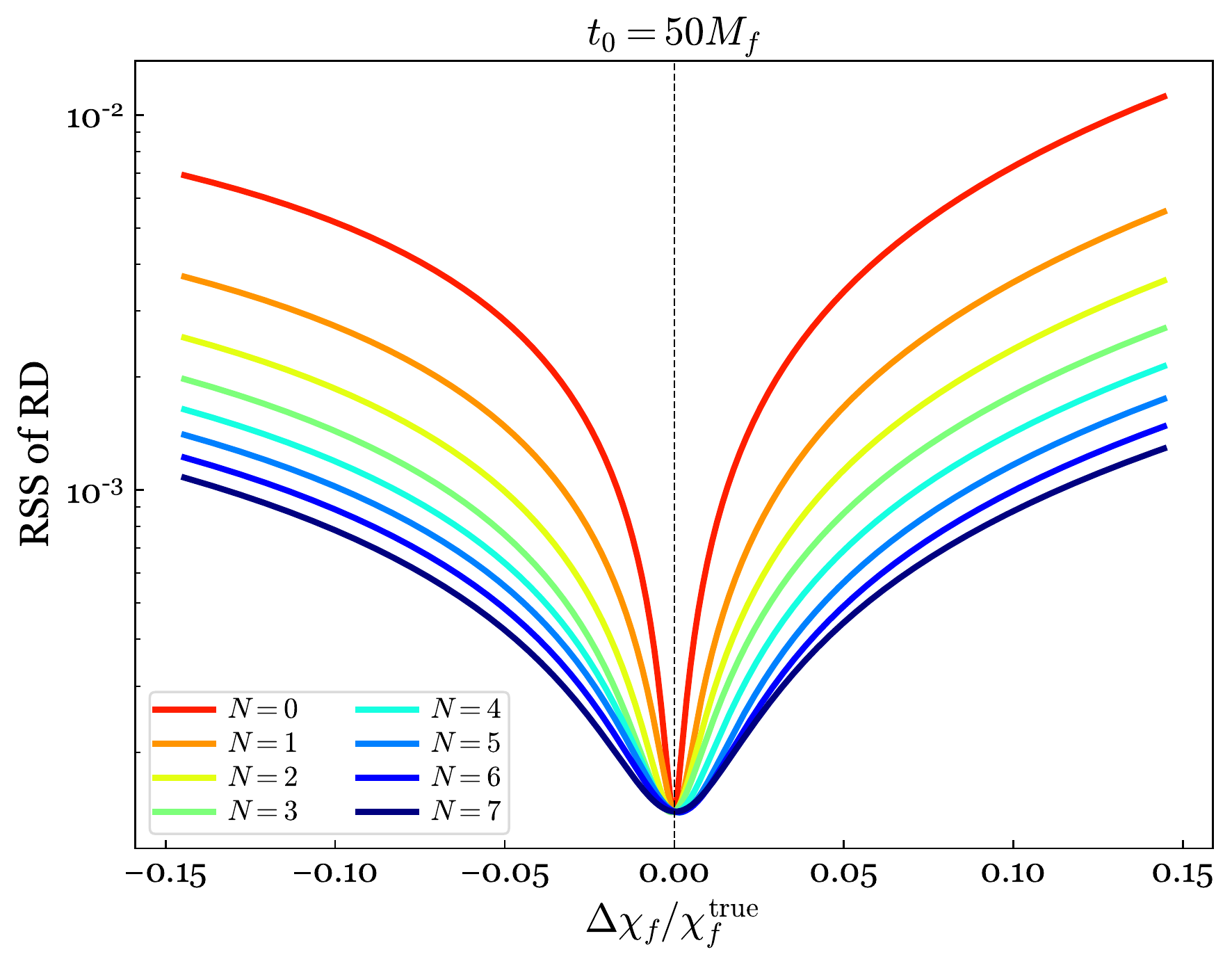}
  \caption{The ringdown RSS of the filtered waveform as a function of $\chi_f$. The \ind~waveform is used. The six panels correspond to different choices of the start time, i.e., $t_0$ in Eq.~\eqref{SNR_filter}. In each panel, different colors indicate the results from removing different numbers of overtones. When $t_0$ is large $(\sim 50M_f)$, the true value of the spin $\chi^{\rm true}_f=0.692$ leads to the smallest RSS no matter how many overtones are removed. However, if we push $t_0$ to an early time, enough overtones need to be removed to obtain the true value. On the other hand, the RSS depends strongly on $\chi_f$: a $2\%$ change in $\chi_f$ can result in around two orders of magnitude change in the RSS, when $t_0$ and $N$ are fixed to their true values.}
 \label{fig:remnant_properties_spin}
\end{figure*}

\begin{figure}[htb]
    \includegraphics[width=\columnwidth,clip=true]{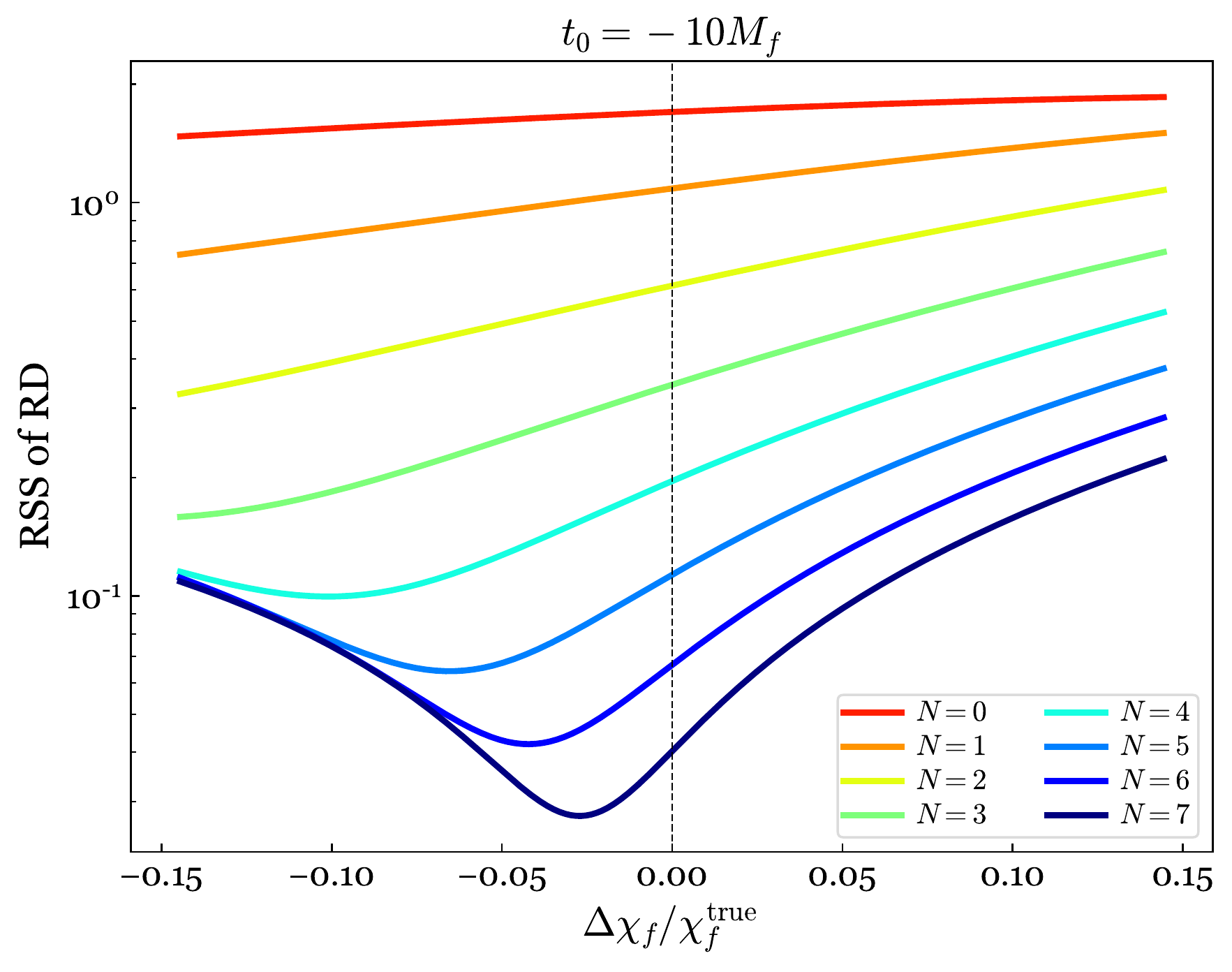}
  \caption{Continuation of Fig.~\ref{fig:remnant_properties_spin}, except that the onset of the ringdown window $t_0$ is set to $-10M_f$.}
 \label{fig:detection_-10}
\end{figure}

\begin{figure*}[htb]
        \subfloat[$t_0=0,N=2$ (zoom in)\label{fig:remnant_properties_all_t0_n2_zoom_in}]{\includegraphics[width=\columnwidth,height=7.3cm,clip=true]{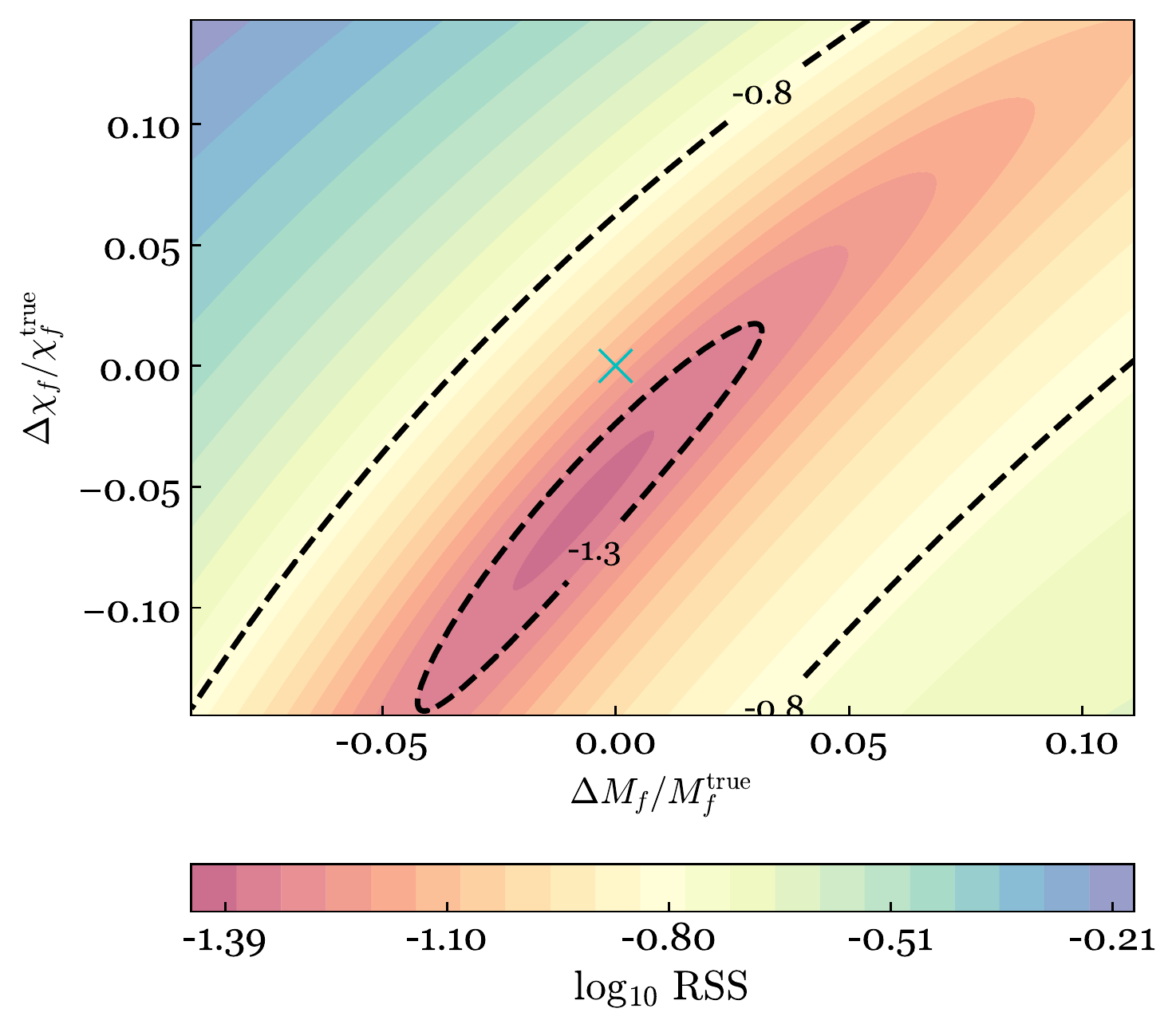}}
        \subfloat[$t_0=0,N=7$ (zoom in)\label{fig:remnant_properties_all_t0_n7_zoom_in}]{\includegraphics[width=\columnwidth,height=7.3cm,clip=true]{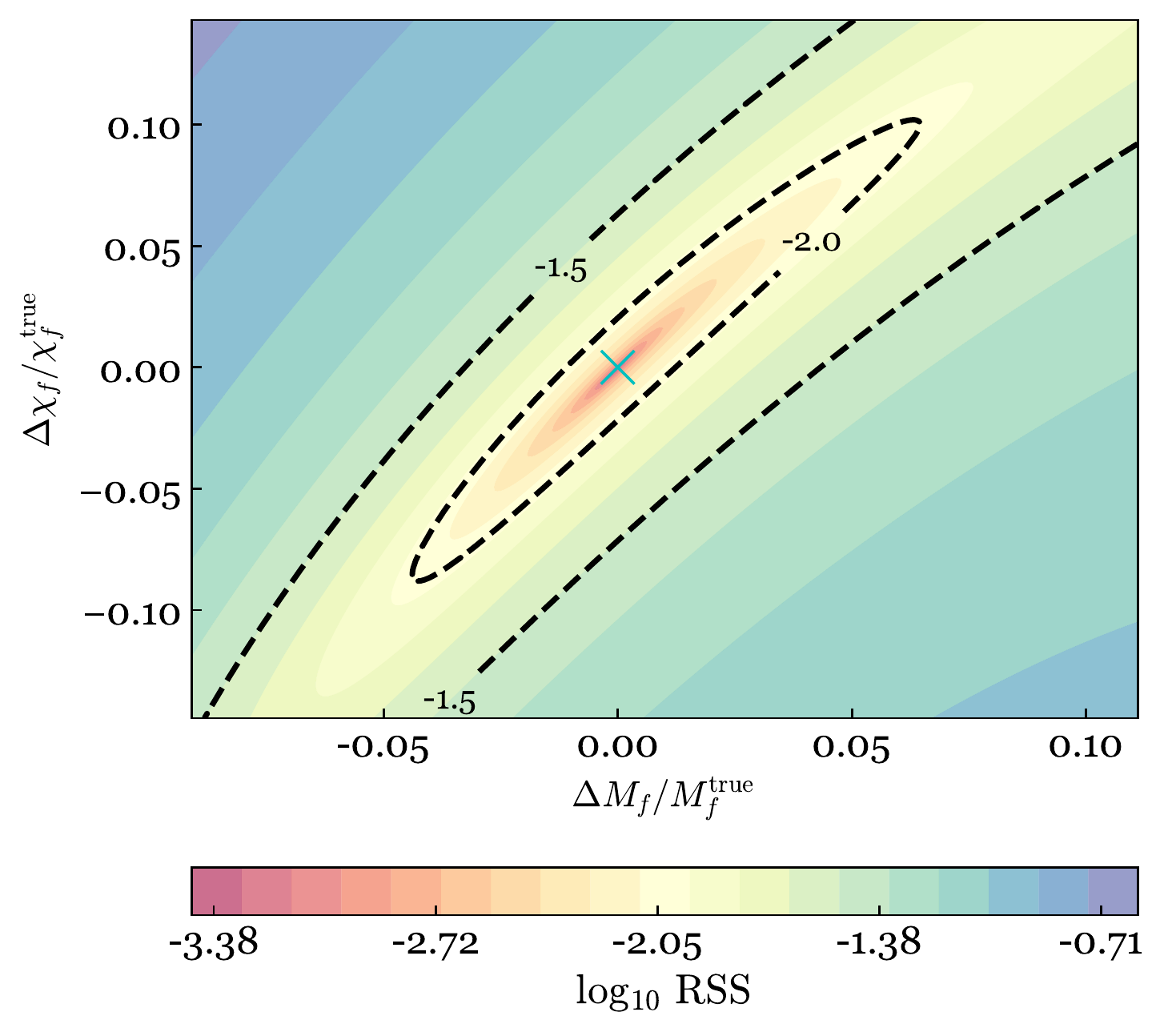}} \\
        \subfloat[$t_0=0,N=2$ (zoom out)\label{fig:remnant_properties_all_t0_n2_zoom_out}]{\includegraphics[width=\columnwidth,height=7.3cm,clip=true]{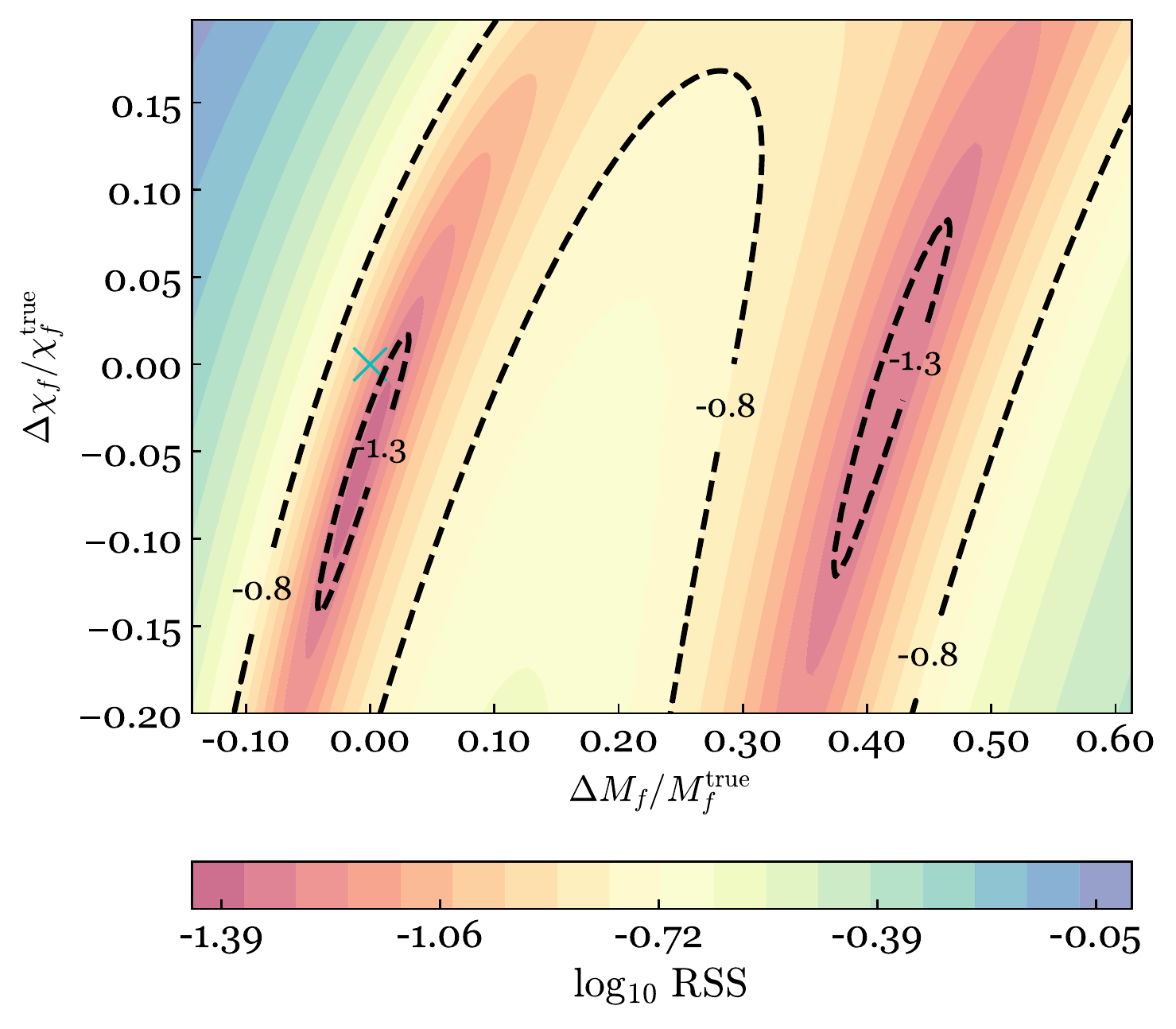}}
        \subfloat[$t_0=0,N=7$ (zoom out)]{\includegraphics[width=\columnwidth,height=7.3cm,clip=true]{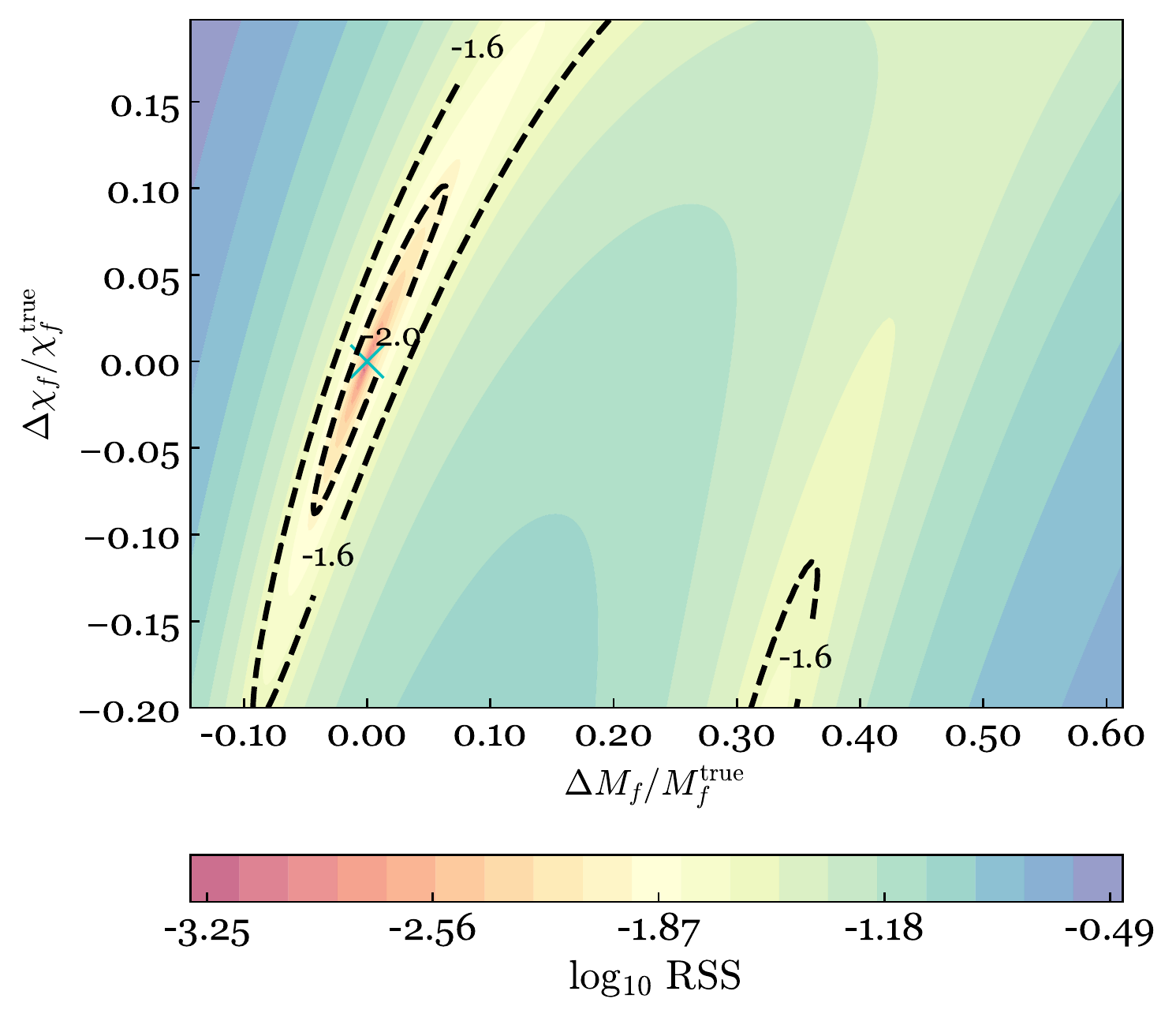}}
  \caption{Contours of RSS with varying $M_f$ and $\chi_f$. To avoid redundancy, we set $t_0$ to 0 and choose $N=2$ (left panels) and $N=7$ (right panels). In the top row, we explore the parameter space near the true remnant properties, whereas in the bottom row we investigate a larger area. The true remnant mass and spin are marked with a cross. The effects of $M_f$ and $\chi_f$ are degenerate --- their difference is more constrained than their sum. In addition, we find there is a second local minimum in Fig.~\ref{fig:remnant_properties_all_t0_n2_zoom_out}.}
 \label{fig:remnant_properties_all}
\end{figure*}

Next, we turn on the perturbation. Since Cheung \etal \cite{Cheung:2021bol} found the spectral instability with $b$ varied, in Fig.~\ref{fig:wronskian-a}, we first compute the filter with two choices of $b$, while fixing $\epsilon=0.1$. We find the modification to the original signal is negligible. The major change is a series of echoes with an interval of $\Delta t\sim 2b$ --- well separated from the original signal (in the plot we only show the first one or two echoes). Meanwhile the amplitude of the echo is independent of $b$. We remark that the $\delta$-function is removed from echoes since the TS coefficient $|D_{lm}/C_{lm}|\to0$ as $\omega\to\infty$. Then in Fig.~\ref{fig:wronskian-b}, we fix the value of $b$ to $200M_f$ but vary $\epsilon$. Again, the perturbation has little impact on the original signal, and the amplitude of the echo scales linearly with $\epsilon$. Compared to the recent work by Berti \etal \cite{Berti:2022hwx}, our studies include not only the fundamental mode, as Berti \etal \cite{Berti:2022hwx} did, but also more overtones. Nevertheless, the qualitative features in our results are the same as theirs. Finally, Figure \ref{fig:wronskian_freq} shows the real part of $\mathcal{F}^{D~{\rm ECO}}_{22}(\omega)$ in the frequency domain for completeness.

\section{Inferring remnant properties from the rational filter}
\label{sec:remnant_depedence}
% \B{true}
% \B{cite \cite{Bhagwat:2019dtm}!!!}
We have shown that our rational filter $\mathcal{F}_{lmn}$ is able to remove a specific QNM $\omega_{lmn}$ from the ringdown regime and reduce the root sum square (RSS) of the ringdown. In particular, the ringdown signal can be almost completely removed if we apply a filter with a series of corresponding modes. Since the mode frequencies $\omega_{lmn}$ are determined by the mass $M_f$ and spin $\chi_f$ of the remnant BH, in this section, we investigate how the ringdown RSS decreases depending on the choices of $M_f$ and $\chi_f$.

\begin{figure}[htb]
    \includegraphics[width=\columnwidth,clip=true]{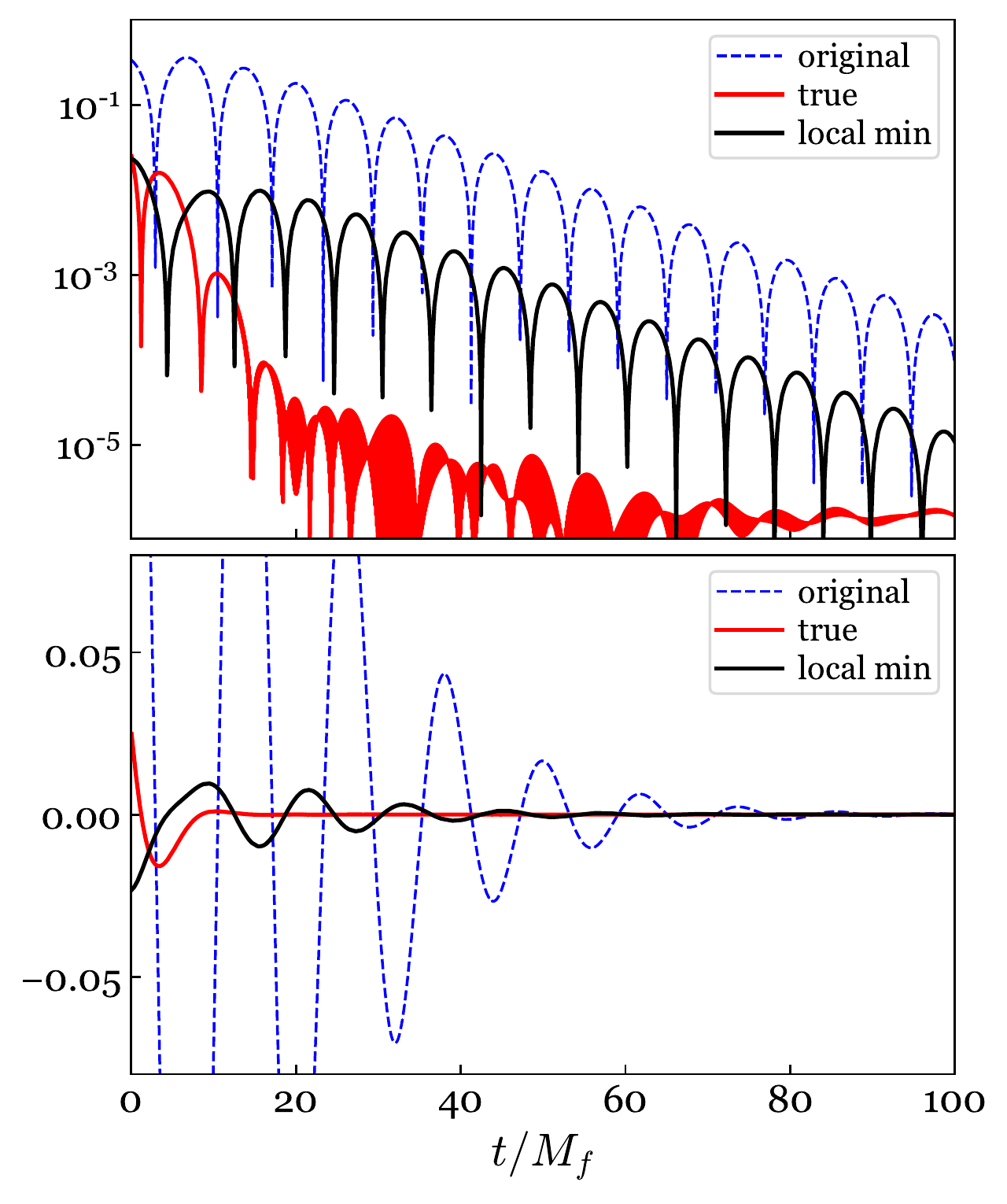}
  \caption{An explanation for the second local minimum in Fig.~\ref{fig:remnant_properties_all_t0_n2_zoom_out}. The blue dashed line corresponds to the original harmonic $h_{22}$ of \ind. Using the true remnant properties, the corresponding QNMs are removed (red curve). However, it has a larger amplitude at around 0. This is because adjacent overtones contribute destructively to the original waveform. Fewer QNMs reduce this cancellation and lead to a larger amplitude. On the contrary, using the remnant properties at the second local minimum (black curve), the amplitude of the original waveform diminishes even though the corresponding QNMs are not filtered away. As a result, two systems lead to similar RSS.}
 \label{fig:wrong_mass_spin_example}
\end{figure}

We define the RSS of a filtered harmonic $h_{lm}^f(t)$ within a time interval $[t_0,t_1]$ to be
\begin{align}
    {\rm RSS} = \sqrt{\int_{t_0}^{t_1}|h^f_{lm}(t)|^2dt}. \label{SNR_filter}
\end{align}
We still take the GW150914-like waveform \ind~as an example. We fix $t_1$ to $100M_f$ and let $t_0$ vary. Then we apply a filter:
\begin{align}
    \mathcal{F}_{N}(M_f,\chi_f)=\mathcal{F}_{320}(M_f,\chi_f)\prod_{n=0}^{N} \mathcal{F}_{22n}(M_f,\chi_f),
\end{align}
to the harmonic $h_{22}$. The filter $\mathcal{F}_{N}$ is a function of remnant mass $M_f$ and spin $\chi_f$. It also depends on how many $(l=2,m=2)$ overtones we want to remove. We want to emphasize that our rational filter leads to a time shift, and in Sec.~\ref{sec:applications} we undid it by aligning early inspiral waveforms. However, in this case we find the alignment can pull some non-ringdown signals into the regime that we are interested in $(t>t_0)$ and make our analyses fail. To avoid this, here we do not perform this alignment. A caveat of this compromise is that the time shift itself can reduce the RSS, even though it is a subdominant effect. In this paper, we ignore the contribution due to this time shift, and leave more self-contained studies for future work.

In Fig.~\ref{fig:remnant_properties_spin}, we vary the value of $\chi_f$ with different choices of $N$ and $t_0$ while keeping $M_f$ fixed at the true value. When $t_0$ is large $(\sim 50M_f)$, we see the true value $\chi^{\rm true}_f=0.692$ leads to the smallest RSS (namely the ringdown is mostly removed) regardless of the value of $N$. This is because in the regime of $t \gtrsim 50M_f$, the signal is dominated by the fundamental mode $\omega_{220}$, and removing $\omega_{220}$ alone is enough to reduce the RSS down to roughly the numerical noise level. 
% On the other hand, the signal also obtains a tiny contribution from the first overtone $\omega_{221}$ (see Fig.~\ref{fig:GW150914_h22_final}), filtering out $\omega_{221}$ can lower the RSS by a little amount, but further removing higher overtones does not continue to reduce the RSS.
However, if we push $t_0$ to an early time, failing to filter out enough modes will result in incorrect values of $\chi_f$ when RSS achieves its local minimum --- the value $\chi_f$ is degenerate with the choice of $t_0$. Especially, in the first panel of Fig.~\ref{fig:remnant_properties_spin}, we see that the ringdown RSS depends monotonically on $\chi_f$ when $t_0=0$ and $N=0$; but the local minimum of the RSS does converge to the true value of $\chi_f$ after we include enough overtones. If we continue to go to an earlier regime, such as $t_0=-10M_f$ in Fig.~\ref{fig:detection_-10}, we can see that the inferred spin is biased even when enough overtones are included, because of the presence of non-ringdown signals (e.g., late inspiral and merger). On the other hand, we also investigate the effect of $M_f$. We find that varying the value of $M_f$ (with $\chi_f$ fixed to the true value) leads to a similar impact on the ringdown RSS, and the results are summarized in Appendix \ref{sec:remnant_properties_mass}. 

Our results shown in Figs.~\ref{fig:remnant_properties_spin} and \ref{fig:remnant_properties_mass} are closely related to Fig.~7 of Ref.~\cite{Bhagwat:2019dtm}, in which the authors show how the mismatch varies with deviations from GR and the start time of analyses. Similarly, our results indicate that the residual RSS depends strongly on the choice of $(M_f,\chi_f)$. In our case, a $2\%$ change in $\chi_f$ can result in around two orders of magnitude change in the RSS, when $t_0$ and $N$ are fixed to their ``true'' values.

\begin{figure}[htb]
        % \subfloat[$t_0=10M_f,N=0$]{\includegraphics[width=\columnwidth,height=7.3cm,clip=true]{all_non_kerr_t10n0}}
        \subfloat[$t_0=50M_f,N=0$]{\includegraphics[width=\columnwidth,height=7.3cm,clip=true]{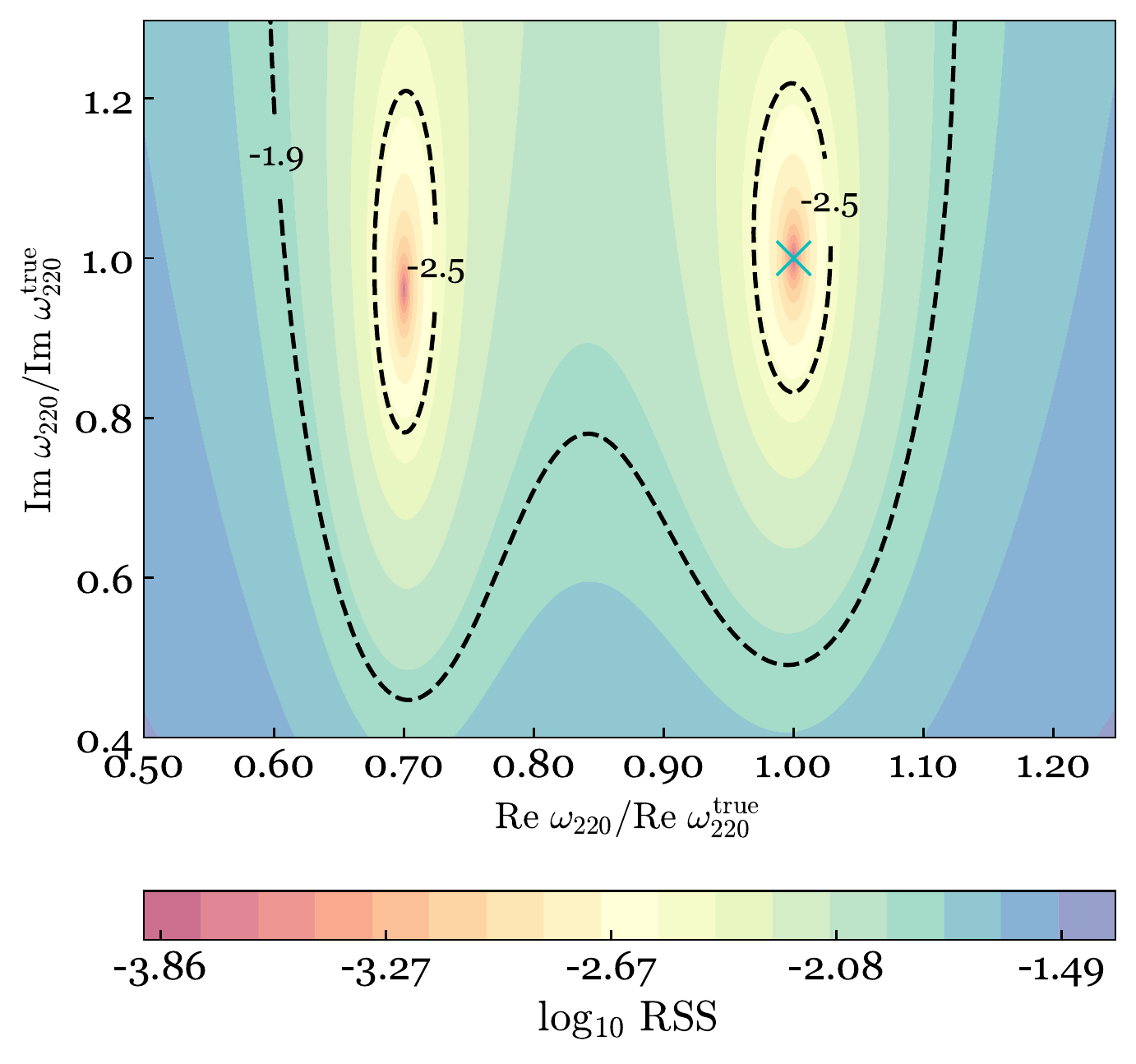}}
  \caption{Same as Fig.~\ref{fig:remnant_properties_all}, except that the real and imaginary parts of the fundamental mode are used as two independent variables. The start time $t_0$ is set to $50M_f$. Similar to Fig.~\ref{fig:remnant_properties_all_t0_n2_zoom_out}, there is a second local minimum.}
 \label{fig:remnant_properties_all_non_kerr}
\end{figure}

After studying the effects of $M_f$ and $\chi_f$ separately, in Fig.~\ref{fig:remnant_properties_all} we provide contours of RSS with varying them together. To avoid redundancy, we set $t_0$ to 0 and focus on two cases: $N=2$ and $N=7$, respectively. If we restrict ourselves to the region near the true remnant properties (Figs.~\ref{fig:remnant_properties_all_t0_n2_zoom_in} and \ref{fig:remnant_properties_all_t0_n7_zoom_in}), the $N=2$ one leads to biases in extracting $M_f$ and $\chi_f$, whereas the latter one can recover the remnant properties (marked with a cross) accurately. In addition, we notice that the effects of $M_f$ and $\chi_f$ are partially degenerated --- their difference $\sim M_f-\chi_f$ is more constrained than their sum $\sim M_f+\chi_f$. This is consistent with Figs. 10 and 11 of Ref.~\cite{Giesler:2019uxc}. On the other hand, if we explore a larger parameter space (zoom out), we find there is a second local minimum in Fig.~\ref{fig:remnant_properties_all_t0_n2_zoom_out}. To explore the reason, in Fig.~\ref{fig:wrong_mass_spin_example} we compare two filtered waveforms with $\chi_f$ and $M_f$ chosen at their true values (red curve) and at the second local minimum (black curve), respectively. Recall that the amplitudes of adjacent overtones are out of phase, e.g., Refs.~\cite{Giesler:2019uxc} and \cite{Ma:2021znq}, they contribute destructively to the final ringdown waveform. Removing some overtones can increase the value of the filtered waveform at an early stage. On the contrary, when $\chi_f$ and $M_f$ are at the second local minimum, even though the corresponding QNMs are not removed, the amplitude of the filtered waveform is reduced by around one order of magnitude. As a result, both cases lead to comparable RSS.

So far, we take $(M_f,\chi_f)$ as two independent variables. The QNM frequencies are obtained by assuming Kerr BHs with GR gravity. In Fig.~\ref{fig:remnant_properties_all_non_kerr}, we relax this assumption and use the real and imaginary parts of a QNM as two independent variables. Here we restrict ourselves to the fundamental mode alone for simplicity, and take $t_0=50M_f$. We find the qualitative feature remains the same --- there is a second local minimum, and the reason is exactly the same as that of Fig.~\ref{fig:remnant_properties_all_t0_n2_zoom_out}.

Our discussions indicate that the filter could serve as a new tool to infer the remnant properties from actual detection data, and we refer the interested reader to our follow-up work \cite{Ma_filter_ligo_data} for more discussions.

% In particular, including the seventh overtone $\omega_{227}$ can further reduce the RSS for $t>0$ by a factor of 2.1, compared to removing overtones up to $\omega_{226}$. This fact supports the existence of $\omega_{227}$ in the early stage of a ringdown signal, consistent with the results shown in Ref.~\cite{Giesler:2019uxc}. 
% We note that in the current case $(t_0=0,N=7)$, the inferred spin $\chi_f$ from the global minimum still deviates from the true value $0.692$ by a factor of $0.04\%$. We attribute this to numerical errors. 

\section{Conclusion}
\label{sec:conclusion}
We have proposed two types of frequency-domain filters that are able to remove QNM(s) from ringdown signals. Our new method serves as a complementary tool to previous studies where the ringdown was analyzed in terms of time-domain fitting (e.g., Ref.~\cite{Giesler:2019uxc}) --- it allows visualizing the existence of subdominant modes without the risk of overfitting. By applying our filter to the waveform of \ind, we find the spherical-spheroidal mixing mode $\omega_{320}$ in harmonic $h_{22}$, the presence of $\omega_{220}$ in $h_{21}$ due to the gravitational recoil, and second-order effects in $h_{44},h_{54}$ and $h_{55}$ due to the quadratic coupling $h_{22}^2$ and $h_{22}h_{33}$. We also find the existence of retrograde modes in waveform \indnew. Our filter leads to an unphysical flipped ringdown prior to the start time of the real ringdown. Consequently, the late-inspiral and merger signals are contaminated. 
% But on the other hand, this feature allows us to estimate the start time of the corresponding QNM.

Although the rational filter in Eq.~\eqref{filter_final} is constructed purely empirically, the full filter $\mathcal{F}^D_{lm}$ in Eq.~\eqref{D_filter_phase} reflects the nature of the BH, and the filtered waveform corresponds to the image wave on the past horizon (Fig.~\ref{fig:image_wave}). Furthermore, in spite of the unstable nature of QNM spectra \cite{Jaramillo:2020tuu,Cheung:2021bol}, we find that the filter $\mathcal{F}^D_{lm}$ is stable in the time domain under the perturbations of the BH potential, in the sense that the original response remains unmodified, while the major correction appears as periodic echoes well-separated from the original BH response. The time interval and amplitude of the echoes depend linearly on the parameters of the perturbation.

Additionally, the rational filter takes the mass and spin of the remnant BH as free parameters. The residual ringdown RSS depends strongly on the choice of these two parameters. The true remnant properties could be recovered accurately from the ringdown of $h_{22}$ as long as one consider a proper number of overtones and the start time of the analysis. 
% These results are consistent with Ref.~\cite{Giesler:2019uxc}.

In this paper, we demonstrate that this new approach is powerful in ringdown analyses and outline a few applications. Future studies could be focused on:

(i) Nonlinearity due to the quadratic couplings. We focused exclusively on a few harmonics of \ind, and exhibited the existence of second-order effects only qualitatively. A more systematic study \cite{Mitman_nonlinear} is needed to investigate quadratic couplings in other BBH systems. We also refer the interested reader to Ref.~\cite{Cheung_nonlinear} for relevant discussions. 
% In addition, it would be also important to compare the results, such as the mode amplitudes, with the prediction of BH second-order perturbation theory \cite{Loutrel:2020wbw,Ripley:2020xby}. 

(ii) Second-order effects in the multipole moments of dynamical horizons. Although Refs.~\cite{Pook-Kolb:2020jlr,Mourier:2020mwa} have shown that the multipole moments might be consistent with the superposition of linear QNMs soon after the formation of the common horizon, it is expected that a majority of nonlinearities are swallowed by horizons \cite{Okounkova:2020vwu}, which in turn should leave imprints on dynamical horizons. It is interesting to study these cases by applying our filters.

(iii) The stability of the two filters. In this work, we considered the stability of the full filter under a simple scenario: the perturbation arising only through a reflective boundary condition at the ECO surface that is very close to the would-be horizon. More sophisticated perturbations, e.g., the ones in Ref.~\cite{Jaramillo:2020tuu,Cheung:2021bol}, could be investigated. In addition, it might also be interesting to study the (in)stability of the rational filter. This requires high-precision calculations of QNMs of the new system. The goal of this trend is to answer: How to distinguish a BH from its mimicker via our filters? And how do the filters reflect the nature of the system?

(iv) Inferring remnant properties from real observational data. Here we restricted ourselves to a particular harmonic $h_{22}$ and found that the remnant properties could be recovered accurately. A possible avenue for future work is to investigate the impact of our filters on the strain that is emitted toward a single angular direction. More importantly, one could apply our filter to real BBH events \cite{Ma_filter_ligo_data} and see whether we could place a tighter constraint on the remnant mass, spin, and also the no-hair theorem \cite{Isi:2019aib}. 

% \ls{Maybe also add some discussion about the limitation of the current analysis (varying mass and spin one at a time), and the extended study of estimating mass and spin at the same time (which is needed for analyzing real events).}

(v) Other filters. In this work, we have studied two related filters. One undesired feature of the rational filter is that it leads to a backward time-shift, which makes it difficult to define the start time of the ringdown in the filtered waveform\footnote{We show that the choice of the start time has a large impact on inferring remnant properties.}. The full filter does not have this problem but is more computationally expensive to obtain. Therefore it might be interesting to look for other new filters with better properties.

% \B{BH determinant? \cite{Denef:2009kn}}

%\section{When the mode frequency is irrelevant}

%\subsection{The sensitivity of the filter}
%As shown in Fig.~\ref{fig:sensitivity}, a $2\%$ deviation in spin magnitude can lead to a factor of 5 change in amplitude.
%
%\begin{figure}[htb]
%        \includegraphics[width=\columnwidth,clip=true]{sensitivity}
%  \caption{ GW190514, after removing the (2,2) pro+retro modes}.
% \label{fig:sensitivity}
%\end{figure}
%
%
%For GW150914, if we apply the $(l=8,m=8,n=8)$ filter twenty times (mildly), there is only a time shift toward early time. If we apply the filter 500 times (aggressively), the inspiral signal keeps moving toward early time, whereas the ringdown signal stays at $t=0$. We can use this fact to infer the starting time of the ringdown.
%\begin{figure}[htb]
%        \includegraphics[width=\columnwidth,clip=true]{Irrelevant}
%  \caption{ GW190514, $(l=8,m=8,n=8)$, 20 times vs. 500 times}.
% \label{fig:Irrelevant}
%\end{figure}

% \subsubsection{\B{$h_{20}$: The memory}}
% \section{Detection}
% In Fig.~\ref{fig:detection_22+44}
% \begin{figure}[htb]
%         \includegraphics[width=\columnwidth,clip=true]{detection_22+44}
%   \caption{ GW150914, $h_{44}+h_{22}$}.
%  \label{fig:detection_22+44}
% \end{figure}
% \begin{figure}[htb]
%         \includegraphics[width=\columnwidth,clip=true]{aggressive_filter_h22}
%   \caption{ GW150914, $h_{22}$}.
%  \label{fig:}
% \end{figure}
%==========================================================================
\begin{acknowledgments}
This work makes use of the Black Hole Perturbation Toolkit. We thank Mark Ho-Yeuk Cheung and Emanuele Berti for sharing their results about second-order effects. We thank Maximiliano Isi, Macarena Lagos, Leo C. Stein, Lam Hui and Saul Teukolsky for productive discussions. We also thank useful discussions with all the attendees at the CCA ringdown workshop. Finally, we are grateful to the anonymous Referee(s) for suggesting the name ``quasinormal-mode filter''. This work was supported in part by the Brinson Foundation, the Simons Foundation (Award Number 568762), the Sherman
Fairchild Foundation and by NSF Grants No.~PHY-2011961, No.~PHY-2011968, and
No.~OAC-1931266 at Caltech, as well as NSF Grants No.~PHY-1912081 and No.~OAC-1931280
at Cornell.
L.S. acknowledges the support of the Australian Research Council Centre of Excellence for Gravitational Wave Discovery (OzGrav), Project No. CE170100004. 
\end{acknowledgments}  

%==========================================================================
\appendix
\section{The up-mode solution of an ECO}
\label{sec:eco_up_mode}
\begin{subequations}
Near the ECO surface, Chen \etal \cite{Chen:2020htz} proposed a physical boundary condition via a family of zero-angular-momentum fiducial observers (FIDOs). The tidal tensor field within the rest frame of the FIDOs is given by \cite{Zhang:2012jj}
\begin{align}
    \mathcal{E}_{ij}=h^a_ih^c_j C_{abcd}U^bU^d,
\end{align}
where $C_{abcd}$ is the Weyl tensor, $U^b$ is the four-velocity of the FIDOs, and $h_i^a=\delta^a_i+U^aU_i$ is the projection operator. Chen \etal argues that the tidal response of the ECO, namely the reflection of incident GWs, is proportional to the transverse component of the tidal field:
\begin{align}
    \mathcal{E}_{\rm transverse}\sim-\frac{\Delta}{4r^2}\psi_0-\frac{r^2}{\Delta}\psi_4^*, \label{tidal_tensor}
\end{align}
where $\psi_0$ and $\psi_4$ are the Weyl scalars. The coefficient depends on the nature of the ECO, such as the reflectivity $\mathcal{R}$ in Eq.~\eqref{reflectivity}. By adopting this type of boundary condition, Xin \etal \cite{Xin:2021zir} shows that the ratio between the reflective wave and the incident wave reads \footnote{The additional factor $(-1)^l$ is due to the assumption that the system is invariant under reflection across the $x$-$y$ plane \cite{Boyle:2014ioa}.} [Eq.~(56) of \cite{Xin:2021zir}]:
\begin{align}
    \frac{{\rm Reflective~wave}}{{\rm Incident~wave}}=\frac{(-1)^{l+m+1}}{4}\mathcal{R}\frac{D_{lm}}{C_{lm}}, \label{ECO_boundary_condition}
\end{align}
with
\begin{align}
    & D_{lm}=64(2r_+)^4ik(k^2+4\epsilon^2)\left(-ik+\frac{\sqrt{1-\chi^2}}{r_+}\right), \\
    &|C_{lm}|^2=(Q^2+4\chi\omega m-4\chi^2\omega^2) \notag \\
    &\times[(Q-2)^2+36\chi\omega m-36\chi^2\omega^2]+144\omega^2(1-\chi^2) \notag \\
    &+(2Q-1)(96\chi^2\omega^2-48\chi\omega m), \\
    &{\rm Im}~C_{lm}=12\omega, \\
    &Q=\lambda+s(s+1)=\lambda+2, \\
    &\epsilon=\frac{\sqrt{1-\chi^2}}{4r_+},\\
    &k=\omega-m\Omega_+,
\end{align}
\end{subequations}
where $\lambda$ is the eigenvalue of spin-weighted spheroidal harmonics and $\Omega_+=\chi/(2r_+)$ is the horizon frequency.

As shown in Fig.~\ref{fig:echo}, if we consider a GW emerging from the horizon with a unity amplitude (ignoring any $r_*$ dependent coefficient), it will bounce back and forth within the cavity formed by the ECO surface and the BH potential. In particular, the observer at infinity will see a main transmissive wave with amplitude $1/D_{lm}^{\rm out}$, followed by a series of echoes. Using the boundary condition in Eq.~\eqref{ECO_boundary_condition}, it is straightforward to obtain the amplitude of the $n$th echo: 
\begin{align}
    \frac{1}{D_{lm}^{\rm out}}\left[\frac{(-1)^{l+m+1}}{4}\mathcal{R}\frac{D_{lm}}{C_{lm}}    \frac{D_{lm}^{\rm in}}{D_{lm}^{\rm out}}\right]^n.
\end{align}
By summing them together, we obtain the total transmissive amplitude:
\begin{align}
    &\sum_{n}\frac{1}{D_{lm}^{\rm out}}\left[\frac{(-1)^{l+m+1}}{4}\mathcal{R}\frac{D_{lm}}{C_{lm}}    \frac{D_{lm}^{\rm in}}{D_{lm}^{\rm out}}\right]^n \notag \\
    &=\frac{1}{D_{lm}^{\rm out}}\frac{1}{1-\frac{(-1)^{l+m+1}}{4}\mathcal{R}\frac{D_{lm}}{C_{lm}}    \frac{D_{lm}^{\rm in}}{D_{lm}^{\rm out}}}.
\end{align}
The inverse of the total amplitude corresponds to $\tilde{D}^{\rm out}_{lm}$ in Eq.~\eqref{up-mode-eco}.

\section{$\mathcal{F}^{D~{\rm ECO}}_{22}$ for a spinning ECO}
\label{app:kerr_D}
\begin{figure*}[htb]
    \centering
\subfloat[$\epsilon=10^{-1}$]{\includegraphics[width=1.0\textwidth]{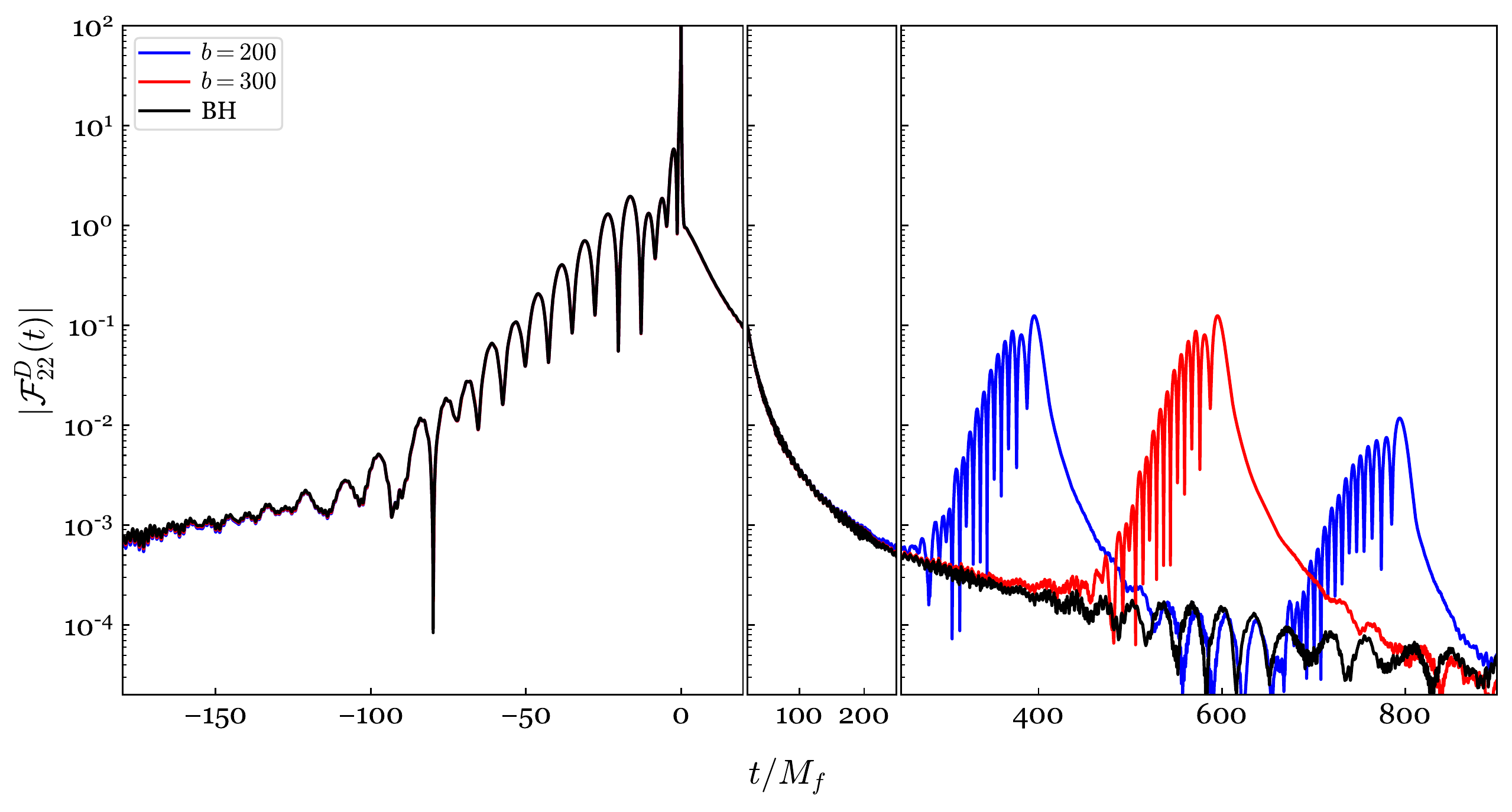}}\\
\subfloat[$b=200M_f$]{\includegraphics[width=1.0\textwidth]{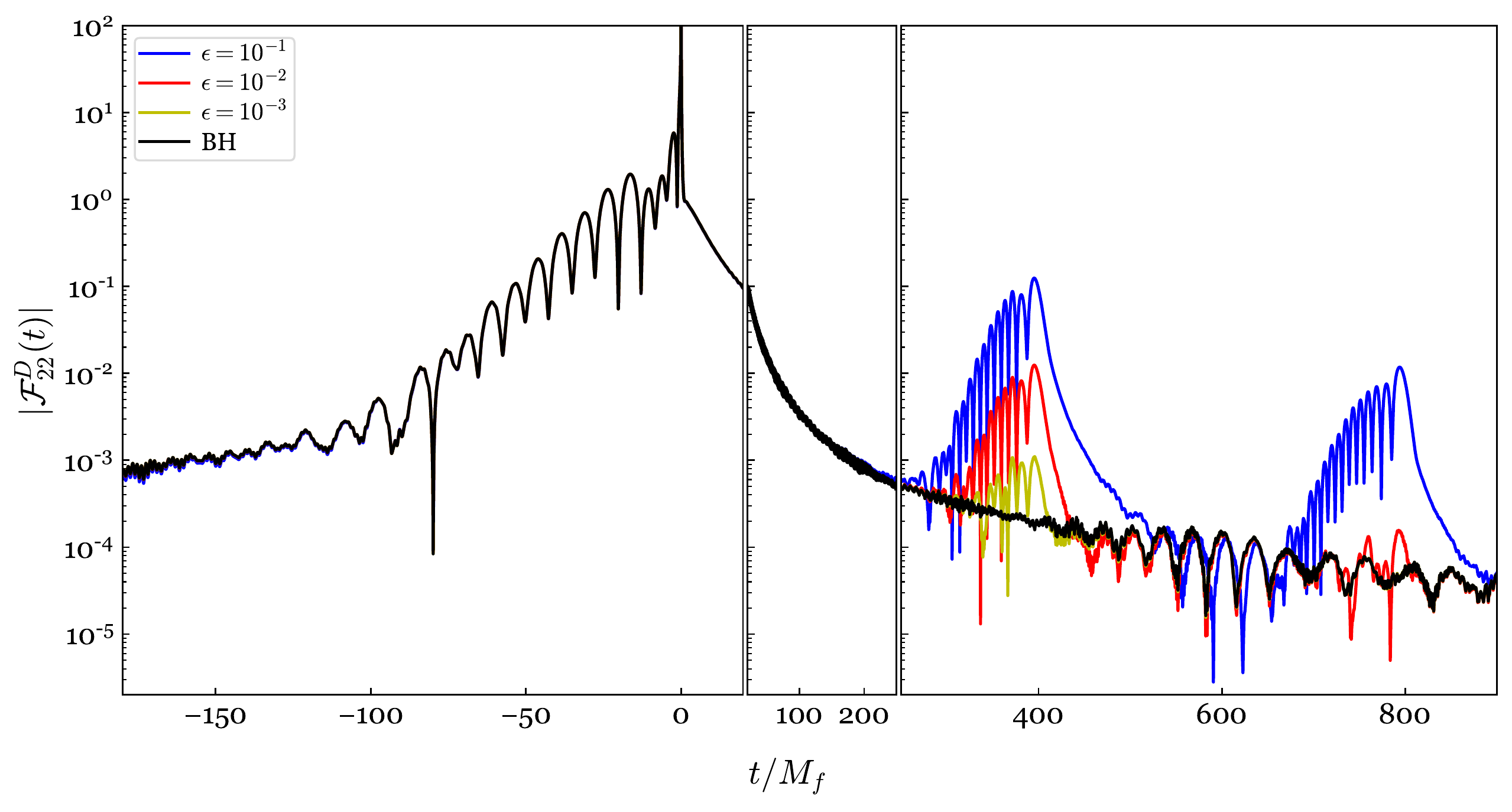}}
  \caption{Same as Fig.~\ref{fig:wronskian}, but for a spinning ECO with $\chi_f=0.692$.}
 \label{fig:kerr_wronskian}
\end{figure*}

\begin{figure*}[htb]
    \centering
\subfloat[$\epsilon=10^{-1}$]{\includegraphics[width=1.0\textwidth]{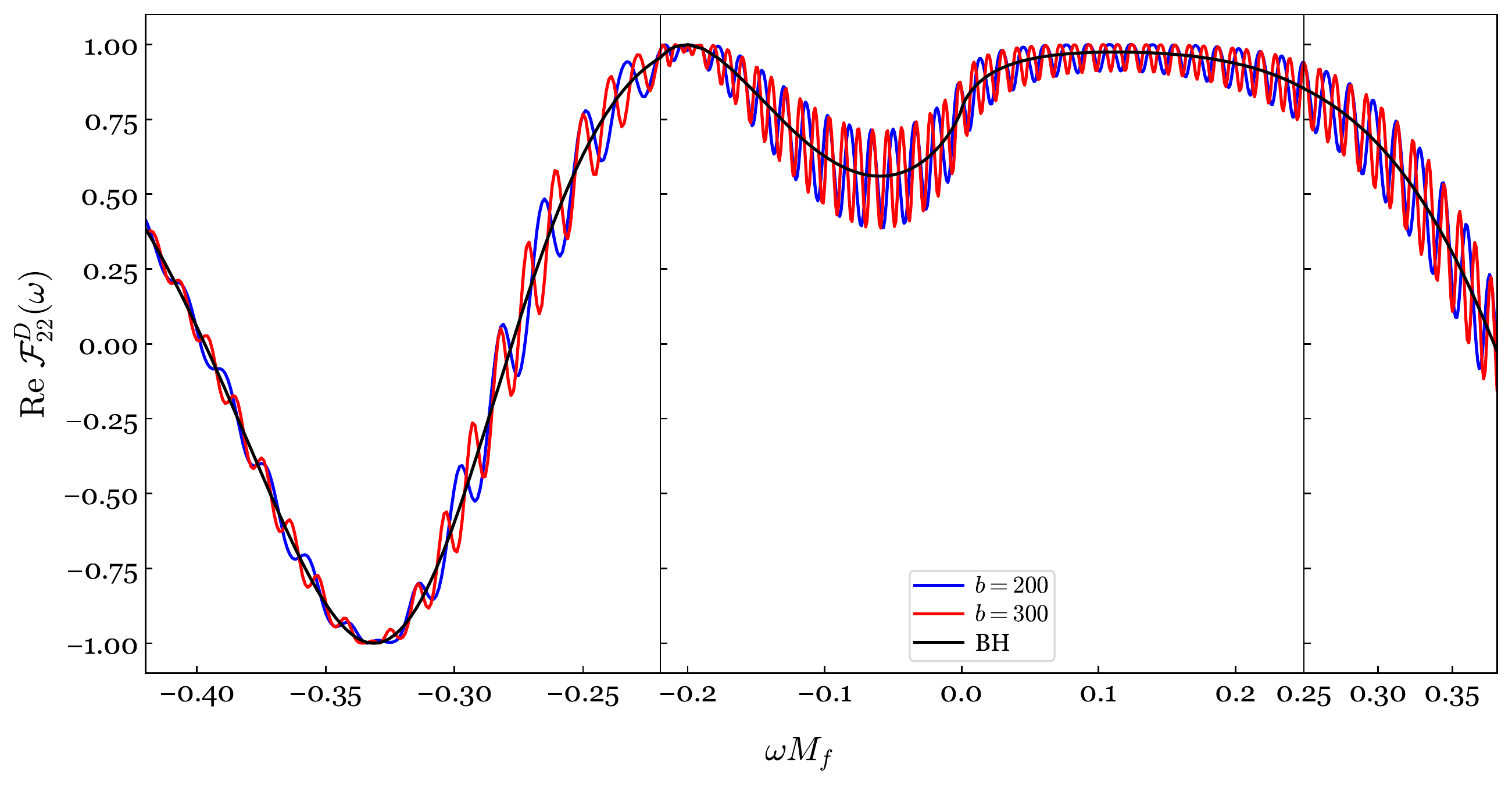}}\\
\subfloat[$b=200M_f$]{\includegraphics[width=1.0\textwidth]{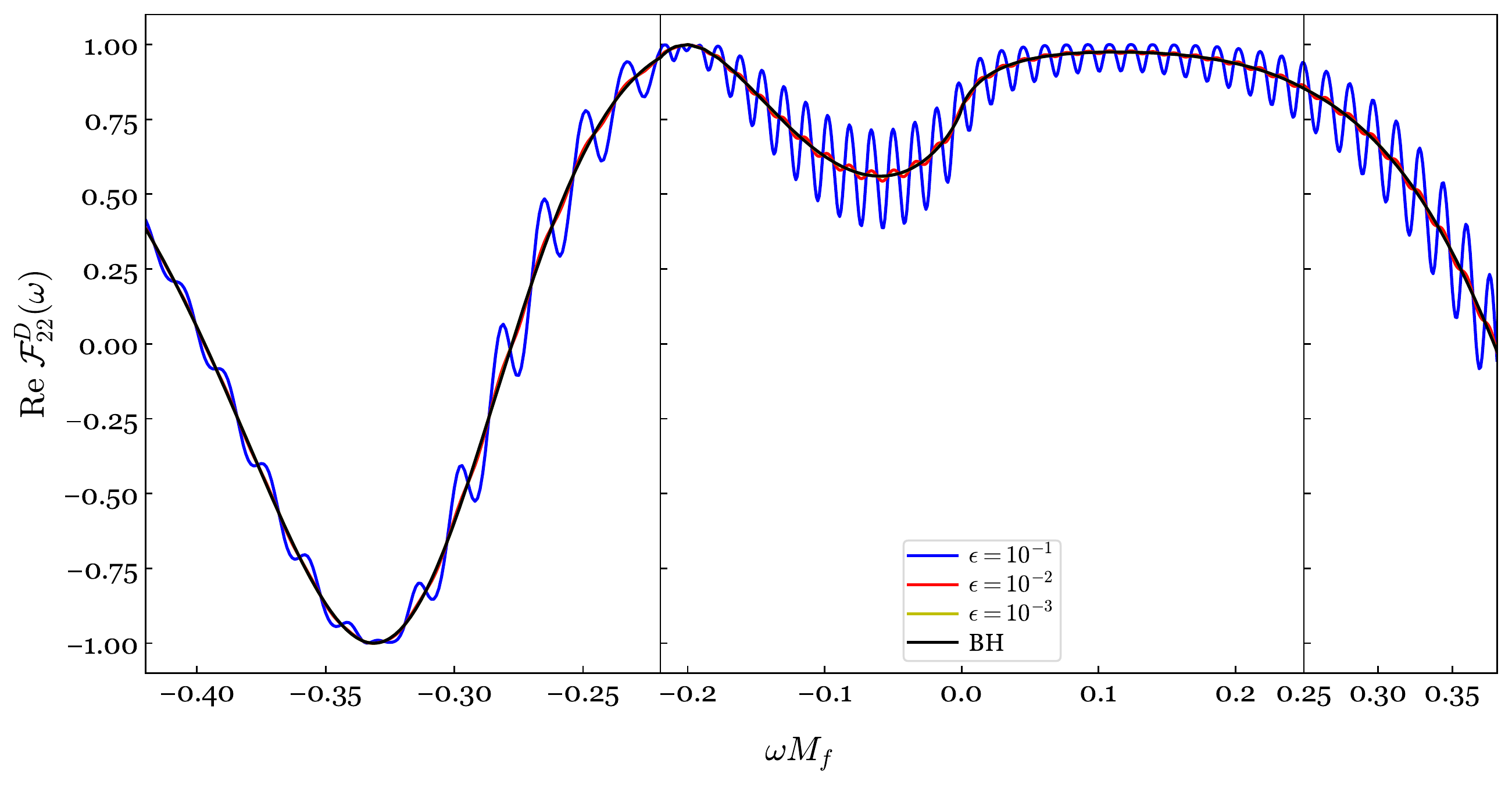}}
  \caption{Same as Fig.~\ref{fig:wronskian_freq}, but for a spinning ECO with $\chi_f=0.692$.}
 \label{fig:kerr_wronskian_freq}
\end{figure*}

Figures \ref{fig:kerr_wronskian} and \ref{fig:kerr_wronskian_freq} show the filter $\mathcal{F}^{D~{\rm ECO}}_{22}$ in the time and frequency domain. The spin of the ECO is $\chi_f=0.692$. It has the same qualitative features as that of the nonspinning ECO (Fig.~\ref{fig:wronskian}). 

\section{$M_f$ and RSS}
\label{sec:remnant_properties_mass}
\begin{figure*}[htb]
        \includegraphics[width=\columnwidth,clip=true]{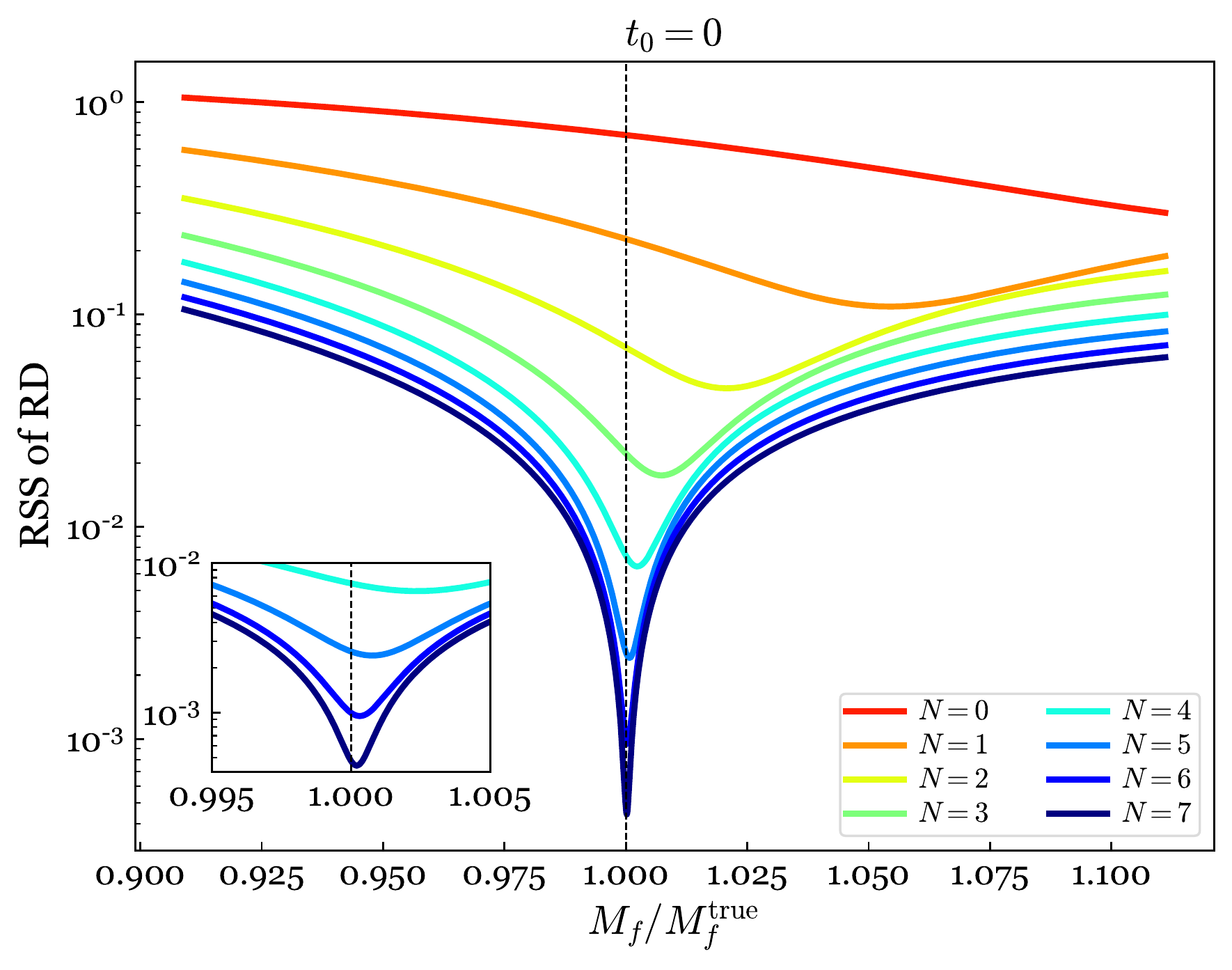}
        \includegraphics[width=\columnwidth,clip=true]{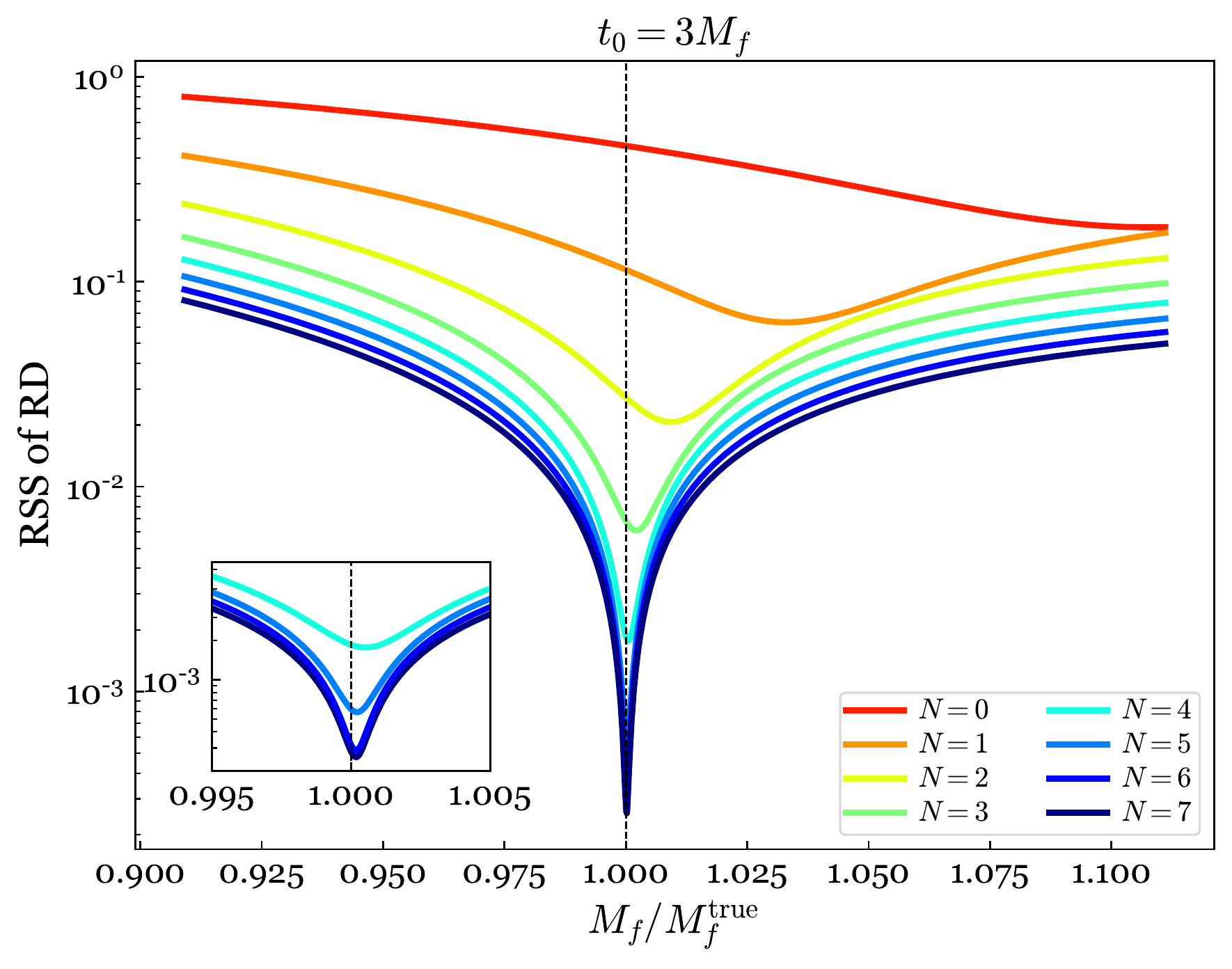}
        \includegraphics[width=\columnwidth,clip=true]{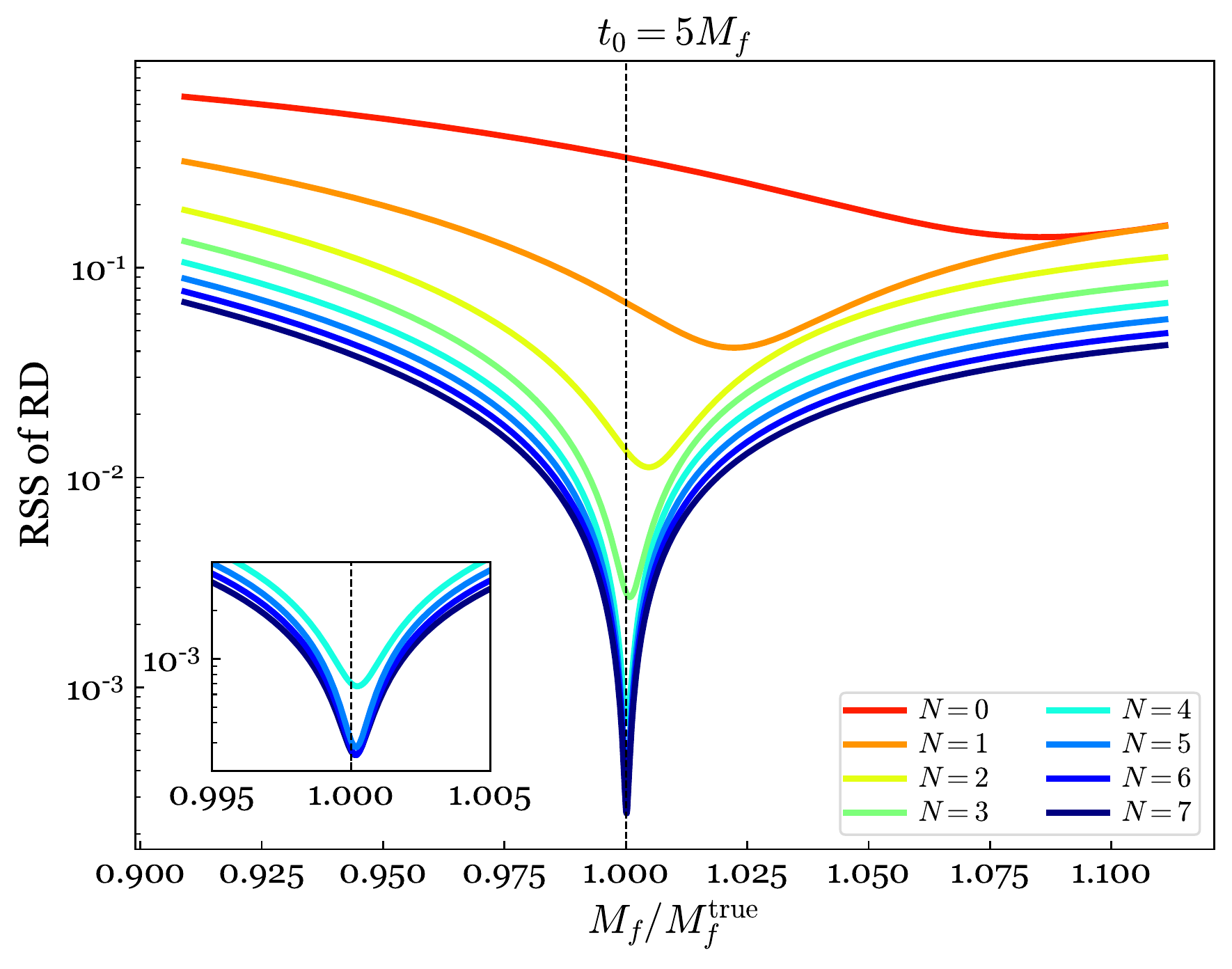}
        \includegraphics[width=\columnwidth,clip=true]{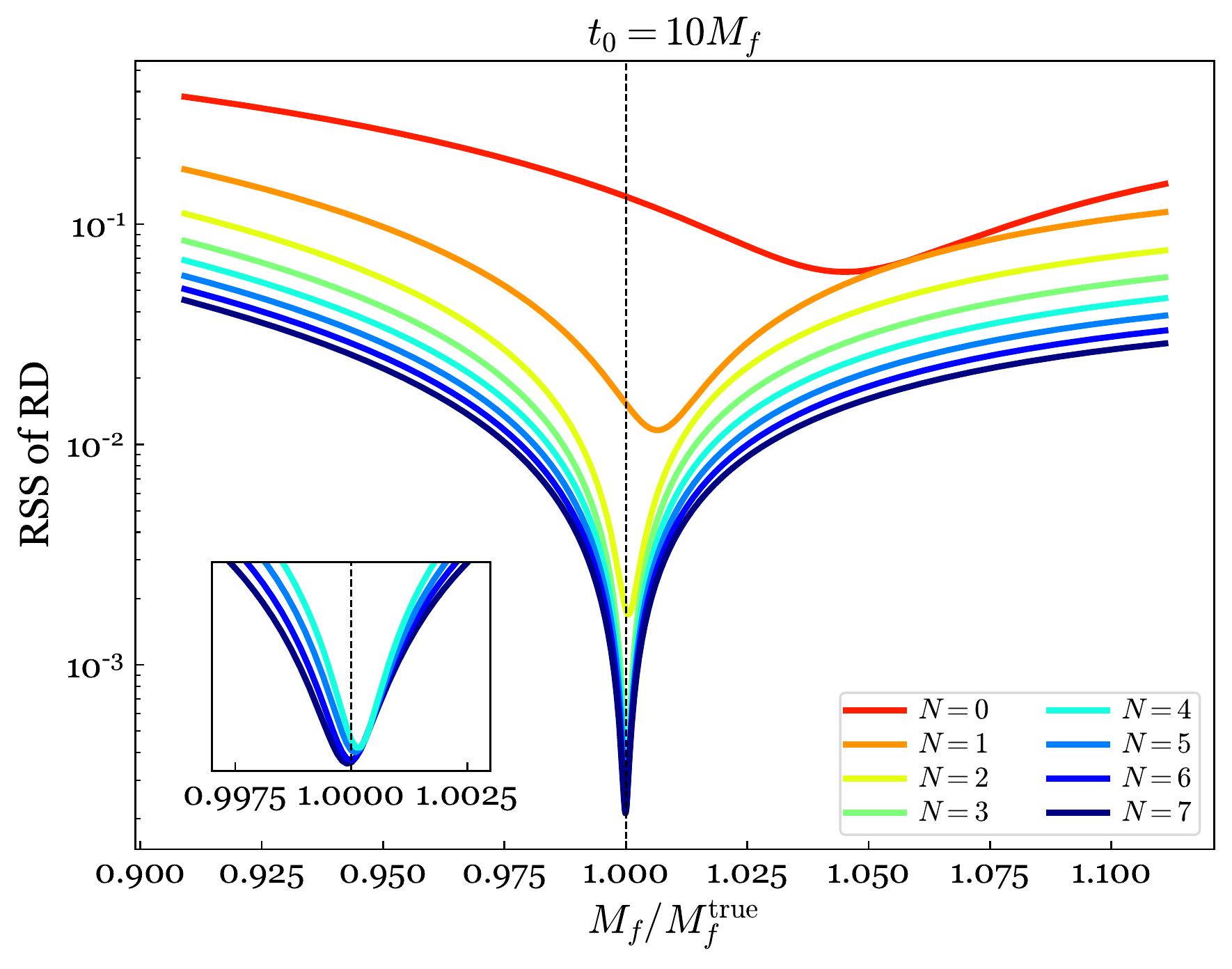}
        \includegraphics[width=\columnwidth,clip=true]{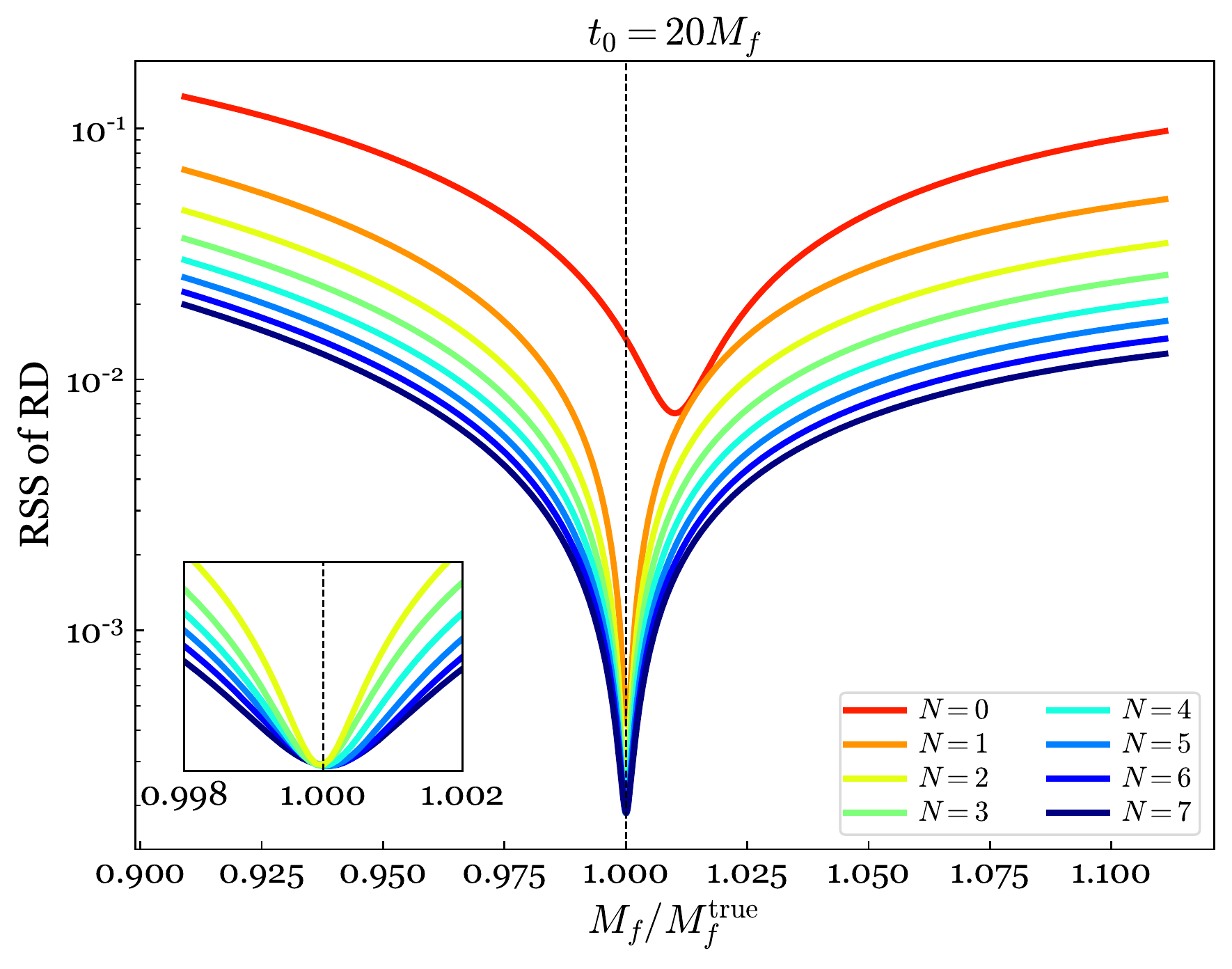}
        \includegraphics[width=\columnwidth,clip=true]{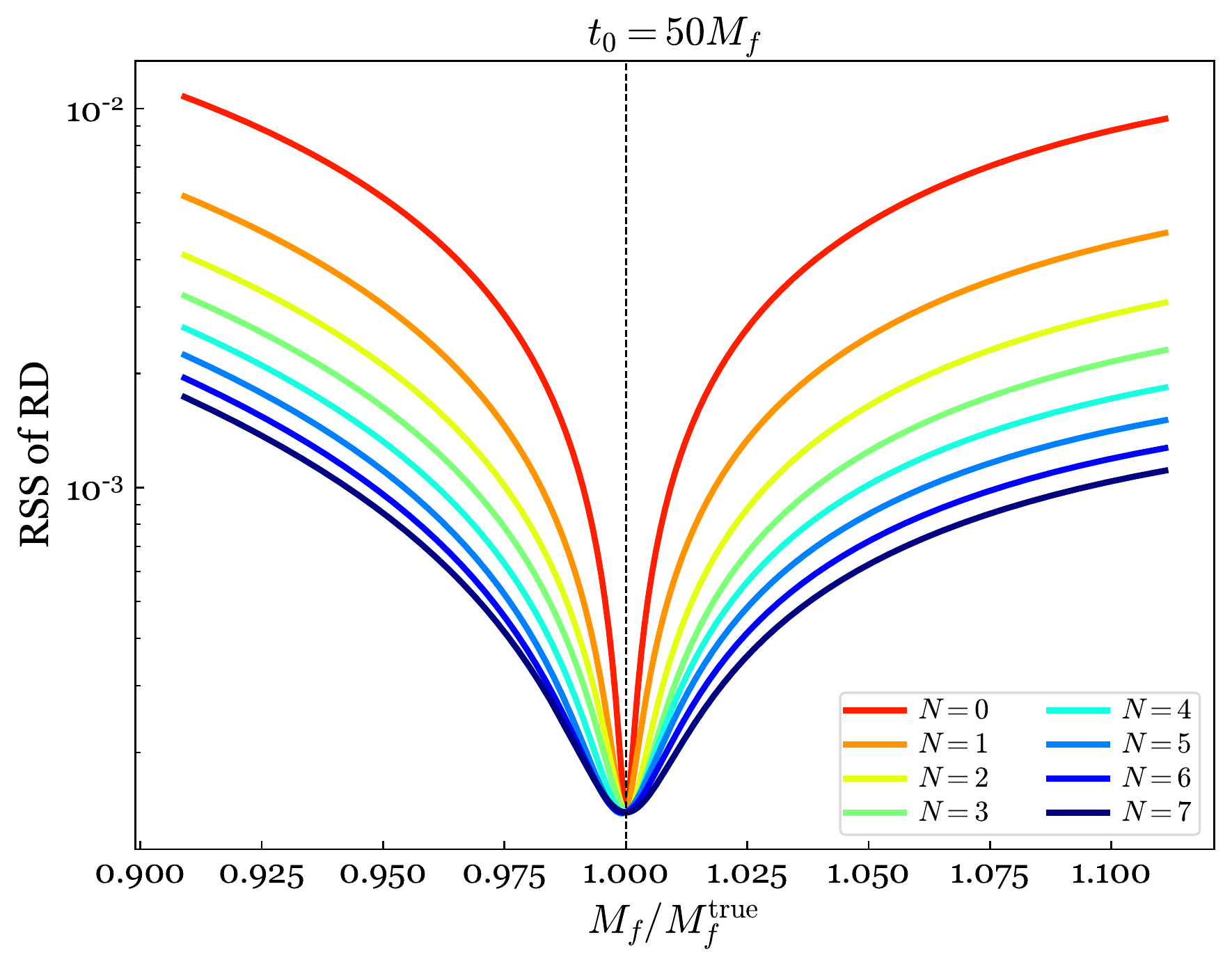}
  \caption{Same as Fig.~\ref{fig:remnant_properties_spin}, but with varying $M_f$ and fixed $\chi_f$.}
 \label{fig:remnant_properties_mass}
\end{figure*}

In Fig.~\ref{fig:remnant_properties_mass}, we plot the ringdown RSS of the filtered waveform as a function of the remnant mass $M_f$, using waveform \ind.

\clearpage
%%%%%%%%%%%%%%%%%%%%%%%%%%%%%%%%%%%%%%%%%%%%%%%%%%%%%%%%%%%%%%%%%%%%%%%%%%%%%%%
\def\bibsection{\section*{References}}
%%%%%%%%%%%%%%%%%%%%%%%%%%%%%%%%%%%%%%%%%%%%%%%%%%%%%%%%%%%%%%%%%%%%%%%%%%%%%%%
\bibliography{References}
\end{document}